\documentclass[onecolumn,10pt]{IEEEtran}
\usepackage[utf8]{inputenc}
\usepackage{hyperref}
\usepackage{cleveref}
\usepackage{amsthm, amssymb,amsmath}
\usepackage{mathtools}
\usepackage{pgfplots}
\usepackage{todonotes}
\usepackage{dsfont}
\usepackage{eurosym}
\usepackage[caption=false, font=normalsize, labelfont=sf, textfont=sf]{subfig}
\newcommand{\dd}{\ensuremath{\,\mathrm{d}}}
\newcommand{\R}{\ensuremath{\mathbb{R}}}

\newcommand{\rr}{\text{r}}
\newcommand{\nr}{\text{nr}}
\newtheorem*{proofsketch}{Sketch of proof}
\title{Integration of Information Patterns in the Modeling and Design of Mobility Management Services}
\author{Alexander~Keimer, Nicolas~Laurent-Brouty, Farhad~Farokhi, Hippolyte Signargout,\\ Vladimir~Cvetkovic, Alexandre~M. Bayen, Karl~H.~Johansson%
\thanks{A. Keimer and A. Bayen are with the Institute of Transportation Studies, University of California, Berkeley, CA, USA. Emails: \{keimer,bayen\}@berkeley.edu}%
\thanks{N. Laurent-Brouty is with Universit\'e C\^ote d’Azur, Inria, CNRS, LJAD and Ecole des Ponts ParisTech, Champs-sur-Marne, France. Email: nicolas.laurent-brouty@inria.fr}%
\thanks{F. Farokhi is with the Department of Electrical and Electronic Engineering, University of Melbourne, Parkville, Victoria, Australia. Email: ffarokhi@unimelb.edu.au}%
\thanks{H. Signargout is with \'Ecole Normale Sup\'erieure de Lyon, Lyon, France. Email: hippolyte.signargout@ens-lyon.fr}%
\thanks{V. Cvetkovic is with the School of Architecture and the Built Environment, KTH Royal Institute of Technology, Stockholm, Sweden. Email: vdc@kth.se}%
\thanks{K. H. Johansson is with the School of Electrical Engineering, KTH Royal Institute of Technology, Stockholm, Sweden. Email: kallej@kth.se}
}

\date{\today}
\newtheorem{Definition}{Definition}

\newtheorem{remark}{Remark}
\newtheorem{example}{Example}
\newtheorem{theorem}{Theorem}
\newtheorem{challenge}{Challenge}
\newtheorem{prop}{Proposition}

\DeclareMathOperator{\modd}{mod}
\pgfplotsset{compat = 1.14}
\begin{document}

\maketitle
\begin{abstract}
The development of sustainable transportation
infrastructure for people and goods, using new technology and business models can prove beneficial or detrimental for mobility, depending on its design and use. The focus of this article is on the increasing impact new mobility services
have on traffic patterns and transportation efficiency in general. 
Over the last decade, the rise of the mobile internet and the usage
of mobile devices has enabled ubiquitous traffic information. With the increased adoption of specific smartphone applications,
the number of users of routing applications has become large enough to 
disrupt traffic flow patterns in a significant manner. Similarly,
but at a slightly slower pace, novel services for freight transportation
and city logistics improve the efficiency of goods transportation and change the use of road infrastructure. 

The present article provides a general four-layer framework for modeling these new trends. The main motivation behind the development is to provide a unifying formal system description that can at the same time encompass system physics (flow and motion of vehicles) as well as  coordination strategies under various information and cooperation structures. To showcase the framework, we apply it to the specific challenge of modeling and analyzing the integration of routing applications in today's transportation systems. In this framework, at the lowest layer (flow dynamics) we distinguish app users from non-app users. A distributed parameter model based on a non-local partial differential equation is introduced and analyzed. 

The second layer incorporates connected services (e.g., routing) and
other applications used to optimize the local performance of the system. As inputs to those applications, we propose a third layer introducing the
incentive design and global objectives, which are typically varying
over the day depending on road and weather conditions, external events etc. The
high-level planning is handled on the fourth layer taking social
long-term objectives into account.

We illustrate the framework by considering its ability to model at two different levels. Specific to vehicular traffic, numerical examples enable us to demonstrate the links between the traffic network layer and the routing decision layer. With a second example on optimized freight transport, we then discuss the links between the cooperative control layer and the lower layers. The congestion pricing in Stockholm is used to illustrate how also the social planning layer can be incorporated in future mobility services. 
\end{abstract}

\section{Introduction}\label{sec:intro}
Traffic congestion is increasing at alarming rates in cities
worldwide~\cite{schrank,itf,ec}. 
%
%
Computing, communication, and
sensing technologies are transforming the transportation infrastructure and
have enabled the engineering community to provide new services leveraging vehicular and
information technologies~\cite{proc_ieee07}. 

\subsection{Motivation}
One of the most un-anticipated impact of information technology on the transportation system has come from routing services
through in-vehicle navigation devices or smartphones. Coverage of
the road network by these ``apps'' has expanded dramatically~\cite{herrera}, leading to massive adoption of services like
Google (with Waze) and Apple~\cite{buczkowski}. 
These apps also being massively used by Mobility-as-a-Service
companies (like Uber and Lyft), accelerate the phenomenon.
The rise of ubiquitous transportation information available to both the public and companies has disrupted numerous suburban areas, leading to various unexpected outcomes. 

\begin{figure}[t]
\centering
\begin{minipage}{0.475\textwidth}
\begin{tikzpicture}
\node[inner sep=0pt] (A) at (0,0){
\includegraphics[width=1\textwidth]{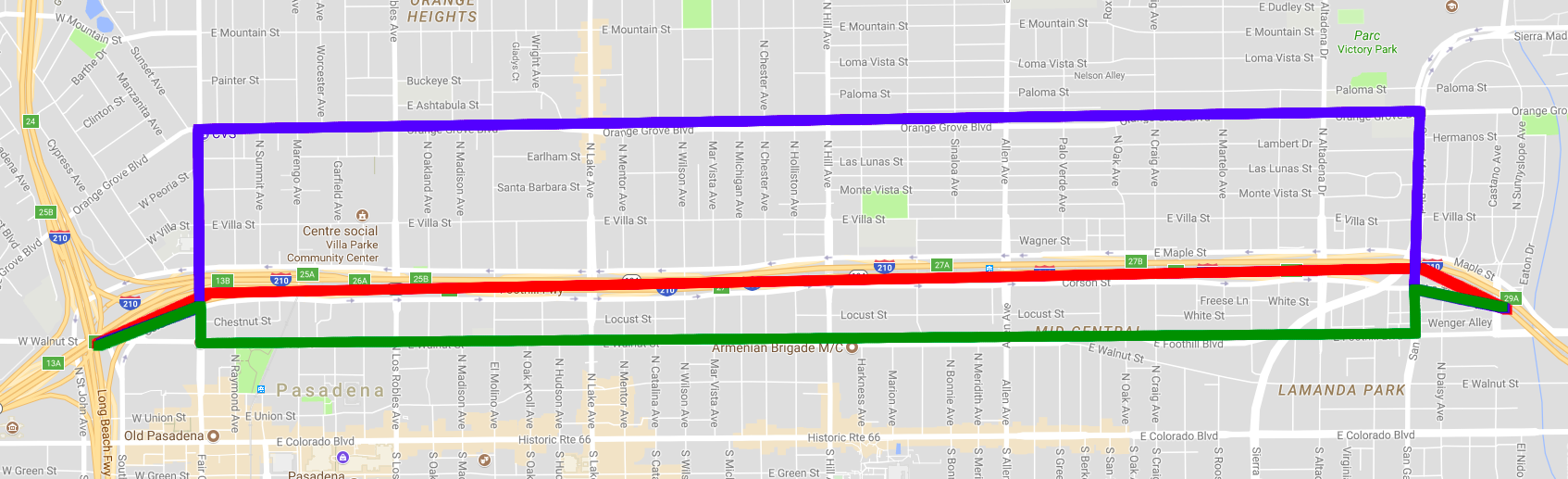}};
\draw (0,0) node {\textbf{I-210}};
\draw (0,1) node {\textbf{Deviation route 1}};
\draw (0,-0.85) node {\textbf{Deviation route 2}};
\end{tikzpicture}
\end{minipage}
\begin{minipage}{0.475\textwidth}
\begin{tikzpicture}
\node[inner sep=0pt] (A) at (0,0){
\includegraphics[width=1\textwidth]{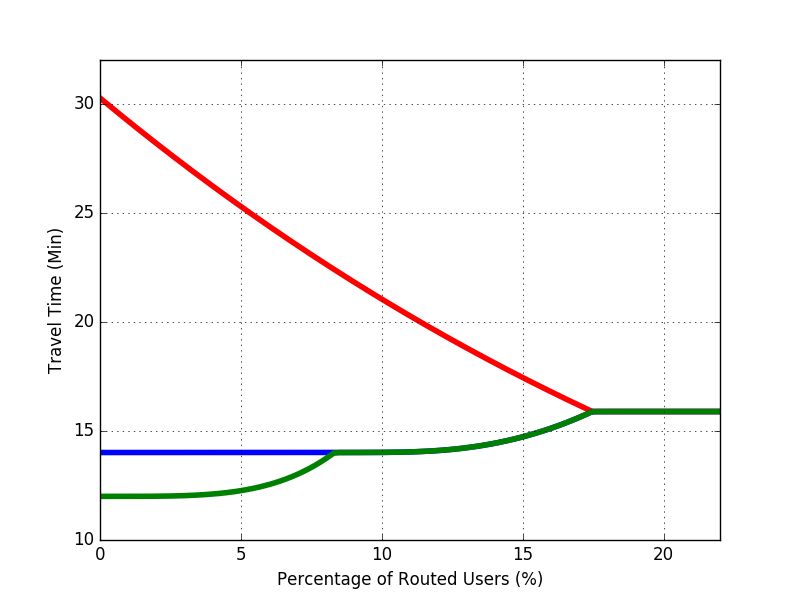}};
\draw (0,0.35) node {\textbf{I-210}};
\draw (-1.5,-1.25) node {\textbf{Deviation route 1}};
\draw (0.5,-2) node {\textbf{Deviation route 2}};
\end{tikzpicture}
\end{minipage}
\caption{Motivating example to illustrate progressive steering of traffic flows to a Nash equilibrium with increased app usage. \textbf{Left:} City of Pasadena with I-210 (red) and north (blue) and south (red) arterial roads. \textbf{Right:} Travel time over I-210 and arterial roads as function of percentage getting rerouted to north and south arterials. Demand corresponds to $27\,500\,\text{veh/h}$.}
\label{f:NashConvergence}
\end{figure}
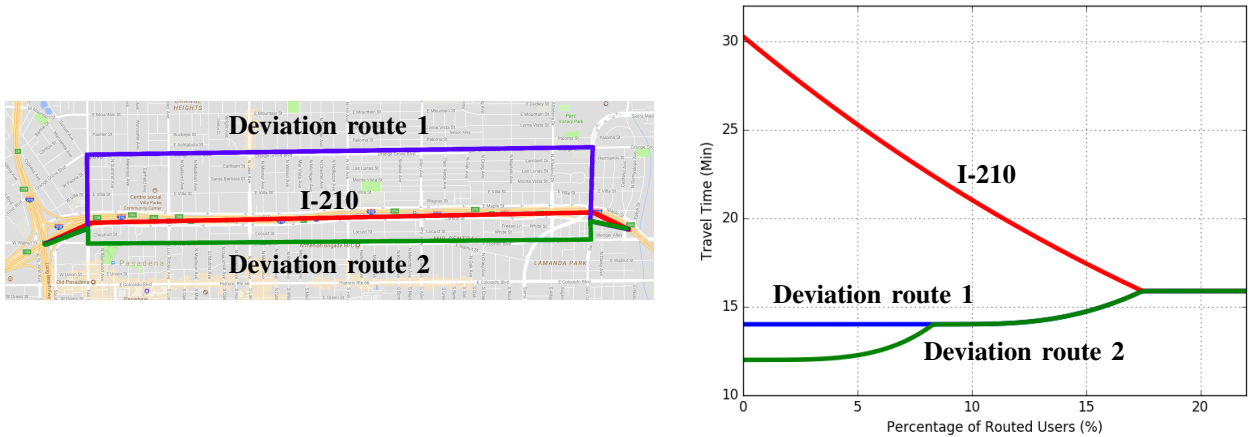

Congestion patterns that never existed before have emerged. The displacement of traffic flows is a major motivation for this article, and is illustrated by a simple motivational example in Figure~\ref{f:NashConvergence}. In this example, as shown in the left figure, we study three parallel paths possible for motorists to drive along I-210 in the LA basin: on the I-210 freeway, or on one of two parallel arterials commonly used for detours. The simulation model relies on an extension of the \textit{user equilibrium}~\cite{patriksson} to multi-class flows integrating users enabled by routing apps \cite{thai2016negative}. The figure shows the convergence of travel time to a single value as the proportion of motorists using the routing app grows. At 0\% usage (representative of the situation around 2005 when no routing information was available), most of the traffic would stay on the freeway, leading to high freeway travel time and low arterial travel times. As the proportion of app usage increases, travel times get ``equalized'' among possible routes as more traffic is diverted onto the arterial roads, leading to a {Nash} equilibrium at around 17\% of app usage. 
In general, the increased adoption of apps leads to the growth of the  number of vehicles progressively routed
outside of the freeways through so-called traffic ``shortcuts.'' While this might in some cases decongest the freeways, it contributes to transferring flow to arterial roads, which are less efficient in processing traffic due to urban infrastructure (lights, stop signs, etc.), never specifically designed for such traffic flow. Because individual
motorists are essentially given a ``selfish'' route (i.e., their own
shortest path) by the routing service, the process is progressively steering the system
towards an equilibrium that might be a Nash equilibrium, but not
efficient from a social viewpoint. This phenomenon is commonly observed in suburban areas in the US, and frequently appears in the news~\cite{geha,hendrix,thornton,associated}. It presents a key motivation of this article, as the design of novel mobility services needs to systematically integrate traffic flow dynamics with decisions made by individual users as well as higher-level forms of resource allocations and cooperations. 

\subsection{Contributions}
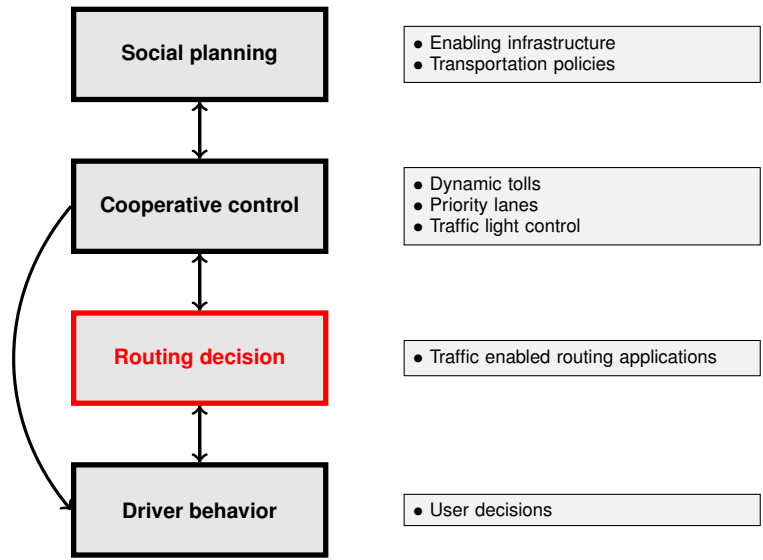
\begin{figure}
\centering
\begin{tikzpicture}
[node distance = 1cm, auto,font=\footnotesize,
every node/.style={node distance=2cm},
comment/.style={draw, fill=black!5, text width=4.5cm, node distance=0.25cm, font=\scriptsize\sffamily},
force/.style={rectangle,line width=2pt, draw, fill=black!10, inner sep=5pt, text width=3cm, text badly centered, minimum height=1.2cm, font=\bfseries\footnotesize\sffamily}] 
\node [force] (cooperative) {Cooperative control};
\node [force, above=0.75cm of cooperative] (social) {Social planning};
\node [color=red,force, below =0.75cm of cooperative] (selfish) {Routing decision};
\node [force, below =0.75cm of selfish] (traffic) {Driver behavior};

\node [comment,right=1cm of social] (comment-com) {$\bullet$ Enabling infrastructure\\
$\bullet$ Transportation policies};
\node [comment, right=1cm of cooperative] (cooperative-com) {$\bullet$ Dynamic tolls\\
$\bullet$ Priority lanes\\
$\bullet$ Traffic light control};

\node [comment, right=1cm of selfish] (selfish-com)
{$\bullet$ Traffic enabled routing applications};
\draw[very thick,->] (cooperative.west) to [bend right=40] (traffic.west);
\node [comment, right=1cm of traffic] (traffic-com)
{$\bullet$ User decisions};
\path[->,very thick] 
(social) edge (cooperative)
(cooperative) edge (social)
(cooperative) edge (selfish)
(selfish) edge (cooperative)
(selfish) edge (traffic)
(traffic) edge (selfish);
\end{tikzpicture} 
\caption{Considered four-layer decision framework. The red ``Routing decision'' layer has drastically changed traffic behavior in some major US cities over the last years. Systematically integrating cooperative control and social planning into the design of mobility services has the potential of improving traffic behavior in the future.}
\label{fig:app-usage}
\label{fig:framework}
\end{figure}

The contributions of this article are structured around the four-layer decision diagram in \Cref{fig:framework}. The top layer ``Social planning'' represents the implementation of transportation policies and design of incentives introduced by enabling infrastructure systems, such as the introduction of novel traffic management. 
``Cooperative control'' aims at the actual control of traffic for the social benefit of everyone. It uses control mechanisms like dynamic tolls, priority lanes, or traffic light control. This layer has a direct influence on the actual traffic situation, while ``Social planning'' implements long-term strategies. The layer ``Routing decisions'' describes the usage of routing apps and other mobility management services. They offer individuals more information on the actual status of traffic. This information can be used to route, and to reroute if necessary, and is thus influencing the ``Driver behavior'', which represents the decisions taken by individual users. 

A key contribution of this article is focused on the ``Routing decision'', indicated red in \Cref{fig:framework}. The currently observed phenomenon of non-cooperation among the users of the same app is due to this layer. 
All users are given a selfish ``shortest route'', basically an enhanced version of shortest path~\cite{skiena,hart1968}. Note that mobility services that provide these routes today do not cooperate among each other. 
Selfish routing algorithms are known to not provide a solution to congestion in general and to not lead to the convergence to social optimum in neither the static case~\cite{nash1950,patriksson,wardrop1952road,friesz1993variational}, nor in specific time-dependent dynamic cases~\cite{nash1951,krichene}.
To tackle this problem, we make several contributions. (1) At the lower level, we propose a mathematical model to describe the flow dynamics on the traffic network. A novel use of non-local terms in a transport partial differential equation (PDE) makes it for instance possible to link flow models and freight transport. This new model demonstrates the applicability of our framework.  (2) It is shown how cooperation can be integrated into the mathematical model. Consequently, as for any new model, it becomes necessary to prove existence and uniqueness of the solutions to the model. The complexity of the proofs increases with the ability for the model to incorporate specific features, such as information-aware routing, and, of course, flow characteristics. 
Several fundamental scientific questions must be addressed to understand the issue of incorporating information patterns in the design of mobility management services. Some of the key issues are addressed in this article. (3) We consider the modeling challenge of incorporating both users of routing applications and those who do not use these applications into the same non-local PDE framework on a network. 
(4) We present different approaches on how decisions of routed and non-routed drivers are made. Thereby, those decisions are described in an abstract and broadly applicable framework.
(5) To incorporate these approaches in a given network model, we introduce time dependent split functions at the junctions of the network which represent the routing. For every choice of splits, we prove existence and uniqueness of a solution on the network.
Then, these results allow to interpret the splits as a function dependent on the load
of parts of the network, network scenario, congestion of outgoing roads, and actual travel time. This set of considerations can be
expanded further as the model matures. 
With the time dynamics included, the model allows for reliable forecasting, making route choices more reasonable and providing a more precise description of reality.
(6) Another contribution in this article is on describing an actual mobility service utilizing the ``Cooperative control'' layer of Figure~\ref{fig:framework}. In particular, as partial or full automation of vehicular and freight traffic provides significant opportunities for optimization of transportation efficiency, we consider a service based on collaborative routing algorithms and platooning of
heavy-duty vehicles, which for instance can be exposed to freight signal prioritization at
traffic intersections. 
We show that by integrating traffic information with private fleet management data, it is possible to optimize the overall system operation in real-time, following a recent proposal for automated coordination of truck platoons for long-haulage transport~\cite{bess+16procieee}. It is also shown that such services can be linked locally to an optimal control problem defined for the introduced continuous traffic model, or over larger geographical areas to a discrete model of a higher level of abstraction. 
(7) As a final contribution of the article, we discuss how several mechanisms exist to promote cooperation opportunities in an urban setting, emphasizing the need of the systematic integration of the ``Social planning'' layer. Road tolling, congestion pricing, and incentivization through rewards are examples of such mechanisms. The discussion focuses on the current and future cordon pricing system in Stockholm. 



\subsection{Related work}
There is a significant amount of work on modeling transportation networks. It is impossible in an article like the present one to review the entire literature on this topic. Thus, we chose to only mention classes of models with a few references for each, to give a sense of where we fit in the landscape of published work. Specific to the problems of interest, one can classify the published work in four categories:
\begin{enumerate}
\item \underline{Micro-simulation models}. This framework enables the description of vehicle by vehicle dynamics at scales which can include second by second movement. With such a framework, embodied by numerous commercial software tools such as Aimsun~\cite{aimsun}, VISSIM~\cite{vissim}, not much analytical work can be done to characterize the problems described in the present algorithm. However, heuristic and experimental approaches are commonly used by practitioners for simulation purposes, see in particular~\cite{chiu} for an overview.
\item \underline{Meso-simulation models}. This framework enables the description of platoons of vehicles through a network, while conserving some level of description of their individual dynamics. While analytical approaches to this framework are possible, they are more involved, see in particular~\cite{tascikaraoglu} and references therein for an overview.
\item \underline{Macroscopic models}. This approach has traditionally been at the core of the transportation engineering community, since the seminal work of Lighthill-Whitham-Richards (LWR)~\cite{lighthill1955kinematic,richards1956shock}. Hundreds of articles were written on these models, their application to networks, and their discretization. Many of these formulations lend themselves to theory, which namely consists in (1) existence and uniqueness proofs for solutions of their equations (as in the present article), (2) controllability, optimality, observability results when problems are well-posed, (3) proper discretizations with proofs of convergence, with the most notable scheme written by Godunov in 1957~\cite{godunov}. 
\item \underline{Operations research models}. Further discretization of network models at an aggregate level can be broadly characterized as operations research models, and encompass queuing networks, delay networks, Jackson networks, and many specific models developed for particular applications, in particular the seminal Merchant-Nemhauser model~\cite{merchant1978model}. These models have very broad scopes of application which range from air traffic control to supply chains (and  traffic in particular). 
\end{enumerate}
The work presented in this article falls mainly in the category of macroscopic models above. For a complete review of this class of models, the reader is referred to the seminal book by Garavello and Piccoli~\cite{garavello2016}. Note that while this framework enables elegant treatment of aggregate flows, some of the issues linked to dynamic traffic assignment for these flow models are still unresolved and open (as will appear in the way we enunciate these problems in our approach). 
Although our framework is different from most existing macroscopic models in several aspects (e.g., multi-class flow and time-dependent routing at junctions), existing traffic network modeling approaches inspired our work significantly. The specific flow model we consider first in this article was introduced in~\cite{armbruster2}. Mathematical properties of simplified versions or archetypes on bounded domains were studied in \cite{wang,shang}. An extension to multi-class frameworks was presented in \cite{keimer2}. In \cite{keimer3} and \cite{goatin2}, non-local conservation laws on unbounded domains were considered.
For an introduction to general traffic flow models, we refer the reader to the monograph \cite{garavello2006traffic}, which gives an exhaustive overview of traffic modeling using networks of PDE's (mostly extensions of the LWR model). Note that for the four categories of models presented above, also routing choices have been the topic of a vast amount of literature. For example, routing behavior has been modeled through a variety of so called logit functions, e.g., \cite{Ben-Akiva1999,mcfadden1978modeling,burrell1968multiple,Manski1977}.

The integration of mobility services and app usage with detailed traffic flow models is a new research area and thus has not been thoroughly studied. The advantage of taking such a \emph{cross-layer} modeling approach was recently explored for control and coordination of a large fleet of heavy-duty vehicles that exploits the benefits of vehicle platooning~\cite{bess+16procieee}. The development of this specific freight transport service was motivated by the concept of an automated highway system~\cite{250509,871301}, in which vehicles are organized in platoons to increase traffic flow under strict safety guarantees. Vehicle platooning is widely being considered as an important automated vehicle  technology, see~\cite{546270} for an overview.  

Finally, as should appear with the broad scope of the four categories above, the variety of models available for this type of problems is significant. The reason why we chose to focus our approach on a new traffic model, was (1) to show the generality of the overall framework, and (2) to demonstrate that the complexity of the model (i.e., non-local conservation laws) can be handled in the framework.

\subsection{Outline}
The outline of this article is as follows. A general framework for cooperative transportation systems was presented earlier in the decision diagram in \Cref{fig:framework}. It relies on a detailed traffic flow model, which is introduced in 
\Cref{subsec:traffic_network}. How to integrate information patterns from routed and non-routed traffic flows is also presented. The determination and implementation of actual routing policies are given in \Cref{sec:routing}, both for local and global routing information. \Cref{sec:Freight} presents a mobility service for cooperative freight transportation based on these models. The section discusses how traffic flow patterns can be controlled over individual links, for instance, for optimizing opportunities for vehicle platooning. Then it is described how such local control can be combined with global coordination to form a cooperative freight transportation system, even under given data privacy guarantees. Existing and evolving cordon pricing strategies in Stockholm is briefly presented in \Cref{sec:case}. Finally, 
\Cref{sec:concl} gives the conclusions and outlines a few items for future work.

\section{Modeling traffic on networks}\label{subsec:traffic_network}
This section builds on traditional network notations for flow networks, and uses the resulting framework to later build an entire description of the PDE flow models used throughout the article. The goal of this model, beyond its intrinsic use, is to demonstrate the generality of the presented framework.
\subsection{Network and dynamics on the network}
We define a network as follows. 
\begin{Definition}[Network]\label{defi:network_notation}
 We call a graph $(\mathcal{V},\mathcal{A})$ with $\mathcal{V}\subset\mathbb{N}_{\geq1}$ the set of nodes and $\mathcal{A}\subseteq\mathcal{V}\times\mathcal{V}$ the set of links a network, if the graph is directed and connected.
 In addition, for $v\in\mathcal{V}$ we define
 \begin{align*}
  \mathcal{A}_{\text{in}}(v)&:=\{a\in\mathcal{A}: a\text{ ends in } v\}\\
  \mathcal{A}_{\text{out}}(v)&:=\{a\in\mathcal{A}: a\text{ originates in } v\}.
 \end{align*}
\end{Definition}
Equipped with this definition, we can now define the fundamental notation for any dynamics on the network. We do this in a general way, without explicitly defining the traffic dynamics, since our approach is applicable to any time-dependent traffic flow.

\begin{figure}
\begin{center}
\begin{tikzpicture}[scale=1.5]
\path (0,0) node(A) [draw=none]{$u_{a_{1}}(t)$}
      (2,0) node(B) [circle,draw]{$v$}
      (4,0.75) node(C) [draw=none]{$y_{a_{2}}(t)$}
      (4,-0.75) node(D) [draw=none]{$y_{a_{3}}(t)$};
\draw[very thick,->] (A) -- (B) node[below,pos=0.5]{$a_{1}$} node[ above,pos=0.5]{$\rho_{a_{1}}(t)$};
\draw[very thick,->] (B) -- (C) node[below,pos=0.5]{$a_{2}$}
node[above,pos=0.5]{$\rho_{a_{2}}(t)$};
\draw[very thick,->] (B) -- (D) node[below,pos=0.5]{$a_{3}$}
node[above,pos=0.5]{$\rho_{a_{3}}(t)$};
\draw[very thick, dotted,->] (1.65,-0.5)--(1.7,-0.1);
\draw (1.75,-0.5) node[below left]{$y_{a_{1}}(t)$};
\draw[very thick, dotted,->] (2.5,-1)--(2.25,-0.1);
\draw (2.5,-1) node[below right]{$u_{a_{3}}(t)=\theta^{v}_{a_{3}}(t)y_{a_{1}}(t)+s^{v}_{a_{3}}(t)$};
\draw[very thick,dotted,->] (2.5,1)--(2.25,0.1);
\draw (2.5,1) node[above right]{$u_{a_{2}}(t)=\theta^{v}_{a_{2}}(t)y_{a_{1}}(t)+s^{v}_{a_{2}}(t)$};
\end{tikzpicture}
\end{center}
\caption{Illustration of traffic network dynamics at the level of node $v\in\mathcal{V}$.}
\label{fig:network_explanations}
\end{figure}
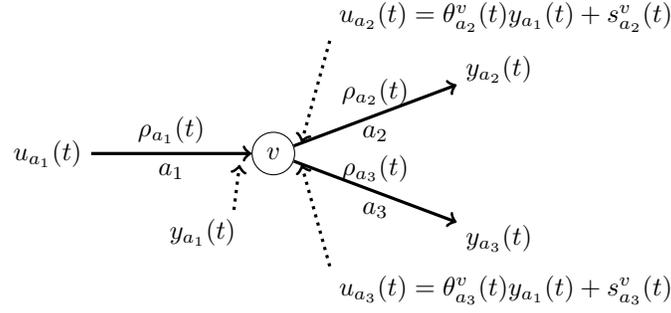

Consider \Cref{fig:network_explanations} as an illustration for the notation around a single node. 
Let $T\in\mathbb{R}_{>0}$ be a finite time horizon.
On every link $a\in\mathcal{A}$ we introduce time- and space-dependent traffic dynamics $\rho_{a}:(0,T)\times(0,1)\rightarrow\R_{\geq0}$. Thereby, $\rho_{a}(t,x)$ can be interpreted as the density of vehicles at position $x\in(0,1)$ at time $t\in(0,T)$ on link $a\in\mathcal{A}$. We have thus assumed, without loss of generality, that every link is scaled to unit length. 
\begin{Definition}[Inflow and outflow of a link]\label{defi:in_out}
 For every link $a\in\mathcal{A}$ in the network, we call $u_{a}:(0,T)\rightarrow\R_{\geq0}$ the flow of vehicles entering link $a$ and $y_{a}:(0,T)\rightarrow\R_{\geq0}$ the flow of vehicles exiting the link.
\end{Definition}
To describe the topology of the network and to implement a possible routing at junctions, we define time-dependent splits over the entire network. 
\begin{Definition}[Set of splits $\boldsymbol{\Theta}$]
Consider a network as described above. Let the set of possible splits at the junctions be denoted
\begin{equation}
 \boldsymbol{\Theta}:=\left\{\theta_{a}^{v}\in L^{\infty}((0,T);\R_{\geq0}): \sum_{\tilde{a}\in \mathcal{A}_{\text{out}}(v)}\theta_{\tilde{a}}^{v}(t)=1 ,\ v\in\mathcal{V}, a\in\mathcal{A}_{\text{out}}(v), \ t\in(0,T) \text{ a.e.}\right\},\label{defi:btheta}
\end{equation}
where $\boldsymbol{\theta}:=\left(\theta_{a}^{v}\right)_{v\in\mathcal{V},a\in\mathcal{A}_{\text{out}}}$ and  $\theta_{a}^{v}:(0,T)\rightarrow\mathbb{R}$ represents the ratio of the flow, entering node $v$ and leaving for link $a\in\mathcal{A}_{\text{out}}(v)$.
\end{Definition}
Hence, the set $\boldsymbol{\Theta}$ represents the routing in the network and guarantees conservation of flow over the nodes.
\begin{remark}
 In case that $|\mathcal{A}_{\text{out}}(v)|=1$ for a $v\in\mathcal{V}$, there is obviously no routing necessary, which also follows from the definition of $\boldsymbol{\Theta}$. 
\end{remark}

We next define the sources in the network, i.e., the flow entering the network. For notational simplicity, we assume to start with only one destination. Flow can enter the network at every junction, so we define the set of sources in the following way. 
\begin{Definition}[Set of sources $\boldsymbol{S}$]\label{defi:bS}
 For every node $v\in\mathcal{V}$ of the network, the set of sources is denoted
 \begin{equation}
  \boldsymbol{S}:=\Big\{s_{a}^{v}\in L^{\infty}((0,T);\R_{\geq0}), \ v\in\mathcal{V}, a\in\mathcal{A}_{\text{out}}(v)\Big\}
 \end{equation}
with element $\boldsymbol{s}:=\left(s_{a}^{v}\right)_{v\in\mathcal{V},a\in\mathcal{A}_{\text{out}}}$.
\end{Definition}
Then, over each junction, the following constraints have to be satisfied:
\begin{Definition}[Flow conservation at nodes]
The flow node constraints at a junction for given $\boldsymbol{s}\in\boldsymbol{S}$ and $\boldsymbol{\theta}\in\boldsymbol{\Theta}$ over a network satisfy
\begin{equation}
s^{v}_{a}(t)+\theta^{v}_{a}(t)\cdot\sum_{\tilde{a}\in \mathcal{A}_{\text{in}}(v)} y_{\tilde{a}}(t)=u_{a}(t)  \ \forall v\in\mathcal{V},\ a\in\mathcal{A}_{\text{out}}(v),\ t\in(0,T) \text{ a.e..}
\end{equation}
\end{Definition}
This flow conservation states that all flow entering the junctions has to be the same as all flow exiting the junctions, adding the flow that  departures at the junction. 
Depending on the splits $\boldsymbol{\theta}$ in the network and the entering flows $\boldsymbol{s}$, traffic is distributed in the network according to the traffic dynamics, which will be detailed next.

\subsection{Non-local PDE model}\label{subsec:pde_1}
As discussed earlier, non-local PDE models are relatively new in the transportation engineering literature. The reason for the selection of such a model here is to illustrate the generality of our framework.
In the following, we assume that traffic flow can be modeled as a fluid; hence, the choice of a {macroscopic} model. We assume for reasons of simplicity that there is only one destination, which will be generalized later. 

The PDE model we consider is a so called \textit{non-local conservation law} as introduced in \cite{armbruster} and studied with regard to uniqueness and regularity of solutions in \cite{wang}. It is specifically considered in \cite{goatin2} for modeling traffic flow. A multi-commodity extension on networks is presented and studied in \cite{keimer2}. 
Conservation laws are frequently used for modeling traffic flow \cite{garavello2006traffic}. These models have an intrinsic velocity so that they encode travel time naturally.

On a given link $a\in\mathcal{A}$ with density $\rho_{a}:(0,T)\times(0,1)\rightarrow\mathbb{R}_{\geq0}$, the model is given by
\begin{align}
\tfrac{\partial}{\partial t}{\rho_{a}}(t,x)+ \tfrac{\partial}{\partial x}\left(\lambda_{a}\Big(t,\int_{b(x)}^{d(x)}\rho_{a}(t,y)\dd y\Big){\rho_{a}}(t,x)\right)&=0    && (t,x)\in (0,T)\times(0,1),\ a\in\mathcal{A}\label{eq:pde_1_1}\\
\rho_{a}(0,x)&=\rho_{a,0}(x)                                            && x\in (0,1),\ a\in\mathcal{A}\label{eq:pde_1_1_a}\\
\lambda_{a}\Big(t,\int_{b(0)}^{d(0)}\rho_{a}(t,y)\dd y\Big)\rho_{a}(t,0)&=u_{a}(t)                              && t\in(0,T),\ a\in\mathcal{A}\\
\lambda_{a}\Big(t,\int_{b(1)}^{d(1)}\rho_{a}(t,y)\dd y\Big)\rho_{a}(t,1)&=y_{a}(t)                              &&
t\in(0,T),\ a\in\mathcal{A}.\label{eq:pde_1_4}
\end{align}
Note that an initial density at $t=0$ is prescribed for every $x\in(0,1)$ in Equation~\eqref{eq:pde_1_1_a}. 
The function $\lambda_{a}$ represents the velocity of the traffic flow at time $t\in(0,T)$. We assume that it is a strictly positive continuously differentiable function and that it only depends on the average density of $\rho_{a}$ in a given neighborhood of $x\in(0,1)$. Potentially, this neighborhood can be the whole link.
Naturally, $\lambda_{a}$ is chosen to be a monotonically decreasing function, since the speed of propagation of the flow should decrease if the average flow increases, as the higher the traffic flow the lower the speed of the flow. A common choice is $\lambda_{a}(t,y)=1/(1+y)$ \cite{armbruster2}.
At $(t,x)\in(0,T)\times (0,1)$ the term $\int_{b(x)}^{d(x)}\rho(t,s)\dd s$ represents the average density of the flow between $b(x)$ and $d(x)$ with $b,d\in C^{1}([0,1];[0,1])$. However, more sophisticated velocity or flux functions can be used, but have to be calibrated by available data.
For simplicity, sometimes it is assumed that $b\equiv0$ and $d\equiv 1$, i.e., that the average is carried out over the entire domain.
\begin{remark}[Extension of the non-local term]
It is possible to introduce an integral kernel $\gamma:(0,T)\times(0,1)^{2}\rightarrow\R$ so that the non-local term is for $(t,x)\in(0,T)\times(0,1)$ instead given by
\[
 \int_{b(x)}^{d(x)}\gamma(t,x,y)\rho(t,y)\;\mathrm{d}y,
\]
which makes the application even broader. However, we will not detail this here.
\end{remark}
Due to the assumptions on $\lambda_{a}$, the propagation speed over the entire network can never reach zero and the model does not have shocks; in particular, the weak solution of the model is unique without any additional entropy condition~\cite{keimer3}. 
\begin{remark}[Non-local vs. local modeling]
The fact that the model cannot capture shocks when compared with other PDE models (like the LWR model and related models) might be interpreted as a weakness of the proposed model. Nevertheless, the model is simpler to implement on a complex network with given controllable splits $\boldsymbol{\theta}$ than classical PDE models. The model is still able to capture many phenomena; in particular, since travel time on a link, the velocity, is a function of the average density.
\end{remark}
To illustrate the dynamics in detail, we present an example. 
\begin{example}[Non-local conservation law on one link]\label{ex:single_lane_pde}
The dynamics of the model are illustrated with nonzero initial density data in the top two plots of Figure~\ref{fig:pde_1_2}
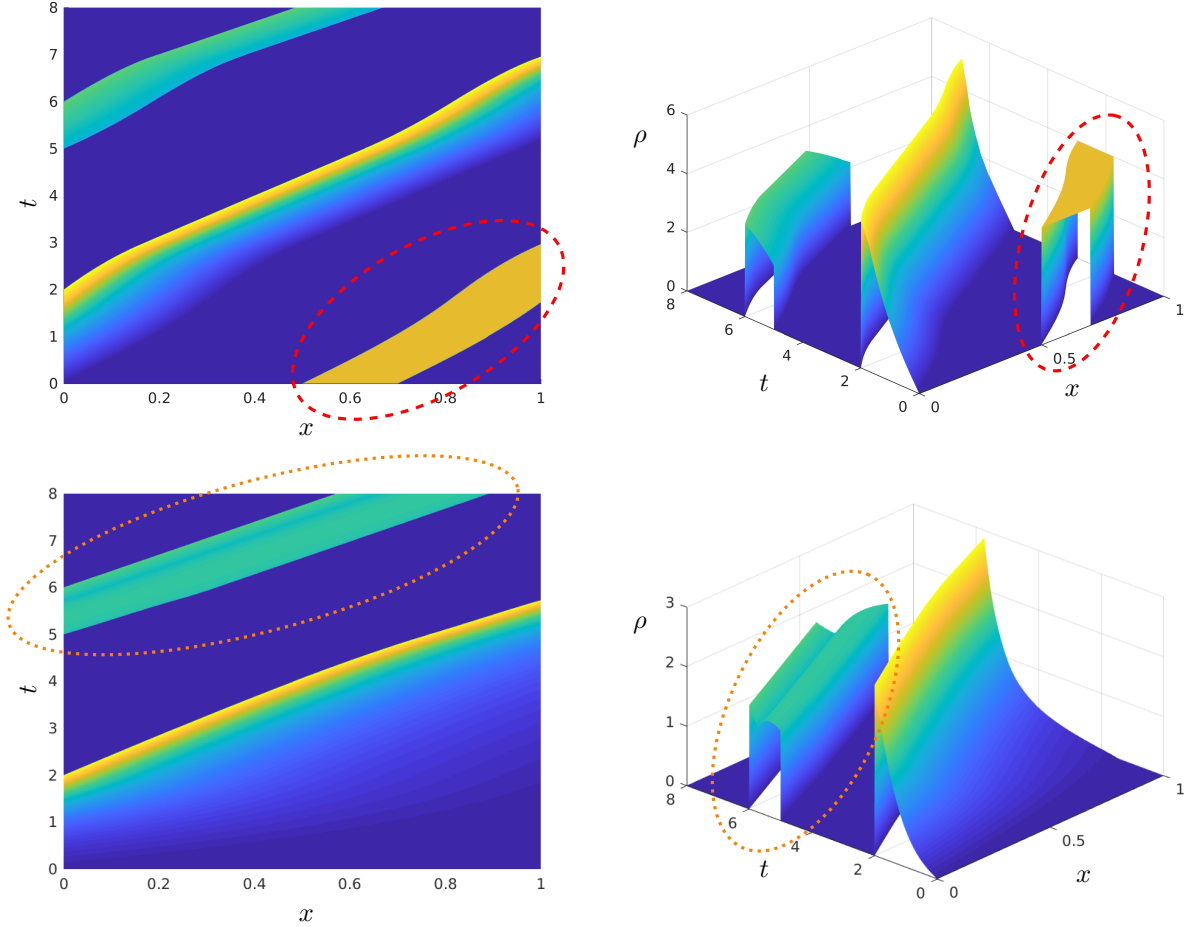
\begin{figure} 
\centering
\begin{tikzpicture}
\node[inner sep=0pt] (A) at (0,0){
\includegraphics[width=0.45\textwidth]{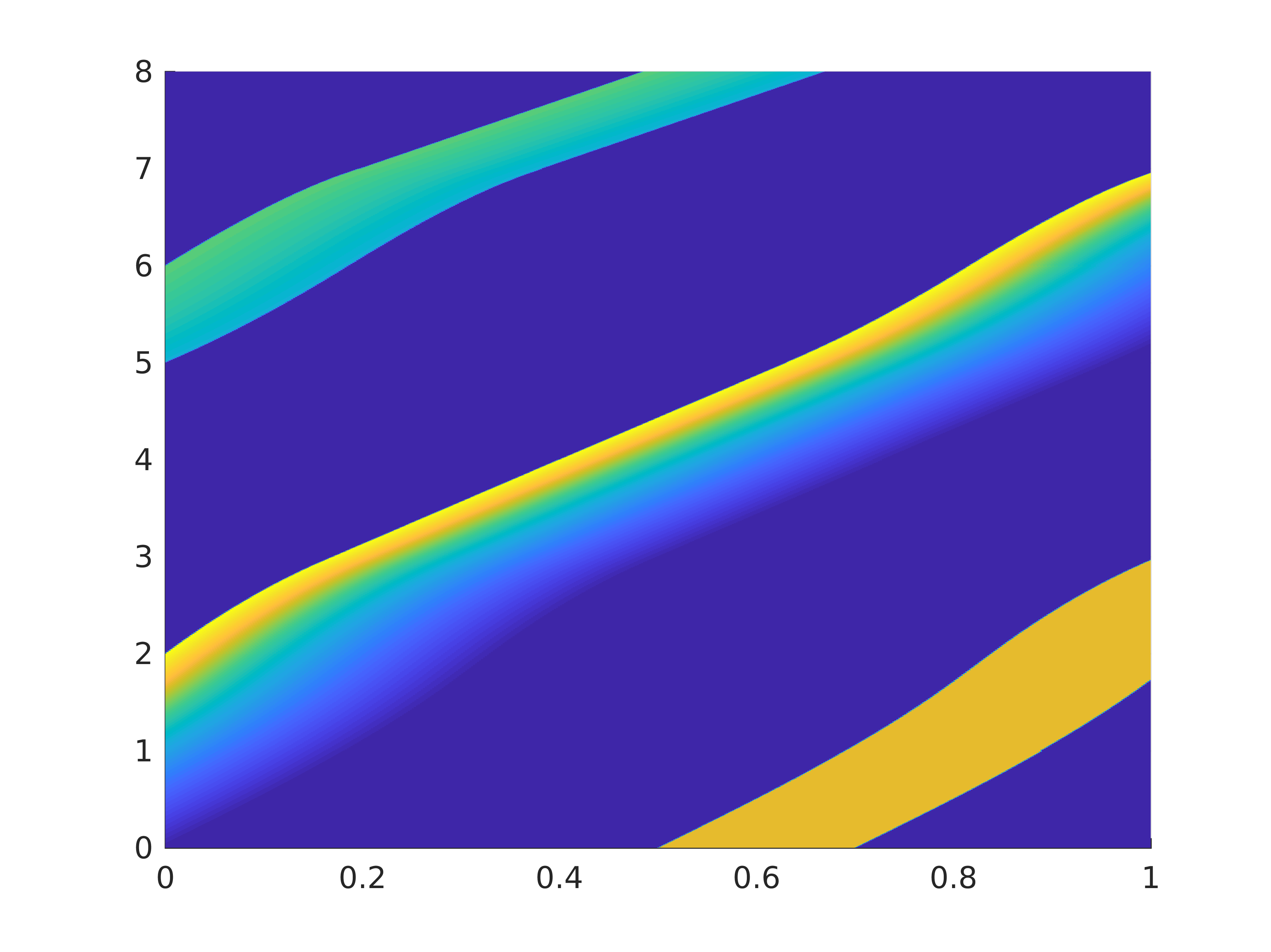}};
\draw (0.2,-3) node {$x$};
\draw(-3.5,0) node[rotate=90] {$t$};
\draw[color=red,rotate=30,very thick,dashed] (0.8,-2.25) ellipse (2cm and 1cm);
\end{tikzpicture}
\begin{tikzpicture}
\node[inner sep=0pt] (B) at (0,0){
\includegraphics[width=0.45\textwidth]{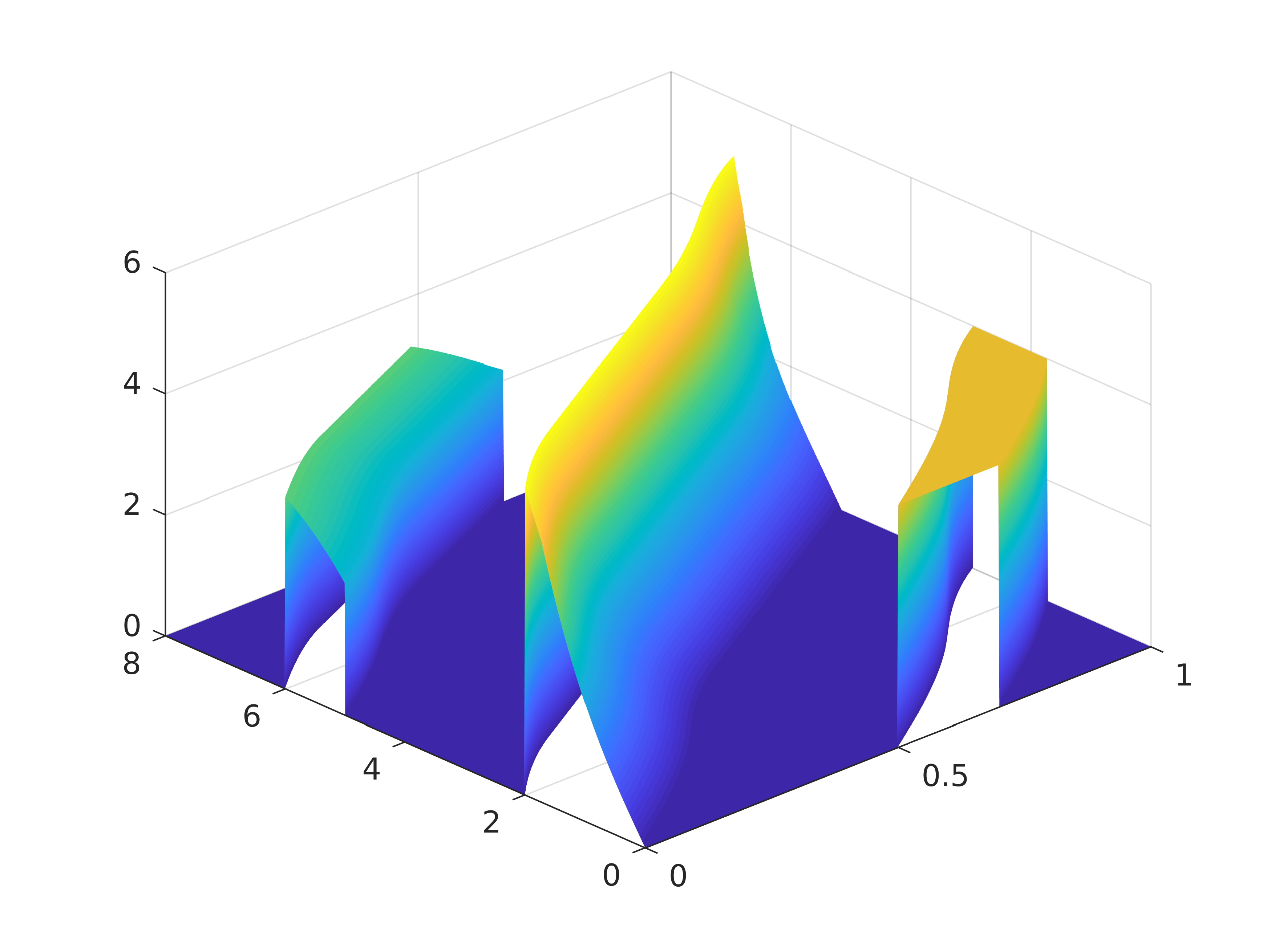}};
\draw (2.1,-2.35) node {$x$};
\draw(-2,-2.25) node {$t$};
\draw (-3.65,1) node {$\rho$};
\draw[color=red,rotate=75,very thick,dashed] (0.2,-2.25) ellipse (1.75cm and 0.8cm);
 \end{tikzpicture}

\begin{tikzpicture}
\node[inner sep=0pt] (A) at (0,0){
\includegraphics[width=0.45\textwidth]{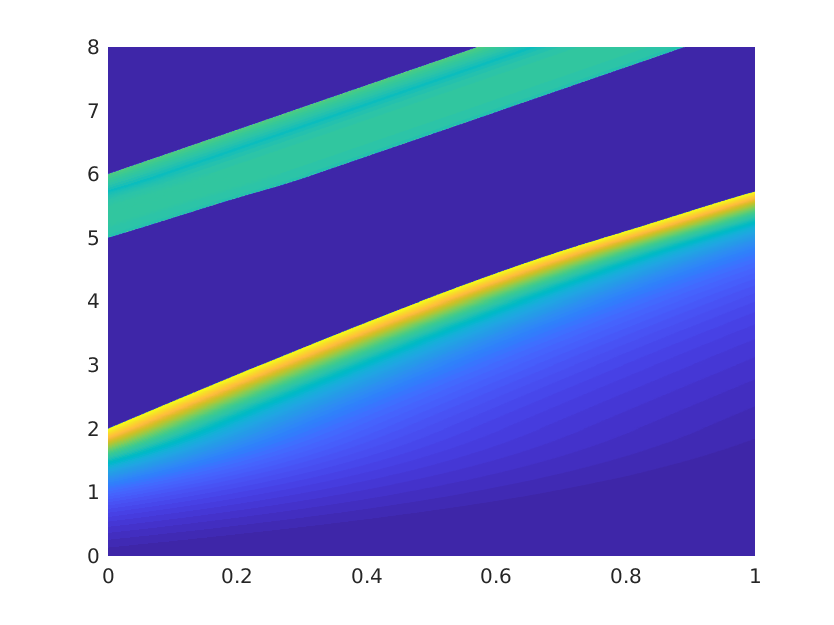}};
\draw (0.2,-3) node {$x$};
\draw(-3.5,0) node[rotate=90] {$t$};
\draw[color=orange,rotate=15,very thick,dotted] (0.1,1.825) ellipse (3.5cm and 1cm);
\end{tikzpicture}
\begin{tikzpicture}
\node[inner sep=0pt] (B) at (0,0){
\includegraphics[width=0.45\textwidth]{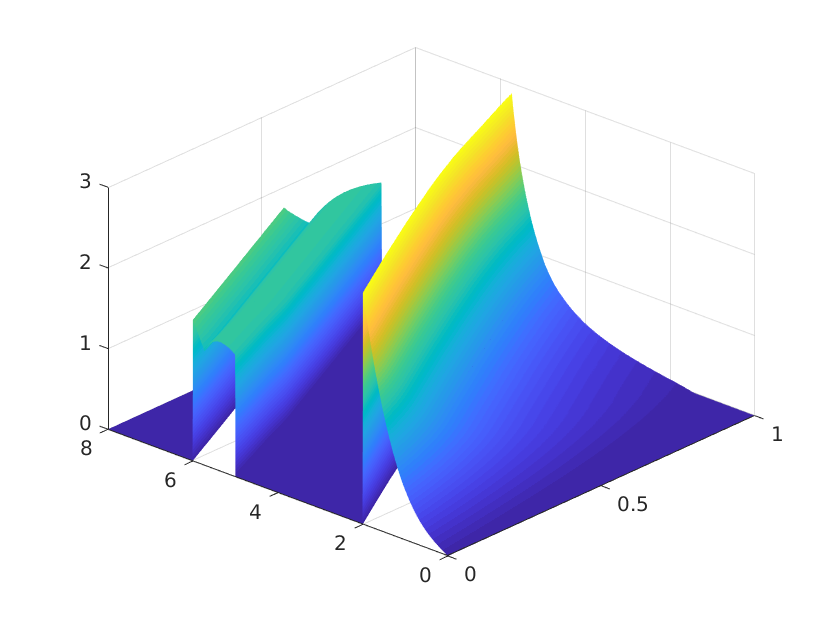}};
\draw (2.25,-2.35) node {$x$};
\draw(-2,-2.25) node {$t$};
\draw (-3.65,1) node {$\rho$};
\draw[color=orange,rotate=65,very thick,dotted] (-0.75,1.25) ellipse (2cm and 1cm);
\end{tikzpicture}
\caption{Non-local conservation law with two different initial conditions for the density $\rho$ over a single link. In both cases, $b\equiv0$ and $d\equiv 1$. Different perspectives. \textbf{Top:} Solution $\rho(t,x)$ for $\lambda(W)={1}/{(1+5W)}$ with influx $u(t)=\tfrac{t}{3}\cdot\mathds{1}_{[0,2]}(t)+\tfrac{1}{2}\cdot\mathds{1}_{[5,6]}(t)$ and initial condition $\rho_{0}(x)=4\cdot\mathds{1}_{[0.5,0.7]}(x)$, where $\mathds{1}$ is the indicator funtion. In red dashed the evolution of the initial datum is marked.  \textbf{Bottom:} Solution for the same $\lambda$ and influx $u$, but initial condition $\rho_{0}(x)\equiv0$. Due to a change of the average density, both examples are quite different even if they satisfy the same boundary datum. In orange dotted the evolution of the boundary datum emanating at $t\in [5,6]$ is marked.}\label{fig:pde_1_2} 
\end{figure}
and with zero initial data in the bottom two plots. Due to the non-local behavior, the initial data influence the propagation speed of the density. In the top plots, the density entering at $t\in[0,2]$ moves slower than the corresponding density for the lower plots. This is due to the fact that there is more average traffic on the road causing a lower speed than when the initial density is equal to zero. When there is no change in the density (as at $t\in[7,8]$), the velocity is constant.
The same can be observed for $t>6$.
The numerical method used to compute the solution in this example is presented in the end of this section. 
\end{example}
The following result states that for the non-local PDE we obtain a unique solution for given routing and inflow functions. The solutions of the PDE are interpreted in a weak sense \cite{evans}, making it possible to solve the system even for boundary data which are not differentiable or not even continuous. Given the novelty of the presented model, a well-posedness proof is necessary to justify the use of the model.
\begin{prop}[Existence and uniqueness of a solution on a single lane for given boundary and initial data]\label{prop:existence_uniqueness}
Let $T\in\R_{>0}$ and $p\in[1,\infty)$ be given and suppose that $b$ and $d$ are continuously differentiable, i.e., $b,d\in C^{1}([0,1];[0,1])$, and that boundary datum $u\in L^{p}((0,T))$ and initial datum $\rho_{0}\in L^{p}((0,1))$ are given. Suppose that in addition $\lambda\in C^{1}([0,T]\times\R;\R_{>0})$ is strictly positive.
Then, the PDE
\begin{align*}
\tfrac{\partial}{\partial t}{\rho}(t,x)+ \tfrac{\partial}{\partial x}\left(\lambda\Big(t,\int_{b(x)}^{d(x)}\rho(t,y)\dd y\Big){\rho}(t,x)\right)&=0    && (t,x)\in (0,T)\times(0,1)\\
\rho(0,x)&=\rho_{0}(x)                                            && x\in (0,1)\\
\lambda\Big(t,\int_{b(0)}^{d(0)}\rho(t,y)\dd y\Big)\rho(t,0)&=u(t)                              && t\in(0,T),
\end{align*}
admits a unique weak solution $\rho\in C([0,T_{1}];L^{p}((0,1)))$ on a sufficiently small time horizon $T_{1}\in(0,T]$. In addition, if $u\in L^{\infty}((0,T))$ and $\rho_{0}\in L^{\infty}((0,1))$, the solution is essentially bounded, i.e., $\rho\in L^{\infty}((0,T_{1})\times(0,1))$.
\end{prop}
\begin{proofsketch}
The proof for the simpler case $b\equiv0$ and $d\equiv1$ can be found in
\cite{shang}. The more general case will be omitted here because of lack of space.
\end{proofsketch}
\begin{remark}[Existence of a solution on every finite time horizon]
In case $b$ and $d$ are not explicitly space dependent, we obtain the result of existence and uniqueness on any finite time horizon. 
\end{remark}
\begin{theorem}[Existence and uniqueness of a solution on the network]\label{theo:pde_1}
Assume that we have an acyclic network. For any time horizon $T\in\R_{>0}$, $p\in[1,\infty)$, initial data $\rho_{0,a}\in L^{\infty}((0,1);\R_{\geq0})$, $\lambda_{a}\in C^{1}([0,T]\times\R;\R_{>0})$, $\boldsymbol{\theta}\in\boldsymbol{\Theta}$, source terms $\boldsymbol{s}\in\boldsymbol{S}$, and fixed bounds $b,d\in [0,1]$, the model in Equations~\eqref{eq:pde_1_1}--\eqref{eq:pde_1_4} admits a unique weak solution $\rho_{a}, a\in\mathcal{A}$ with
\[
\rho_{a}\in C([0,T];L^{p}((0,1)))\cap L^{\infty}((0,T);L^{\infty}((0,1))) \ \forall a\in\mathcal{A}.
\]
\end{theorem} 
\begin{proofsketch}
First, we recall that for given initial and boundary data on a given link, the solution of the non-local PDE exists and is unique as shown in Proposition~\ref{prop:existence_uniqueness}.

Since the network is acyclic, the next step is to use an induction argument: Suppose the solution is given on all incoming edges $a\in\mathcal{A}_{\text{in}}(v)$ of a node $v\in\mathcal{V}$. Due to the regularity of the solution on those incoming edges, meaning $\rho_{a}\in C([0,T];L^{p}((0,1)))$, the density can be evaluated at $x=1$ and we obtain as outflux $y_{a}\equiv\lambda(\cdot,\int_{b}^{d}\rho_{a}(\cdot,y))\rho_{a}(\cdot,1)\in L^{p}((0,T))$.

The routing functions $\theta_{\tilde{a}}^{v}$ are for $\tilde{a}\in\mathcal{A}_{\text{out}}(v)$ by the assumption on $\boldsymbol{\Theta}$ essentially bounded and so is the source $s_{\tilde{a}}^{v}$ by the assumption on $\boldsymbol{S}$, so that the new boundary data for the outgoing edges $\tilde{a}\in\mathcal{A}_{\text{out}}(v)$ satisfies on $t\in(0,T)$ a.e.
\begin{equation}
u_{\tilde{a}}(t)=s_{\tilde{a}}^{v}(t)+\theta_{\tilde{a}}^{v}(t)\cdot\sum_{a\in\mathcal{A}_{\text{in}}(v)}y_{a}(t)
\end{equation}
and by a simple H\"older estimate
\begin{align*}
\|u_{\tilde{a}}\|_{L^{p}((0,T))}&\leq \|s^{v}_{\tilde{a}}\|_{L^{p}((0,T))}+\|\theta^{v}_{\tilde{a}}\|_{L^{\infty}((0,T))}\cdot \sum_{a\in\mathcal{A}_{\text{in}}(v)}\|y_{a}\|_{L^{p}((0,T))}\\
&\leq T^{\tfrac{1}{p}}\|s^{v}_{\tilde{a}}\|_{L^{\infty}((0,T))} +\sum_{a\in\mathcal{A}_{\text{in}}(v)}\|y_{a}\|_{L^{p}((0,T))}.
\end{align*}
The right-hand side is bounded (even for $p=\infty$), so that we can conclude that the entering boundary datum $y_{\tilde{a}}$ is a $L^{p}$ function. Now, we again use the existence and uniqueness of the PDE for given boundary and initial data in Proposition~\ref{prop:existence_uniqueness}. This procedure can be iterated until we have exhausted all links in the network.

The uniqueness of the solution on the network directly follows by the uniqueness of the solution on a given link.
Thus, we obtain for any routing $\boldsymbol{\theta}\in\boldsymbol{\Theta}$ and any source $\boldsymbol{s}\in\boldsymbol{S}$, the existence and uniqueness of the solution on the network.\qed
\end{proofsketch}
\begin{remark}[Networks with cycles]
One might wonder why the statement of the theorem is not necessarily true for networks with cycles, since the hyperbolic character of the solution should lead to a finite propagation speed of boundary and initial data so that for sufficiently small time one could decouple the entire network, meaning that exiting boundary data would not be a function of the entering boundary data on every given link. This could be iterated so that one would obtain uniqueness on the entire network for every finite time horizon. Although the proposed non-local dynamics make the right-hand side boundary data not explicitly dependent on the left-hand side boundary data for small time, due to the non-local term, the speed of propagation, which would also affect the exiting boundary data, is implicitly a function of the entering boundary data at every time.

The problems with cycles can be avoided by assuming that the lower boundary of the non-local term satisfies $b>0$.
Another solutions is to consider a fixed-point formulation in the boundary data of a given cycle. However, this requires some mathematical technicalities, which we do not detail here.
\end{remark}

\subsection{Integrated information patterns}\label{subsec:app_enabled_dynamics}
In this section, we model traffic dynamics on networks while taking into account multiple destinations and the impact of the potential use of navigational applications and routing algorithms on decision processes. For this, we distinguish between different groups of drivers and use the following terminology: We call \textit{routed flow} or \textit{routed users} the category of drivers that has access to real-time traffic information, for instance, provided by GPS-enabled devices or smartphone applications. By contrast, the category of drivers who do not use these devices and mostly follow street signs, etc. to head to their destination are called \textit{non-routed flow} or \textit{non-routed users} \cite{thai2016negative}.
These commodities will be denoted by a superindex $\rr$ or $\nr$ and are defined as follows:
\begin{Definition}[Routed and non-routed flows with multiple destinations]
 Consider a network and define its destinations as $\mathcal{D}\subseteq \mathcal{V}$. We distinguish between routed and non-routed users $\{\rr,\nr\}$.
 When we consider any commodity or class of drivers, we mean all pairs $k\in\{\rr,\nr\}\times\mathcal{D}$.
\end{Definition}
Since routed and non-routed users might head to the same destination, it is reasonable to take as superindex $k$ to distinguish between them.
In the following, we have to keep track of all these different flows, routed and non-routed, with different destinations in the network over time, which requires the addition of dynamics for every pair $k\in\{\nr,\rr\}\times\mathcal{D}$ on every link.

Introducing routed and non-routed traffic as well as multiple destinations requires a change in the traffic network notation outlined in \Cref{subsec:traffic_network}. We replace the source $s_{a}^{v}$ by $s_{a}^{v,k}$ but still write $\boldsymbol{S}$ for the set of all sources, analogously to Definition~\ref{defi:bS}. The splits $\theta_{a}^{v}$ are replaced by $\theta_{a}^{v,k}$ for $k\in\{\nr,\rr\}\times\mathcal{D}$ but we still write $\boldsymbol{\Theta}$, as in Definition~\ref{defi:btheta}. 
The inflow $u_{a}$ is replaced by $u_{a}^{k}$ and the exiting flow by $y_{a}^{k}$. 
If a destination $k$ cannot be reached through a link $a\in\mathcal{A}_{\text{out}}(v)$ for a given junction $v$, we set $\theta_{a}^{v,k} = 0$.
Although this might seem difficult to determine, it only has to be done one time for the network to obtain all possible routes between all the given origin--destination pairs. 
\begin{remark}[Necessity of defining every class on every link]
On links where there is no actual route to the destination, the dynamics for that class can be omitted. For a unified presentation, however, we keep them in the following discussion.
\end{remark}

The proper conditions on the splits $\boldsymbol{\theta}$ will be imposed, restricting the set of admissible routes. However, these restrictions will not change the principle properties of $\boldsymbol{\Theta}$.

We present in the multi-destination and routed and non-routed framework the dynamics as follows:
\begin{allowdisplaybreaks}
\begin{align}
\sum_{k\in\{\nr,\rr\}\times\mathcal{D}}\rho_{a}^{k}(t,x)&=:\rho_{a}(t,x)  &&  (t,x)\in(0,T)\times(0,1),\ a\in\mathcal{A}\label{eq:pde_1_routed_1}\\
\tfrac{\partial}{\partial t}\rho_{a}^{k}(t,x)+ \tfrac{\partial}{\partial x}\left(\lambda_{a}\Big(t,\int_{b(x)}^{d(x)}\rho_{a}(t,y)\dd y\Big)\rho_{a}^{k}(t,x)\right)&=0    && (t,x)\in (0,T)\times(0,1),\ a\in\mathcal{A},\ k\in\{\nr,\rr\}\times\mathcal{D}\\
\rho_{a}^{k}(0,x)&=\rho_{a,0}^{k}(x)                                            && x\in (0,1),\ a\in\mathcal{A},\ k\in\{\nr,\rr\}\times\mathcal{D}\\
\lambda_{a}\Big(t,\int_{b(0)}^{d(0)}\rho_{a}(t,y)\dd y\Big)\rho_{a}^{k}(t,0)&=u_{a}^{k}(t)                              && t\in(0,T),\ a\in\mathcal{A},\ k\in\{\nr,\rr\}\times\mathcal{D}\\
\lambda_{a}\Big(t,\int_{b(1)}^{d(1)}\rho_{a}(t,y)\dd y\Big)\rho_{a}^{k}(t,1)&=y_{a}^{k}(t)                              &&
t\in(0,T),\ a\in\mathcal{A},\ k\in\{\nr,\rr\}\times\mathcal{D}.\label{eq:pde_1_routed_5}
\end{align}
\end{allowdisplaybreaks}
The velocity function $\lambda_{a}\in C^{1}([0,T]\times\R;\R_{>0})$ only depends on $a\in\mathcal{A}$ but not on commodity or routed/non-routed flow, since all vehicles on a given road must have the same velocity, regardless of their destination or their use of navigation tools.

The velocity also depends on the summarized average flow as stated in Equation~\eqref{eq:pde_1_routed_1}, which is a reasonable assumption: the speed of the flow will be determined by the entire flow on a given link.
\begin{remark}[Commodity-dependent velocities $\lambda_{a}$]
The model allows velocities dependent on the commodity $k\in\{\rr,\nr\}\times\mathcal{D}$ without changing any of the following results. It also allows different boundaries of the non-local terms. A straightforward interpretation of both is that different commodities might drive with different velocities. However, for multi-destinations and routed and non-routed users this does not make sense.
For coordination of truck platoons and for being able to distinguish between trucks and regular cars (as discussed in \Cref{sec:Freight}), this is a reasonable assumption.
\end{remark}
Taking into account multiple destinations and distinguishing routed and non-routed flow, we obtain an analogous result as that in \Cref{theo:pde_1} for a single commodity.
\begin{theorem}[Existence and uniqueness of the solution for routed and non-routed classes and multiple destinations]\label{theo:ex_uni_network}
Assume that we have an acyclic network. For any time horizon $T\in\R_{>0}$, $p\in[1,\infty)$, initial data $\rho_{0,a}^{k}\in L^{\infty}((0,1);\R_{\geq0})$, $\lambda_{a}\in C^{1}([0,T]\times\R;\R_{>0})$, $\boldsymbol{\theta}\in\boldsymbol{\Theta}$ and source terms $\boldsymbol{s}\in\boldsymbol{S}$ and fixed bounds $b,d\in [0,1]$, the model as defined in Equations~\eqref{eq:pde_1_routed_1}--\eqref{eq:pde_1_routed_5} admits a unique weak solution $\rho_{a}^{k}, a\in\mathcal{A}$ on the network so that
\[
\rho_{a}^{k}\in C([0,T];L^{p}((0,1)))\cap L^{\infty}((0,T);L^{\infty}((0,1))) \ \forall a\in\mathcal{A}, \forall k\in\{\rr,\nr\}\times\mathcal{D}.
\]
\end{theorem}
\begin{proofsketch}
 The only difference to the proof of Theorem~\ref{theo:pde_1} is that we have to consider multi-commodities and destinations. In the case the PDE is well-posed and admits a unique weak solution, the same reasoning as above will complete the proof. The result of existence for $b\equiv0$ and $d\equiv 1$ is provided in \cite{keimer2}.\qed
\end{proofsketch}
\subsection*{Numerical realization of the non-local model}\label{subsubsec:numerics}

We present for a specific case a numerical realization of the non-local model. For simplicity, we now assume that $b\equiv0$ and $d\equiv 1$.
Then, following \cite{keimer2} in the given framework the solution can be presented in terms of characteristics. Define the integral equality for $k\in\{\rr,\nr\}\times\mathcal{D}$ on a given link $a\in\mathcal{A}$ for $t\in[0,T]$
\begin{align}
 \xi^{k}_{a}(t)=\int_{0}^{t} \lambda_{a}\left(s,\sum_{k\in\{\rr,\nr\}\times\mathcal{D}\}}\int_{0}^{s} u_{a}^{k}(y)\mathrm{d} y +\int_{0}^{1-\xi^{k}_{a}(s)}\rho_{a,0}^{k}(y)\mathrm{d} y\right)\mathrm{d} s.\label{eq:fixed-point_numeric}
\end{align}
The solution of the integral equality which can be shown to be unique and continuously differentiable is the characteristic (the major part of the proof of Theorem~\ref{theo:ex_uni_network} consists of showing this), and with that characteristic it is possible to define the solution explicitly as
\begin{equation}
 \rho_{a}^{k}(t,x)=\begin{cases}
		      \rho_{a,0}^{k}(x-\xi_{a}^{k}(t))	& x\geq \xi_{a}^{k}(t)\\
		      \frac{u_{a}^{k}\left(\left(\left(\xi^{k}_{a}\right)^{-1}(\xi_{a}^{k}(t)-x\right)\right)}{{\xi_{a}^{k}}'(\left(\xi^{k}_{a}\right)^{-1}\left(\xi_{a}^{k}(t)-x)\right)} & x<\xi_{a}^{k}(t)
                   \end{cases}\label{eq:42}
\end{equation}
for sufficiently small time $(t,x)\in(0,T)\times(0,1)$. 
Therefore, one can solve Equation~\eqref{eq:fixed-point_numeric} on a given time--space by a canonical fixed-point iteration, and then state the solution as given in Equation~\eqref{eq:42} above. The procedure can now be iterated for sufficiently many time intervals to arrive at any finite time.

\section{Routing behavior for different information patterns}
\label{sec:routing}
Having presented and studied the dynamics on the network for a given routing policy, the next step is to define how that routing is actually determined and implemented. We provide a routing framework capable of integrating the status of the network.

\subsection{Local routing choices}
\label{subsec:wardrop}
\begin{figure}
\centering
\begin{tikzpicture}
[node distance = 1cm, auto,font=\footnotesize,
every node/.style={node distance=2cm,line width=2pt},
comment/.style={rectangle, fill=black!10,inner sep= 5pt, text width=2cm, very thick, node distance=0.25cm, font=\scriptsize\sffamily},
force/.style={rectangle, draw, fill=black!10, inner sep=5pt, text width=3cm, text badly centered, minimum height=1.2cm, font=\bfseries\footnotesize\sffamily}] 
\node [force] (rr) {Routing decisions, information enabled};
\node [force, below=1cm of rr] (routed) {Routed traffic};
\node [force,left=1cm of routed] (non) {Non-routed traffic};
\node [force, right=2cm of routed] (network) {Traffic network};
\node [force, above=1cm of non] (routing_non) {Routing decisions, information disabled};
\node [force, above=1cm of rr] (wardrop) {Routing services};
\draw[very thick,->] (routed.south) to [bend right] (network.south);
\draw[very thick,->] (non.south) to [bend right] (network.south);
\path[->,very thick,dashed] 
(network) edge (wardrop)
(wardrop) edge (rr)
(non) edge (routed)
(routed) edge (non);
\path[->,very thick,dotted]
(rr) edge (routed)
(routing_non) edge (non);
\end{tikzpicture} 
\caption{Illustration of how routing services influence the traffic dynamics and the role of routed and non-routed users. The block diagram corresponds to the lower two layers of Figure~\ref{fig:framework}. \textbf{Solid lines:} Impact, \textbf{Dashed lines:} Information, \textbf{Dotted lines:} Actions.}
\label{fig:routed_non_routed}
\end{figure}
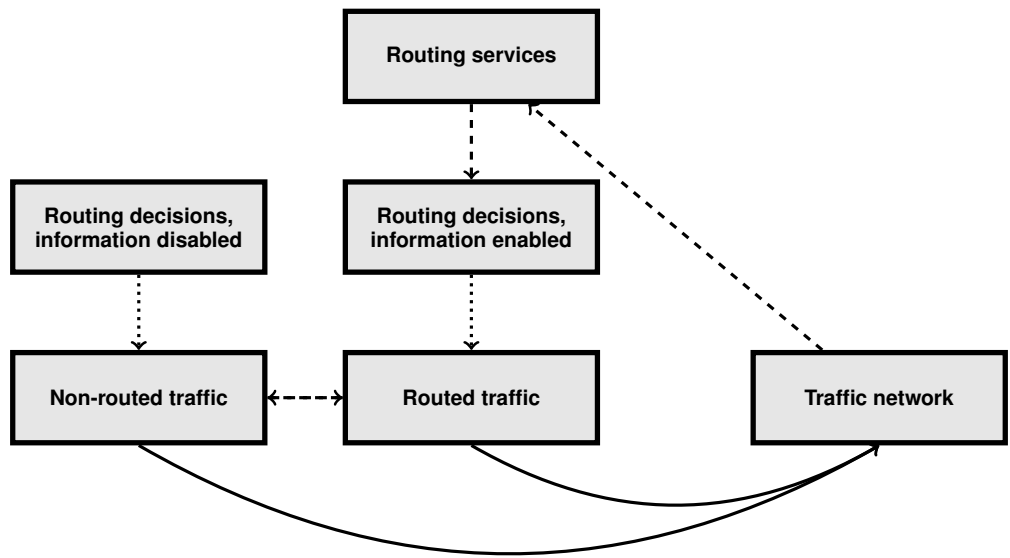

Since the process of choosing a path to one's destination in the network will crucially depend on the information available to users, we will distinguish between routed traffic flow in \Cref{subsubsec:routed} and non-routed traffic flow in \Cref{subsubsec:non_routed}.
This is illustrated in \Cref{fig:routed_non_routed}, where the solid lines indicates physical coupling, the dashed lines information flow, and dotted lines actions. The status of the traffic network is influenced by routed and non-routed traffic. However, while non-routed users make routing decisions not based on the status of the entire network, routed users can use available information to get routed more efficiently. It is required on an abstract level to define how routing is made for routed flow but also how routing decisions of non-routed traffic are made.

\subsection{Routing behavior for routed  traffic flow}\label{subsubsec:routed}
The question of traffic assignment and routing naturally came with the development of the field of traffic engineering. In 1952, Wardrop formally defined two principles in \cite{wardrop1952road} that any traffic assignment may satisfy in order to be optimal. Typically, Wardrop's conditions are used to determine the routing in a given traffic network for a given demand and the corresponding origin--destination pairs \cite{patriksson}. However, those conditions were introduced for stationary traffic flow models and are not easily extended to a dynamic framework.
Wardrop's equilibrium can be formulated verbally as follows.
\begin{Definition}[Wardrop's principles---stationary case]~
\begin{itemize}
\item[\textbf{(1)}] The travel times on all the used paths in the network are equal. In addition, those are less than the travel time on any unused path which would be experienced by a single vehicle changing its route choice.
\item[\textbf{(2)}] The average journey time is a minimum.
\end{itemize}
\label{def:wardropStatic}
\end{Definition}

Wardrop's first principle can also be interpreted in some cases as a Nash equilibrium \cite{patriksson} and is sometimes referred to as user equilibrium. Users choose routes selfishly and do not cooperate.

As can be seen, for the definition of these principles it is crucial to have a notion of travel time.
In the time-dependent case which we consider here, an extension of these principles is necessary. We propose the following:
\begin{Definition}[Extension of Wardrop's principles---time-dependent case]~\label{defi:wardrop_time}
\begin{itemize}
\item[\textbf{(1a)}] At any time $t$, for traffic conditions evaluated at time $t$,  Wardrop's first principle is satisfied, in a stationary sense.
\item[\textbf{(1b)}] The travel times on all the used paths, taking into consideration the time-evolution of traffic conditions, are equal. In addition, they are less than the travel time on any unused path which would be experienced by a single vehicle changing its route choice.
\item[\textbf{(2)}] The average journey time is a minimum considering the entire time horizon.
\end{itemize}
\end{Definition}
This generalization from the stationary case to the time dependent case is not without problems. While \textbf{(1a)} only requires route choices to be based on the current state of traffic, i.e., instantaneous, \textbf{(1b)} needs to know what traffic will look like in the future. This requirement necessitates full information on the network at future times or, in a weaker interpretation, requires a proper forecast.

The second principle \textbf{(2)} can also be interpreted as a system optimal solution or a so called social optimum. It will be addressed in \Cref{subsec:global} as an optimal control problem.

All the introduced Wardrop's conditions can be formulated mathematically, as a set of equations to solve.

When it comes to incorporating routed and non-routed flow, Wardrop's conditions have to be modified, since it can be expected that routed users have more knowledge (and finally) perfect knowledge of the actual state of the traffic network. 

In a more realistic setup, vehicles will actually not use Wardrop's principle individually but rely on service providers, which collect this information and pass it to users in an adapted or conditioned form, not necessarily following Definition~\ref{defi:wardrop_time} \textbf{(1)}. 
This is why we introduce another layer between Definition~\ref{defi:wardrop_time} and the routing choice:
\begin{Definition}[Routing principles for routed traffic flow]\label{defi:routing_options_routed_users}
We define the following principles regarding how routed flow might make their decision:
\begin{itemize}
\item \textbf{Full information:} Routed flow obtains full information to make the best choice, i.e., is able to follow Definition~\ref{defi:wardrop_time} \textbf{(1)}.
\item \textbf{Delayed information:} Routed flow obtains routing suggestions based on the traffic scenario of the near past.
\item \textbf{Incentivized information:} Routed flow obtains routing suggestions where not necessarily travel time is minimized but congestion, or pollution, or other objectives, which are not necessarily known to the flow.
\item \textbf{Data-base driven information:} Routed flow obtains routing suggestions not based on the current status of the network but on a data-base with recorded information and heuristics.
\item \textbf{Simplified forecast information:} Routed flow obtains routing suggestions, where the routing is actually based on a simplified forecast in time.
\end{itemize}
\end{Definition}
As can be seen, except for full information (which corresponds indeed to \textbf{1a)} in Definition~\ref{defi:wardrop_time}) and simplified forecast (which might be interpreted as a weak form of \textbf{1b)}), all the other routing suggestions are ``sub-optimal'' in comparison with having full information at every time.
Note also that in Definition~\ref{defi:routing_options_routed_users}, categories are not exclusive of each other, i.e., information can be full, but delayed, partial and incentivized, etc.


\subsection{Routing behavior for non-routed traffic flow}\label{subsubsec:non_routed}
A key question of interest is how non-routed users make their routing decisions. In response to this question, several reasonable answers can be given, each leading to specific modeling approaches:
\begin{Definition}[Behavior of non-routed traffic flow]\label{princ:non_routed}
We define the following principles regarding how non-routed flow might make their decisions:
\begin{itemize}
\item \textbf{Static:} The non-routed flow follows a predetermined, well-known path between the origin and destination and does not change its routing. The routed flow allocation is then independent of the traffic conditions on the network.
\item \textbf{Ex ante:} The non-routed flow determines its routing based on the state of traffic at the time it enters the network. This routing is then independent of the evolution of traffic conditions.
\item \textbf{Local:} The non-routed flow considers local information and changes its routing dependent on if there is congestion in a close neighborhood ahead of it. 
\item \textbf{Sub-network:} The non-routed flow only considers a sub-network for its routing choices, since it might not be aware of shortcuts, and tries to avoid arterial roads.
\end{itemize}
\end{Definition}
In reality, the routing process is a combination of a subset of above mentioned strategies and also individual driver choices. 
Part of the difficulty in modeling this problem is precisely the heterogeneity of the populations, leading to significant differences in the information patterns used for routing.
However, to understand dynamics better and to be able to identify specific phenomena and patterns, it seems crucial to keep the system as simple as possible.

\begin{remark}[Wardrop's principles]
From the viewpoint of Definitions~\ref{defi:routing_options_routed_users} and~\ref{princ:non_routed}, Wardrop's principles might not seem to be reasonable anymore, since routed users are routed externally. Without an extensive data study there is no way of knowing if external routing really follows Wardrop's principles.
Non-routed users do not have enough information to follow Wardrop's principles in the case there are high fluctuations in the traffic flow.  Thus, the entire flow of routed and non-routed users might instead  follow these new principles.
\end{remark}

\begin{challenge}[Routed and non-routed flow]\label{ch:models_routed_non_routed_network}
We have now introduced several routing choice models. These models can capture many phenomena which we expect to see for routed and non-routed traffic flow. However, the routing choice models and the corresponding traffic flow model have to be tested for validity and realism when it comes to the application to more realistic traffic scenarios.
\end{challenge}
As the routing $\boldsymbol{\theta}\in\boldsymbol{\Theta}$ was arbitrary in~\Cref{theo:ex_uni_network}, we can now consider different realizations and thanks to the results above we always obtain a well-posed and unique solution on the network.

\subsection{Global route choices and time-optimal routing}\label{subsec:global}
An interesting application of the introduced routing behavior principles is the solution of the corresponding minimization problem that provides as a result a social optimum. 
Thereby, we mean with a social optimum a solution which optimizes routing decisions (and, optionally, departure times) to the benefit of everyone over the entire time horizon and, therefore, taking into account the evolution of the network. This is an nonlinear infinite dimensional optimization problem. Dependent on the norm we chose to measure travel time or flux in the network, we will obtain different results. For reasons of simplicity we use the $L^{2}$ norm, but emphasize the presented results do not require this specific setup.
Moreover, since we are in an optimal control framework over time, we can also add as minimization variable the departure time of the flow.

Recalling that the flow entering the network is the time varying source $\boldsymbol{s}$, we assume this demand to be known and given over time by $d^{v,l}_{a}\in\R_{\geq0}$ for $v\in\mathcal{V}$, $a\in\mathcal{V}_{\text{out}}(v)$ and $l\in\{\rr,\nr\}\times\mathcal{D}$ and define all admissible inflows via the set
\begin{equation}
\boldsymbol{\tilde{S}}:=\boldsymbol{S}\cap\left\{s_{a}^{v,k}\in L^{\infty}((0,T);\R_{\geq0}):\ \int_{0}^{T}s_{a}^{v,k}(t)\;\mathrm{d} t=d^{v,k}_{a} \ \forall v\in\mathcal{V}, a\in \mathcal{A}_{\text{out}}(v), k\in\{\rr,\nr\}\times\mathcal{D}\right\}.\label{eq:variable_demand}
\end{equation}
Equation $\int_{0}^{T}s_{a}^{v,k}(t)\;\mathrm{d} t=d^{v,k}_{a}$ for $v\in\mathcal{V}, a\in \mathcal{V}_{\text{out}}(v)$ enables that the departure rate $s_{a}^{v,k}(t)$ varies at time $t\in(0,T)$ as long as it satisfies the demand $d^{v,k}_{a}$ over the entire considered time horizon $(0,T)$. In case only a small quantity of flow leaves first, later in time a higher quantity has to leave to satisfy the demand constraints.

We then have to define an appropriate functional $J$ to formulate an optimal control problem:
\begin{equation}
\begin{split}
\min_{(\boldsymbol{\theta},\boldsymbol{s})\in\boldsymbol{\Theta}\times\boldsymbol{S}} &\qquad J(\boldsymbol{\theta},\boldsymbol{s})\\
&\text{s.t. the dynamics in \Cref{subsec:app_enabled_dynamics} and the network constraints in \Cref{subsec:traffic_network} are satisfied,}\\
&\text{\quad $\boldsymbol{s}\in\boldsymbol{\tilde{S}}$ for the given demand $d^{v,k}_{a}$, $v\in\mathcal{V}, a\in\mathcal{A}_{\text{out}}(v), k\in \{\rr,\nr\}\times \mathcal{D}$,}\\
&\text{\quad } \boldsymbol{\theta}\in\boldsymbol{\Theta} \text{ with }\boldsymbol{\Theta} \text{ defined in Equation~\eqref{defi:btheta}.}
\end{split}\label{eq:optimal_control_problem}
\end{equation}
\begin{remark}[Reduced optimal control problem]
If we do not impose control on the departure time $\boldsymbol{s}$ but only assume that $\boldsymbol{s}$ is given over the entire time horizon, the optimal control problem becomes an optimal routing problem, where the best possible $\boldsymbol{\theta}$ subject to the given objective function is determined.
On the other hand, one can fix the time-dependent routing $\boldsymbol{\theta}$ and only optimize w.r.t. the departure rate $\boldsymbol{s}$ as done for a time discrete model in \Cref{sec:Coordination}.
In both cases, the following Theorem~\ref{theo:existence_social} still remains valid.
\end{remark}

To obtain reasonable routing results, the choice of the objective function $J$ is crucial. Often it should to some extent represent the travel time in the network. We use the following functional
\begin{equation}
J(\boldsymbol{\theta}, \boldsymbol{s}):=\sum_{v\in\mathcal{D}}\sum_{k_{1}\in\{\rr,\nr\}}\int_{0}^{T}\left(\int_{0}^{t}\sum_{\tilde{v}\in\mathcal{V}\setminus\{v\}}\sum_{a\in\mathcal{A}_{\text{out}}(\tilde{v})}d^{\tilde{v},(k_{1},v)}_{a}-\sum_{a\in\mathcal{A}_{\text{in}}(v)}y_{a}^{(k_{1},v)}(s)\dd s\right)^{2}\dd t,\label{eq:objective_social_optimim}
\end{equation}
where $y^{k}_{a}\equiv\rho_{a}^{k}(\cdot,x)\lambda_{a}\left(\cdot,\int\rho^{k}(\cdot,s)\dd s\right)\Big|_{x=1}$. The term $\sum\limits_{\tilde{v}\in\mathcal{V}\setminus\{v\}}\sum\limits_{a\in\mathcal{A}_{\text{out}}(\tilde{v})}d^{\tilde{v},(k_{1},v)}_{a}$ represents the total flow of cars which have to arrive at their destination $v\in\mathcal{D}$ and $\sum\limits_{a\in\mathcal{A}_{\text{in}}(v)}y_{a}^{(k_{1},v)}(t)$ the time dependent flow which finally arrives at destination $v\in\mathcal{V}$ at $t\in(0,T)$. 
\begin{remark}[The proposed objective]
The objective \eqref{eq:objective_social_optimim} is a so called backlog functional which penalizes flow not yet arrived at the destination in the $L^{2}$ norm. Since we only consider vehicles departing, but not the initial load of the network, it is reasonable to assume 
$\rho_{0,a}^{k}\equiv 0\ \forall a\in\mathcal{A}$ and $k\in\{\rr,\nr\}\times\mathcal{D}$. However, by adding initial data in the backlog functional, one would obtain an objective for initial load not equal to zero as well.

The proposed objective prohibits that vehicles remain too long on the road network, since every delay in the arrival time is penalized by the backlog functional \eqref{eq:objective_social_optimim}.
\end{remark}

Note that $\boldsymbol{\theta}$ and $\boldsymbol{s}$ are no explicit variables in Equation~\eqref{eq:objective_social_optimim}, however, they are implicitly encoded due to the network coupling.

The set $\boldsymbol{\tilde{S}}$ defines a convex set of the possible departure times. In addition, the set of routing functions $\boldsymbol{\Theta}$ is a convex set. This allows us to show that this complex optimal control problem has at least one solution assuming sufficient regularity on inflow and routing:

\begin{theorem}[Existence of a social optimum]\label{theo:existence_social}
Suppose that for every $\boldsymbol{s}\in\boldsymbol{S}$ and for every $v\in\mathcal{V}$ and $a\in\mathcal{A}_{\text{out}}(v)$ the split $s^{k}_{a}$ is of bounded variation with a uniform bound. In addition, assume that the analogue is true for any $\boldsymbol{\theta}\in \boldsymbol{\Theta}$. Let the network be acyclic. Then, there exists a solution of the optimal control problem~\eqref{eq:optimal_control_problem}--\eqref{eq:objective_social_optimim}.
\end{theorem}
\begin{proofsketch}
Due to the objective function $J$ being bounded from below, there exists a sequence $(\boldsymbol{\theta}_{l},\boldsymbol{s}_{l})_{l\in\mathbb{N}}$ converging to the infimum of the objective. Due to the uniform bounded variationbound and the convexity, we obtain by means of a compactness argument a strongly convergent subsequence in $L^{p}$ and due to the properties of the solution of the non-local PDE model, a weak convergent subsequence of the solution to the PDEs on the network. Since the product of a weakly convergent sequence and a strongly convergent sequence is again weakly convergent, we find a sequence of minimizers converging to the infimum of the objective function. Due to the convexity of the sets $\boldsymbol{S}$ and $\boldsymbol{\Theta}$ and their closedness, and thanks to the lower-semicontinuity of the objective function, the infimum is actually achieved.\qed
\end{proofsketch}
\begin{remark}[Bounded variation and more regular splits]
The space of bounded variation does not allow arbitrarily many jumps, which is thus a reasonable assumption for any routing and departure policy.
\Cref{theo:existence_social} remains valid if we assume that our routing functions are uniformly Lipschitz-continuous, which is a much stronger condition than the uniform bounded variation bound. In that case, the splits are not allowed to jump over time any more, in particular making the routing at the junctions continuous. 
\end{remark}
One might wonder if the computational effort to solve the introduced optimal control problem seems appropriate, since the computation of such a minimum is numerically hard. It is in particular not straightforward to apply first-order adjoint methods \cite{groeschel}. However, as a benchmark for the previous chapters, it might be worthwhile to compute or at least approximate the solution to be able to compare the results with those obtained by other routing choices as provided in Definition~\ref{princ:non_routed}, i.e., comparing a more social solution with a (purely) selfish one.

\section{A Mobility Service for Cooperative Freight Transportation} \label{sec:Freight}
\label{sec:control}
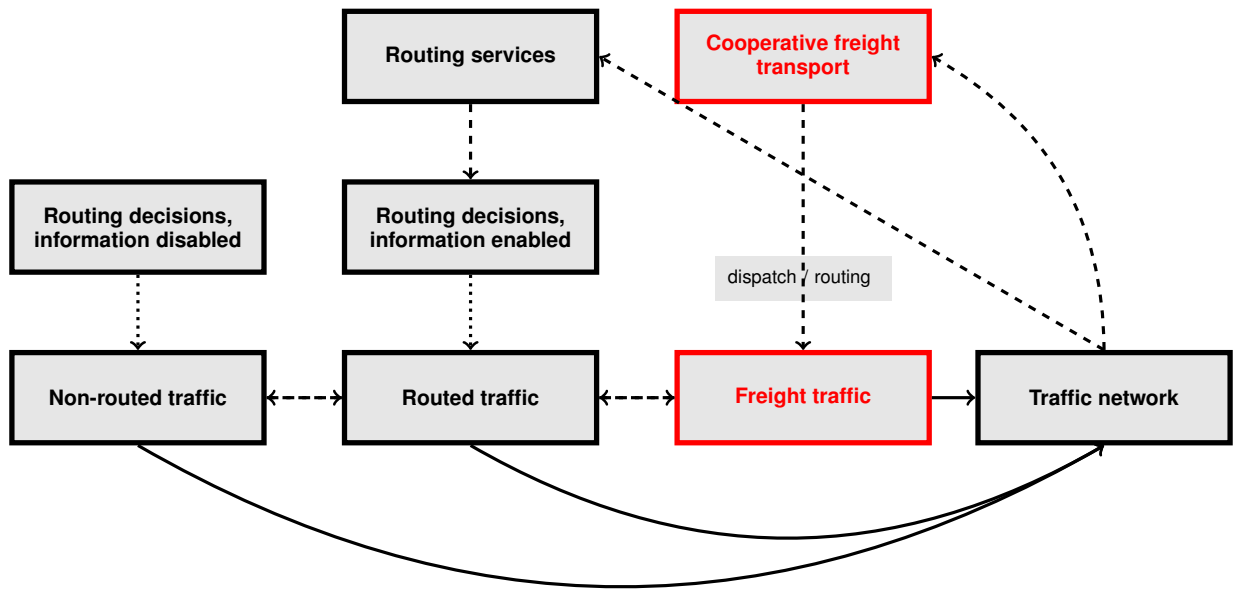
\begin{figure}
\centering
\begin{tikzpicture}
[node distance = 1cm, auto,font=\footnotesize,
every node/.style={node distance=3cm,line width=2pt},
comment/.style={rectangle, fill=black!10,inner sep= 5pt, text width=2cm, very thick, node distance=0.25cm, font=\scriptsize\sffamily},
force/.style={rectangle, draw, fill=black!10, inner sep=5pt, text width=3cm, text badly centered, minimum height=1.2cm, font=\bfseries\footnotesize\sffamily}] 
\node [force] (rr) {Routing decisions, information enabled};
\node [force, below=1cm of rr] (routed) {Routed traffic};

\node [color=red,force,right=1cm of routed] (freight) {Freight traffic};

\node [force,left=1cm of routed] (non) {Non-routed traffic};
\node [force, right=5cm of routed] (network) {Traffic network};
\node [force, above=1cm of non] (routing_non) {Routing decisions, information disabled};
\node [force, above=1cm of rr] (wardrop) {Routing services};
\node [color=red,force, right=1cm of wardrop] (coop) {Cooperative freight transport};
\node [comment, below =2cm of coop] {dispatch / routing};

\path[->,very thick,dashed] 
(network.north) edge (wardrop.east)
(wardrop) edge (rr)
(routed) edge (non)
(non) edge (routed)
(coop) edge (freight)
(routed) edge (freight)
(freight) edge (routed)
(network.north)   edge [bend right] (coop.east);
\path[->,very thick,dotted]
(routing_non) edge (non)
(rr) edge (routed);
\path[->,very thick] 
(freight) edge (network)
(routed.south) edge[bend right] (network.south)
(non.south) edge [bend right=30] (network.south);
\end{tikzpicture} 
\caption{Illustration of how a mobility service for cooperative freight transport can be implemented over the layered framework in \Cref{fig:framework}. \textbf{Solid lines:} Impact, \textbf{Dashed lines:} Information, \textbf{Dotted lines:} Actions.}
\label{fig:freight_routed_non_routed}
\end{figure}
In this section we utilize the modeling framework introduced above to present how a class of mobility services can be developed for cooperating fleets of trucks. Road freight transportation represents an essential part of traffic flow, sometime up to 60\% of traffic on some route like the I-710 in Los Angeles.

By extending the layered model for routing services in the previous section, we end up with the model and information flow illustrated in Figure~\ref{fig:freight_routed_non_routed}. It represents an underlying  multi-commodity traffic flow model with both light and heavy-duty vehicles. 
 Thus, the example of freight illustrates the generality of the proposed framework, and links to the non-local model in \Cref{subsec:pde_1} and \Cref{subsec:app_enabled_dynamics}.

In addition to the selfish routing of the information-disabled and -enabled users, a cooperation layer is introduced for the joint decisions made by fleet owners and freight service providers. We subsequently illustrate the setup on a platooning service, where these decisions are made to maximize the opportunities for trucks to share part of their overall travel from source to destination with other trucks to form fuel-saving vehicle platoons. It is shown for such a system that if global traffic information is available and both routing and velocity decisions can be influenced, a global optimizer can be derived. It is also shown that the coordination can be done over discrete sets of assignments represented by road segments traveled over certain time intervals. The corresponding match-making between trucks with potentially overlapping travel assignments can be efficiently computed even in a distributed manner. We end this section by showing that the service can be implemented under privacy guarantees, i.e., each individual truck or fleet owner does not have to give up information about source--destination pairs, which is an important aspect for commercial deployment. 

\subsection{Optimal control of individual traffic flows}\label{sec:flow_platooning}
We consider a scenario where perfect traffic network information is available and both routing and velocity decisions can be influenced. Such a scenario is already relevant today for truck platooning discussed in this section, and it will be even more relevant in the future with partial or full vehicle automation. For this setup, it is possible to form a control problem optimizing the decisions based on a suitable cost function. Here we study a setup where we are interested in optimizing one specific commodity, namely, the flow of goods traffic. For truck fuel-savings enabled by vehicle platooning, it is shown that an optimal control problem can be solved over the flow model introduced previously. 

We consider one commodity (representing trucks that can platoon with each other) and make its velocity function explicitly space and time dependent. To simplify the presentation, we consider only a single road assumed to be sufficiently long. The velocity of the trucks (or rather the truck flow) depends on position along the road and on time.
On the given road, which is parametrized by $x\in X$ with $X\subset\R$ a given bounded interval, we assume that there is typical traffic modeled by a non-local conservation law as introduced before:
\begin{align}
\tfrac{\partial}{\partial t} \rho(t,x)+ \tfrac{\partial}{\partial x}\Big(v\left(t,\int\rho(t,s)\dd s\right)\rho(t,x)\Big)&=0 && (t,x)\in (0,T)\times X\notag\\
\rho(0,x)   &=\rho_{0}(x) && x\in X \label{eq:freight_remaining_traffic}\\
v\left(t,\int\rho(t,s)\dd s\right)\rho(t,x)\Bigg|_{x=0}&=u^{1}(t) && t\in(0,T),\notag
\end{align}
where the velocity $v\in C^{1}([0,T]\times\R;\R_{>0})$ is assumed to be strictly positive and depends on the average load of the road. Here, 
$\rho_{0}$ is the given initial density on the road and $u^{1}$ the flux entering the road over time. Proposition~\ref{prop:existence_uniqueness} guarantees the existence and uniqueness of a solution.

It is sometimes reasonable to suppose that the number of trucks are relatively small; thus, we assume that the overall traffic is not influenced by the truck dynamics, but the overall traffic influences the trucks. This influence is specified by a non-local term $\int\rho(t,s)\dd s$. Making the truck velocity function explicitly space and time dependent, the dynamics of the truck density $q:(0,T)\times X\rightarrow\R_{\geq0}$ follows the following equations for given boundaries $b,d\in C^{1}(X;X)$
\begin{align}
\tfrac{\partial}{\partial t} q(t,x) +\tfrac{\partial}{\partial x} \left(\lambda\left(t,x,\int_{b(x)}^{d(x)}\rho(t,s)\dd s\right)q(t,x)\right)&=0 && (t,x)\in (0,T)\times X\notag\\
    q(0,x)   &=q_{0}(x) && x\in X \label{eq:freight_truck}\\
\lambda\left(t,0,\int_{b(0)}^{d(0)}\rho(t,s)\dd s\right)q(t,0)&=u^{2}(t) && t\in(0,T).\notag
\end{align}
Here, as pointed out, the velocity $\lambda$ is also space dependent and depends non-locally on the non-local term of the model.
The quantity $q_{0}$ represents the initial density of trucks on the road, and $u^{2}$ the entering flux of trucks.

Now, our aim is to control the velocity function $\lambda$ in a given feasible set in such a way that the density $q$ becomes more and more localized. This  can be interpreted as the continuous analogue to truck platooning, as the density of trucks is concentrated over time. 

We assume trucks only change their speed between speed limits $\lambda_{\text{min}}\in\R_{>0}$ and $\lambda_{\text{max}}\in\R_{>0}$ with $\lambda_{\text{min}}\leq\lambda_{\text{max}}$ and that they will not change speed too fast in space and time (which can be interpreted as a Lipschitz-bound on the space dependency of the velocity function). Then, the set of admissible speeds (the control set) can be denoted by
\begin{equation}
\begin{split}
\Lambda:=&\Big\{\lambda\in C([0,T]\times X\times\R_{\geq0};\R): \lambda_{\min}\leq \lambda(t,x,y)\leq \lambda_{\max},\\ &\quad |\lambda(t,x,y)-\lambda(\tilde{t},\tilde{x},\tilde{y})|\leq L\left(|x-\tilde{x}|+|y-\tilde{y}|+|t-\tilde{t}|\right)\ \forall t,\tilde{t}\in[0,T],\ \forall x,\tilde{x}\in X,\ \forall y,\tilde{y}\in\R_{>0}\Big\}
\end{split}\label{defi:Lambda}
\end{equation}
with $L\in\R_{\geq0}$ given.
\begin{remark}[The set $\Lambda$]
The set $\Lambda$ can be simplified to functions that are only space-dependent and where there is explicit dependency on time or a non-local term.
\end{remark}
\begin{theorem}[Well-posedness of the PDE system]\label{theo:well_posed_pde_1}
For every function $\lambda\in \Lambda$ the PDE system 
\eqref{eq:freight_remaining_traffic}--\eqref{eq:freight_truck}
admits a unique weak solution $\rho,q\in C([0,T];L^{p}(X))$ for initial and boundary data of $L^{\infty}$-regularity ($p\in[1,\infty)$), which is essentially bounded over space and time.
\end{theorem}
\begin{proofsketch}
Since there is only a uni-directional coupling between the PDEs, we know by Proposition~\ref{prop:existence_uniqueness} that the equation in $\rho$ admits a unique, weak solution. 

Since initial and boundary data are essentially bounded, so is $\rho$. Thus, the velocity of the equation in $q$ (the equation for the trucks) is uniformly Lipschitz-continuous. 
This is obvious for the first two components by the assumptions on $\Lambda$, for the third component, i.e., $\int_{b(x)}^{d(x)}\rho(t,s)\dd s$, we can propose a Lipschitz constant by $\left(\|b'\|_{L^{\infty}(X)}+\|d'\|_{L^{\infty}(X)}\right)\|\rho\|_{L^{\infty}((0,T)\times X)}$. 
We obtain that the velocity function $(t,x)\mapsto\lambda(t,x,\int_{b(x)}^{d(x)}\rho(t,s)\dd s)$ is uniformly Lipschitz-continuous by the definition of $\Lambda$ in Equation~\eqref{defi:Lambda} 

This enables us to use the method of characteristics. Define for $(t,x)\in(0,T)\times X$ the integral equation
\begin{equation}
\xi[t,x](\tau)=x+\int_{t}^{\tau}\lambda\left(t,\xi[t,x](s),\int_{b(\xi[t,x](s))}^{d(\xi[t,x](s))}\rho(t,y)\dd y\right)\dd s,\qquad \tau\in[0,T].\label{eq:characteristics}
\end{equation}
As pointed out the function $\xi[t,x](\cdot)$ defines the characteristic for every given $[t,x]\in(0,T)\times X$.
It admits for every $[t,x]\in[0,T]\times X$ a unique solution which is Lipschitz-continuous in all components. Due to the assumption on $\lambda$ being nonnegative, we can also define for sufficiently small time $T_{1}\in (0,T]$ the inverse w.r.t. time, i.e., $\xi[t,x]^{-1}:X\rightarrow[0,T_{1}]$ and can state the solution of the PDE in $q$ as
\begin{equation}
 q(t,x)=\begin{cases}q_{0}(\xi[t,x](0))\cdot \partial_{2}\xi[t,x](0) & x\geq \xi[0,0](t)\\
                \tfrac{u(\xi[t,x]^{-1}(0))}{\lambda(\xi[t,x]^{-1}(0),0)}\cdot\partial_{2}\xi[\xi[t,x]^{-1}(0),x](t) &x> \xi[0,0](t)
            \end{cases},\quad t\in[0,T_{1}],\label{eq:explicit}
\end{equation}
where $\partial_{2}$ denotes the partial derivative of a function w.r.t. the second component.
This expression can be shown to satisfy all required properties to be the unique weak solution of the PDE on a sufficiently small time horizon. Given new initial datum and the shifted boundary datum at $t=T_{1}$, we can reiterate the procedure. Since the characteristcs do not explicitly depend on $q$, we can thus exhaust every finite time horizon.\qed
\end{proofsketch}
Using the developed framework for optimal platooning, for $i\in\{1,2\}$ we consider the  optimal control problem
\begin{equation}
\begin{split}
\min_{\lambda\in \Lambda} J_{i}(q,\rho)\\
\text{s.t. \eqref{eq:freight_remaining_traffic} and \eqref{eq:freight_truck},}
\end{split}\label{eq:optimal_platooning_flow}
\end{equation}
where the objective is given by
\begin{align*}
J_{1}(q)&:=\int_{0}^{T}\int_{X} x^{2}q(t,x)\dd x -\left(\int_{X}xq(t,x)\dd x\right)^{2}\dd t
\intertext{or}
J_{2}(q,\rho)&:=\int_{0}^{T}\int_{X} x^{2}q(t,x)(1+\rho(t,x))\dd x-\left(\int_{X} xq(t,x)(1+\rho(t,x))\dd x\right)^{2}\dd t.
\end{align*}
The objective $J_{1}(q)$ can be interpreted as a measurement for the variance of the truck flow while $J_{2}(q,\rho)$ represents a weighted variance where the weight is inducted by the overall density of traffic on the road. In both cases we aim to minimize the variance, which can be interpreted as continuous platoon formation.
\begin{remark}[The objective function]
It might make sense to weight the cost function by an integral kernel to make the minimization problem more general and reasonable. For reasons of simplicity we do not detail this here.

As pointed out, the solution in $\rho$ is always predetermined and thus acts only as a ``parameter'' for the optimal control problem. 
\end{remark}
Given the well-posedness in Theorem~\ref{theo:well_posed_pde_1} of the PDE system, another question when it comes to infinite dimensional optimization is the existence of a minimizer of the introduced optimal control problem \eqref{eq:freight_truck}.
\begin{theorem}[Existence of an optimizer]\label{theo:freight_existence_optimizer}
Under the assumptions of \Cref{theo:well_posed_pde_1} and over $\Lambda$, the optimal control problem~\eqref{eq:optimal_platooning_flow} admits a solution.
\end{theorem}
\begin{proofsketch}
By Theorem~\ref{theo:well_posed_pde_1} we have the well-posedness of a solution for every velocity $\lambda\in\Lambda$. Since the objective function is bounded from below, we can find a minimizing sequence $(\lambda_{l})_{l\in\mathbb{N}}\subset \Lambda$. The set $\Lambda$ admits a uniform Lipschitz constant and thus, there exists a subsequence $(\lambda_{l_{k}})_{k\in\mathbb{N}}$ strongly converging in $L^{p}((0,T))$ to a limit $\lambda_{*}\in\Lambda$. By Gr{\"o}nwall's Lemma we can show that the corresponding characteristics as given in Equation~\eqref{eq:characteristics} converge uniformly. Using the explicit solution formula in Equation~\eqref{eq:explicit} one can use the uniform convergence of the sequence of characteristics to obtain the weak convergence of the solution. By the lower semi-continuity of the objective function, we obtain as minimizer indeed $\lambda_{*}$.\qed
\end{proofsketch}
\Cref{theo:freight_existence_optimizer} only guarantees the existence of a minimizer, while, depending on initial and boundary data, there might be several local optima. With a canonical optimization framework we can thus improve the situation although there is no guarantee on ending up with a global optimizer. The significance of the optimal control problem~\eqref{eq:optimal_platooning_flow} and the theorem is considerable. It enables the combined use of flow models with different populations and a framework for cooperation. To illustrate the modeling, we present an example. 
\begin{example}[Optimal platooning on a single road]\label{ex:platooning_flow}
For the named optimal control problem, take as objective function $J_{1}$ and as velocity $\lambda(t,x,y)\equiv\tilde{\lambda}(t,x)$ for $(t,x)\in(0,T)\times X$ and $y\in\R$. Then, we do not have to solve for $\rho$ but only for $q$.
The numerics presented in \Cref{fig:example_platooning} are achieved by using an explicit formula of the solution $q$ for given $\lambda$ in terms of characteristics, as presented in the sketch of the proof of~\Cref{theo:well_posed_pde_1}.

As one can see in~\Cref{fig:example_platooning}, we start with initial data $q_{0}(x)=(2.6-x)(x-1)\mathds{1}_{[1,2.6]}(x)$. In the simulations for the upper plots, we have applied a constant velocity $\lambda\equiv 0.75$ (see top right plot). The evolution of the initial condition moves with constant speed to the right, see the top left and middle plots. 

In the bottom plots, we apply an optimization algorithm over space and time on the same initial data with a Lipschitz-condition on the velocity of $L=0.1$ and upper and lower bounds $\lambda_{\min}=0.5$ and $\lambda_{\max}=1$, respectively. The flow now becomes more concentrated in space over time compared to the upper plots, as desirable. This results from the optimized velocity $\lambda$, which speeds up on the left-hand side of the peak and slows down on the right-hand side over time. 
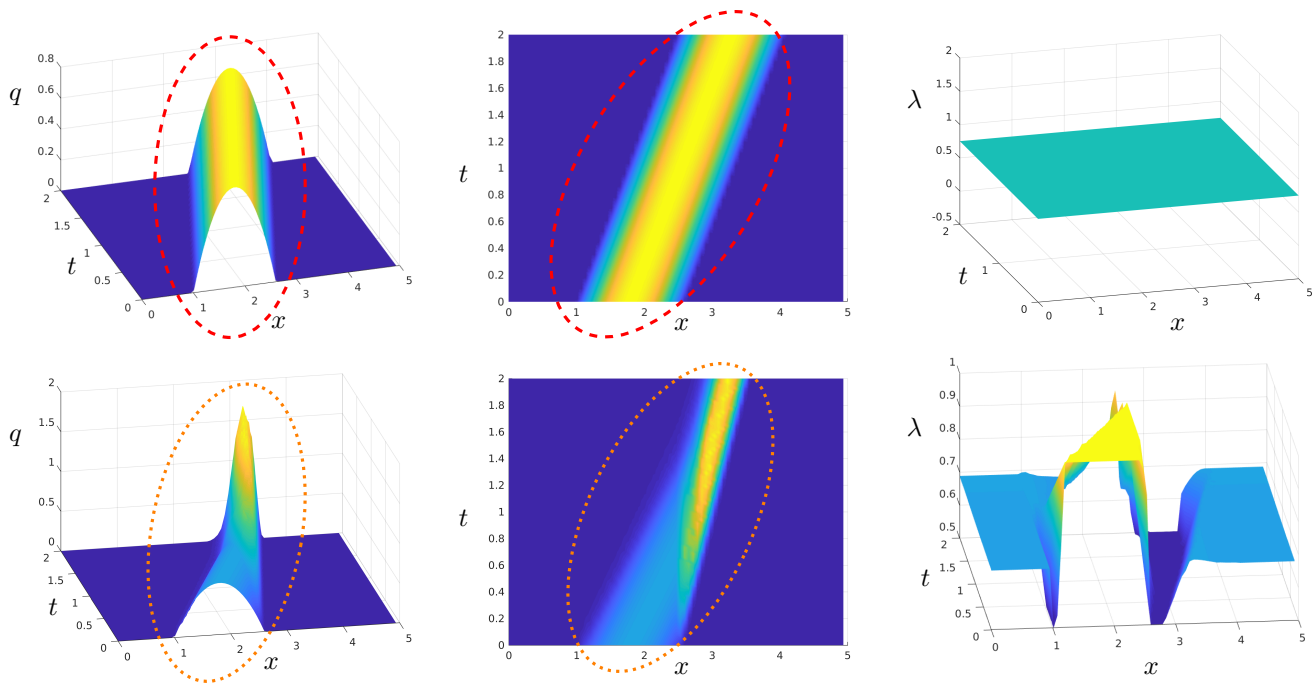
\begin{figure} 
\centering
\begin{tikzpicture}
\node[inner sep=0pt] (B) at (0,0){
\includegraphics[width=0.32\textwidth]{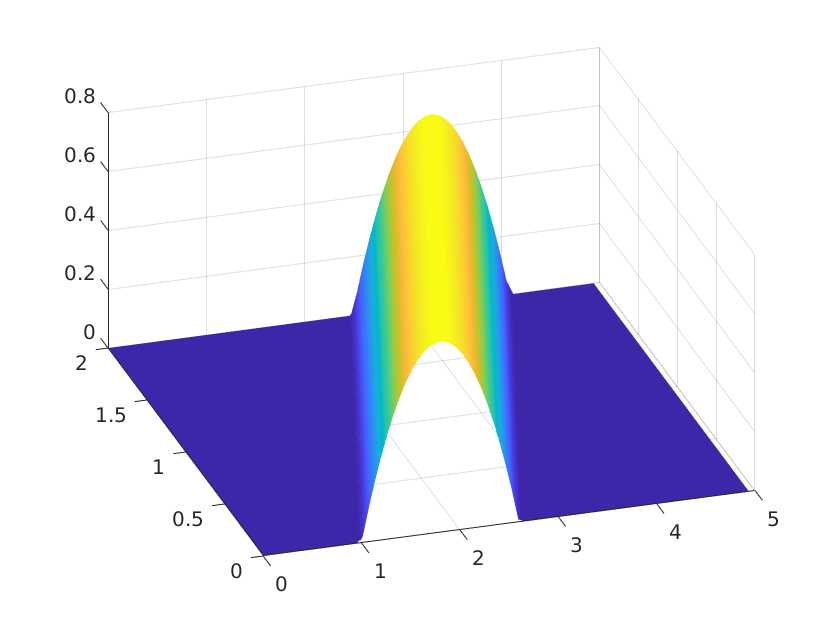}};
\draw (0.75,-2) node {$x$};
\draw(-2,-1.25) node {$t$};
\draw (-2.75,1) node {$q$};
\draw[color=red,very thick,dashed] (0.1,-0.2) ellipse (1cm and 2cm);
\end{tikzpicture}
\begin{tikzpicture}
\node[inner sep=0pt] (B) at (0,0){
\includegraphics[width=0.32\textwidth]{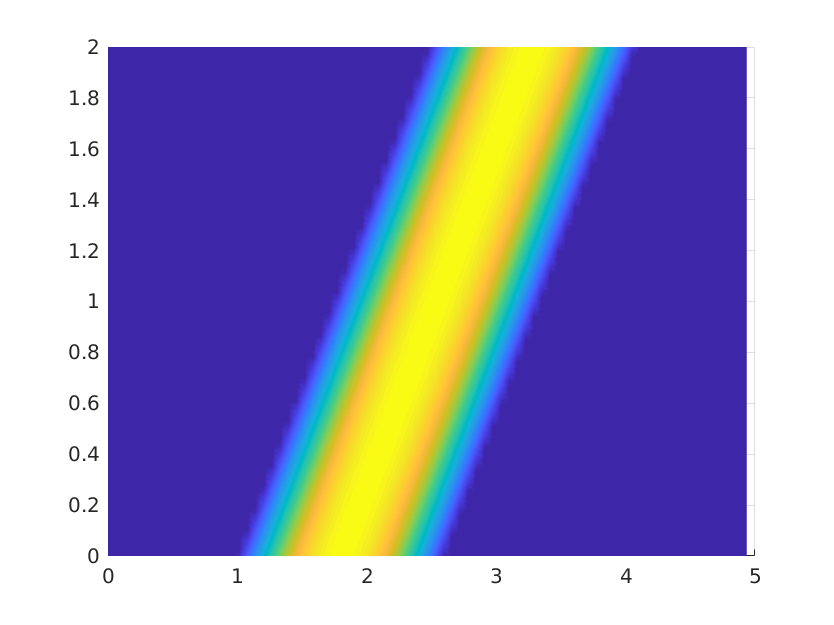}};
\draw (0.15,-2) node {$x$};
\draw(-2.75,0) node {$t$};
\draw[color=red,very thick,rotate=-30,dashed] (0,0) ellipse (1.2cm and 2.4cm);
\end{tikzpicture}
\begin{tikzpicture}
\node[inner sep=0pt] (B) at (0,0){
\includegraphics[width=0.32\textwidth]{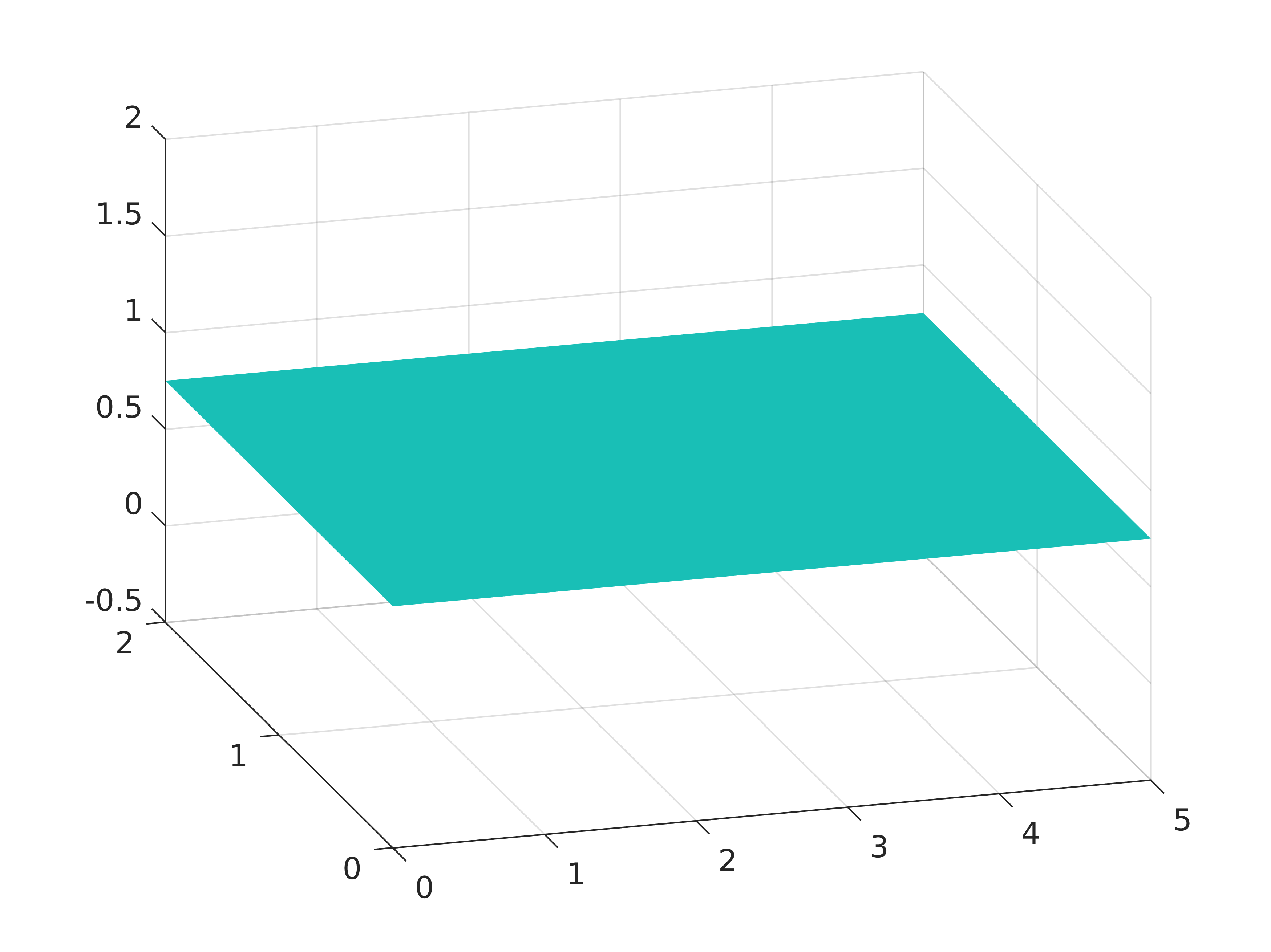}};
\draw (0.75,-2) node {$x$};
\draw(-2.1,-1.35) node {$t$};
\draw(-2.75,1) node {$\lambda$};
\end{tikzpicture}

\begin{tikzpicture}
\node[inner sep=0pt] (B) at (0,0){
\includegraphics[width=0.32\textwidth]{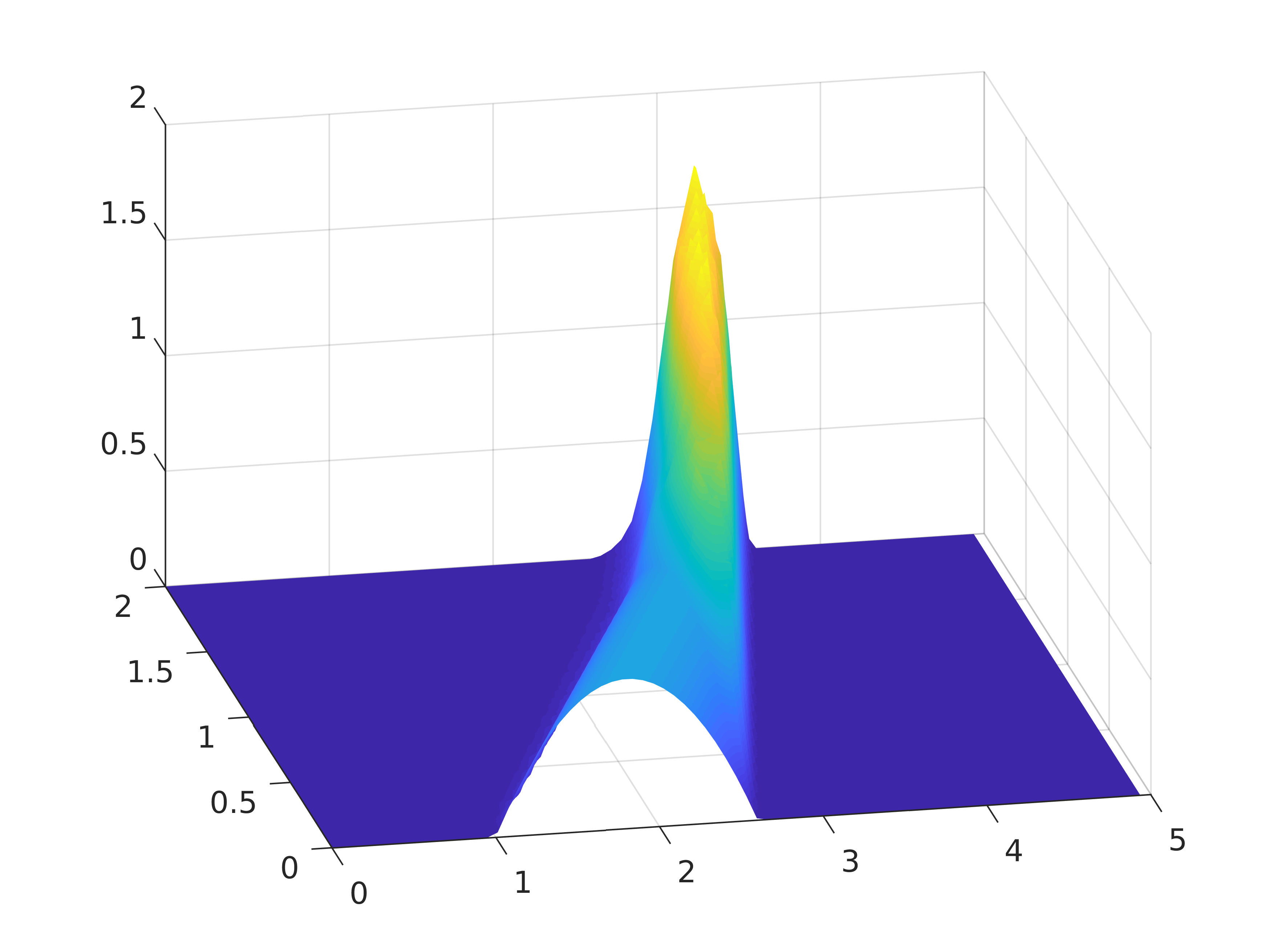}};
\draw (0.65,-2) node {$x$};
\draw(-2.25,-1.25) node {$t$};
\draw (-2.75,1) node {$q$};
\draw[color=orange,very thick,dotted,rotate=-10] (0.1,-0.25) ellipse (1cm and 2cm);
\end{tikzpicture}
\begin{tikzpicture}
\node[inner sep=0pt] (B) at (0,0){
\includegraphics[width=0.32\textwidth]{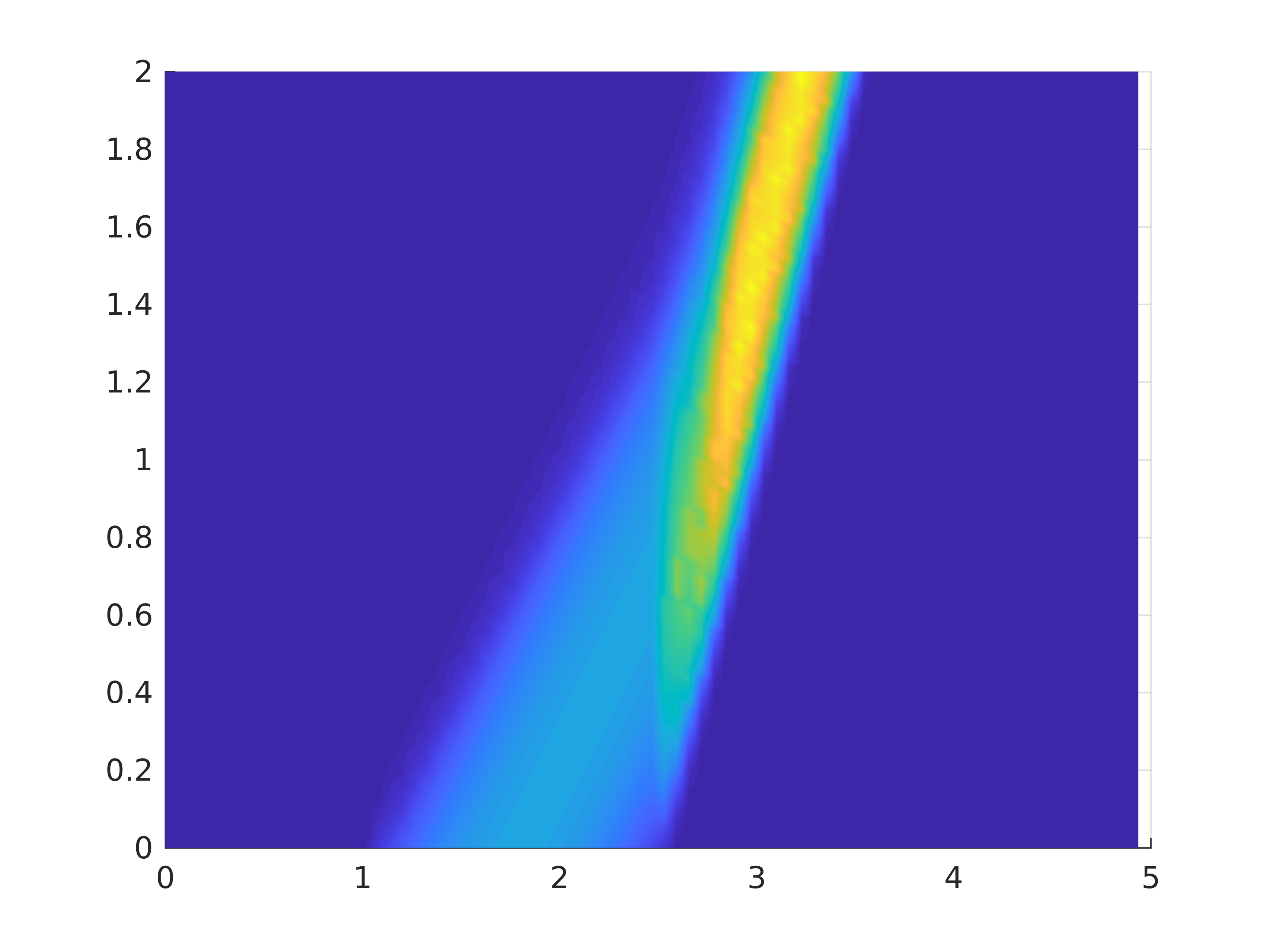}};
\draw (0.15,-2) node {$x$};
\draw(-2.75,0) node {$t$};
\draw[color=orange,very thick,rotate=-25,dotted] (0,-0) ellipse (1.1cm and 2.2cm);
\end{tikzpicture}
\begin{tikzpicture}
\node[inner sep=0pt] (B) at (0,0){
\includegraphics[width=0.32\textwidth]{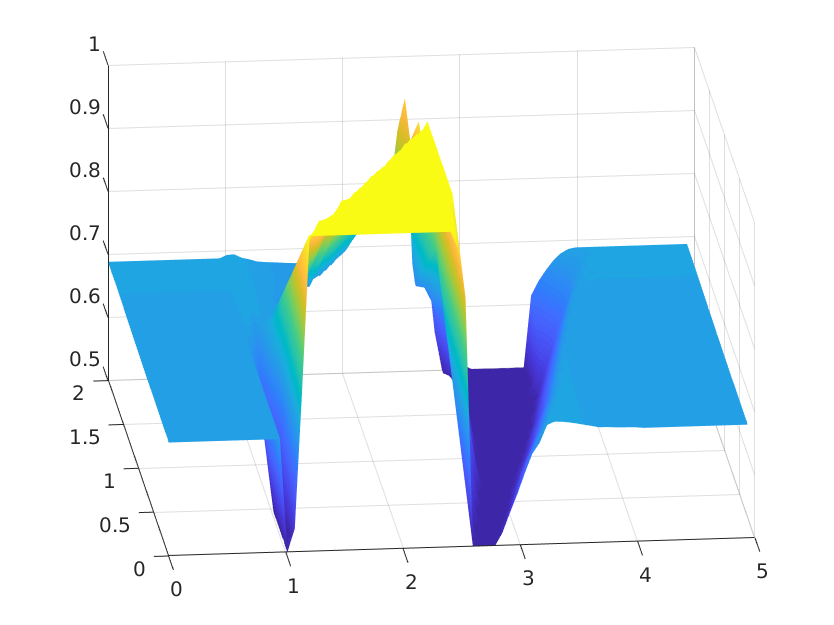}};
\draw (0.4,-2.2) node {$x$};
\draw(-2.6,-1) node {$t$};
\draw(-2.75,1) node {$\lambda$};
\end{tikzpicture}
\caption{The left two pictures show the evolution of flow over space and time $(t,x)\in(0,2)\times (0,5)$. The right two pictures represent velocity. The bottom plots are for initial data $q_{0}(x)=(2.6-x)\cdot(x-1)\cdot\mathds{1}_{[1,2.6]}(x)$, $\lambda_{\min}=0.5,\lambda_{\max}=1$, $L=0.1$ and boundary data equal zero. Note that the flow concentrates the more time passes. The top plots are with constant velocity $\lambda\equiv 0.75$.
In red dashed, the evolution of the initial datum for constant velocity, in orange dotted the evolution for the optimized velocity: Density is concentrated over time.}
\label{fig:example_platooning}
\end{figure}
\end{example}

\subsection{Coordinating the formation of truck platoons}\label{sec:Coordination}
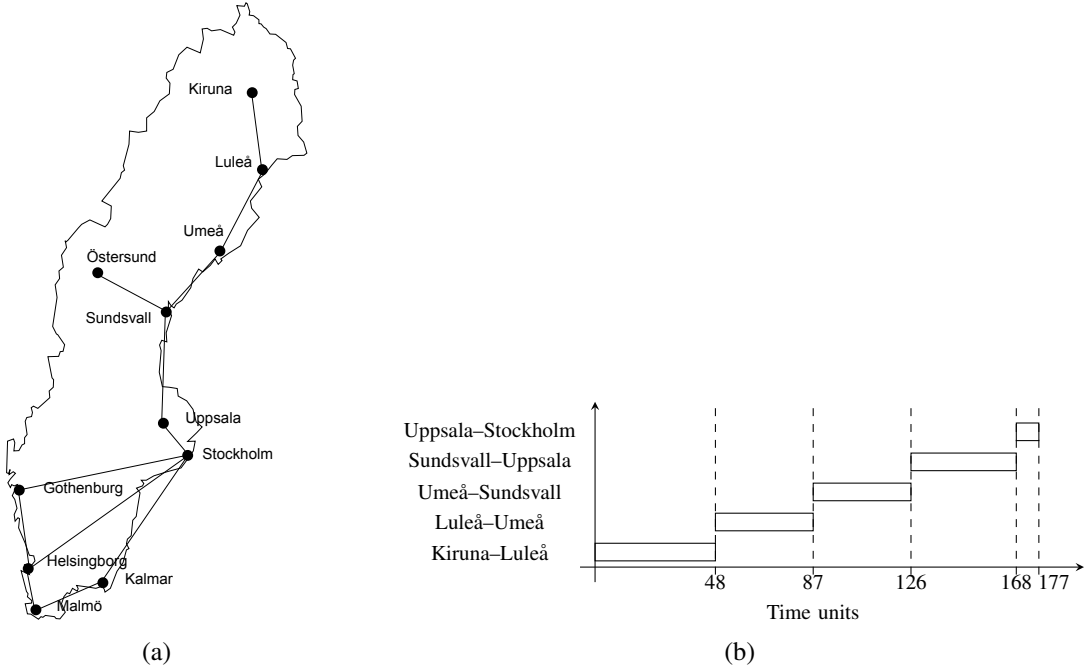
\begin{figure} 
\centering
\begin{tabular}{cc}
\begin{tikzpicture}
\node[] at (0,0) {\includegraphics[width=4cm]{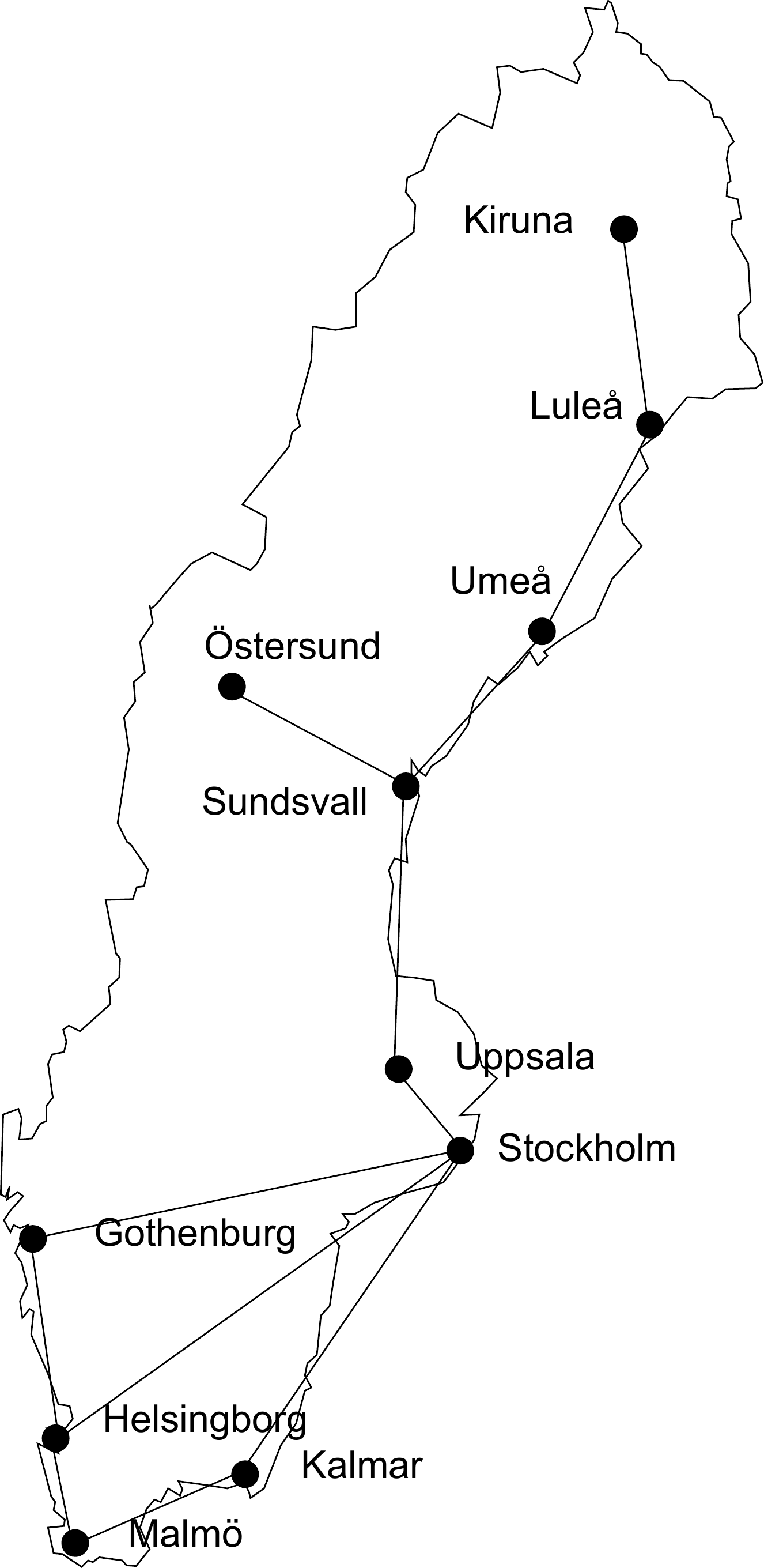}};
\end{tikzpicture}
& \hspace{.2in}
\begin{tikzpicture}[>=stealth]
\draw[->] (-0.2,0) -- (6.5,0);
\draw[->] (0,-0.2) -- (0,2.2);
\node[] at (-1.4,0.2) {\footnotesize Kiruna--Lule{\aa}};
\node[] at (-1.4,0.6) {\footnotesize Lule{\aa}--Ume{\aa}};
\node[] at (-1.4,1.0) {\footnotesize Ume{\aa}--Sundsvall};
\node[] at (-1.4,1.4) {\footnotesize Sundsvall--Uppsala};
\node[] at (-1.4,1.8) {\footnotesize Uppsala--Stockholm};
\node[rectangle,draw,minimum width=1.6cm] at (0.80,0.2) {};
\node[rectangle,draw,minimum width=1.3cm] at (2.25,0.6) {};
\node[rectangle,draw,minimum width=1.3cm] at (3.55,1.0) {};
\node[rectangle,draw,minimum width=1.4cm] at (4.90,1.4) {};
\node[rectangle,draw,minimum width=0.3cm] at (5.75,1.8) {};
\node[] at (+1.6,-0.2) {\footnotesize 48};
\node[] at (+2.9,-0.2) {\footnotesize 87};
\node[] at (+4.2,-0.2) {\footnotesize 126};
\node[] at (+5.6,-0.2) {\footnotesize 168};
\node[] at (+6.1,-0.2) {\footnotesize 177};
\node[] at (+2.9,-0.6) {\footnotesize Time units};
\draw[dashed] (+1.6,-0.1) -- (+1.6,+2.2);
\draw[dashed] (+2.9,-0.1) -- (+2.9,+2.2);
\draw[dashed] (+4.2,-0.1) -- (+4.2,+2.2);
\draw[dashed] (+5.6,-0.1) -- (+5.6,+2.2);
\draw[dashed] (+5.9,-0.1) -- (+5.9,+2.2);
\end{tikzpicture}
\\
(a) & \hspace{.2in} (b)
\end{tabular}
\caption{\label{fig:sweden} (a) An example of a Swedish freight transportation network and (b) the travel plan of vehicles going from Kiruna to Stockholm with sampling time $5\,\mathrm{min}$. }
\end{figure}

The previous section described how to continuously steer trucks to form platoons if they are close in time and space. In this section, we discuss how to coordinate truck departure times, routes, and velocities to maximize the opportunity for such continuous steering to be enabled. 
We illustrate the problem by considering the Swedish road network shown in Figure~\ref{fig:sweden}~(a). It will be shown that it is possible to slightly adjust the velocity of the trucks a long time in advance to improve the chances of platooning on shared links. It can be viewed as possibly changing the departure times of the vehicles so that they can meet over the shared links in various paths at similar times. 
This high-level adjustment of the vehicle schedules can be implemented  through a freight transportation service employed by drivers or fleet owners. 

As the freight transport network considered in this section as on a more abstract level than in previous sections, we introduce the new notation $\mathcal{G}=(\mathbb{V},\mathbb{E})$ for this directed graph. Here, $\mathbb{V}$ denotes the set of vertices, such as transportation hubs and large metropolitan areas, and $\mathbb{E}\subseteq\mathbb{V}\times\mathbb{V}$ denotes directed edges, such as the highway networks spanning across a country or continent. Figure~\ref{fig:sweden}~(a) illustrates a freight transportation network among some major cities in Sweden. This freight network is subsequently used to demonstrate the definitions and the results. 

We consider the planning horizon $T\in\mathbb{N}$, where $\mathbb{N}$ denotes the set of natural numbers. Each vehicle $i\in\mathbb{I}:=\{1,\dots,I\}$, with $I\in\mathbb{N}$ denoting the number of the vehicles, has a transportation assignment over the horizon $\mathbb{T}:=\{1,\dots,T\}$. The transportation assignment is modeled by the time series $e_i:\mathbb{T}\rightarrow \mathbb{E}$, such that the ordered set $(e_i[t])_{t\in\mathbb{T}}$ defines a walk over the graph $\mathcal{G}$ connecting the source of vehicle $i$ to its destination. Note that a vehicle can be on an edge for more than one time step depending on the length of the road corresponding to that edge.
Vehicle $i$ can decide to postpone its travel assignment by $\tau_i\in\{\underline{\tau}_i,\dots,\overline{\tau}_i\}\subseteq\mathbb{T}$; however, this comes at the cost of $h_i(\tau_i)$. The cost could be caused by contract fees, driver salaries, or lost opportunities. Note that this postponement could be an actual delay in the assignment or it could model the case where the vehicle drives slightly
slower for the catch-up to be possible in the future. 



The underlying goal of ensuring that the vehicles use the same road at similar times can be achieved by solving the optimization problem for $\gamma\in\R_{>0}$
\begin{subequations} \label{eqn:optim_coordination_1}
\begin{align} 
\min_{(\tau_i)_{i\in\mathbb{I}}}\quad & \sum_{i\in\mathbb{I}} h_i(\tau_i)-\gamma \sum_{t\in\mathbb{T}}\sum_{e\in\mathbb{E}} w_ef(|\{i\,|\,e_i(t-\tau_i)=e\}|),\\
\mathrm{s.t} \hspace{.23in} & \tau_i\in\{\underline{\tau}_i,\dots,\overline{\tau}_i\},\forall i\in\mathbb{I},
\end{align}
\end{subequations}
where $f(\cdot)$ is a mapping that grows faster than a linear function, e.g., quadratic or exponential, to ensure that the cost function rapidly decreases upon increasing the number of the vehicles on an edge. Furthermore, the constant $w_e>0$ denotes the weight attached to the desire for improving platooning possibility on edge $e$, e.g., it is desirable to form platoons on long freeways.

The optimization problem~\eqref{eqn:optim_coordination_1} can be transformed into a nonlinear integer program by setting $\tau_i=\sum_{j\in\mathbb{T}}jz_{ij}$, where $z_{ij}\in\{0,1\}$ for all $j\in\mathbb{T}$ and $\sum_{j\in\mathbb{T}}z_{ij}=1$. We define vector $\xi_{ije}\in\{0,1\}^{|\mathbb{T}|}$ such that $\ell$-th entry of $\xi_{ije}$, denoted by $\xi_{ije,\ell}$, is equal to one if $e_i(\ell-j)=e$ and zero otherwise. It can be shown that
$|\{i\,|\,e_i(t-\tau)=e\}|=\sum_{j\in\mathbb{T}}\sum_{i\in\mathbb{I}}\xi_{ije,t}z_{ij}$,
and, as a result,
\begin{align*}
\sum_{i\in\mathbb{I}} h_i(\tau_i)-\gamma \sum_{t\in\mathbb{T}}\sum_{e\in\mathbb{E}} f(|\{i\,|\,e_i(t-\tau_i)=e\}|)
&=\sum_{j\in\mathbb{T}}\sum_{i\in\mathbb{I}} h_i(j)z_{ij}-\gamma \sum_{t\in\mathbb{T}}\sum_{e\in\mathbb{E}} f(\sum_{j\in\mathbb{T}}\sum_{i\in\mathbb{I}}\xi_{ije,t}z_{ij}).
\end{align*}
Thus, the optimization problem in~\eqref{eqn:optim_coordination_1} can be transformed into the nonlinear integer program
\begin{align*}
\min_{(\tau_i)_{i\in\mathbb{I}}}\quad & \sum_{j\in\mathbb{T}}\sum_{i\in\mathbb{I}} h_i(j)z_{ij}-\gamma \sum_{t\in\mathbb{T}}\sum_{e\in\mathbb{E}} f(\sum_{j\in\mathbb{T}}\sum_{i\in\mathbb{I}}\xi_{ije,t}z_{ij}),\\
\mathrm{s.t} \hspace{.23in} & z_{ij}\in\{0,1\},\forall i\in\mathbb{I}, j\in\mathbb{T}, \\
& \sum_{j\in\mathbb{T}} z_{ij}=1,\forall i\in\mathbb{I},\\
&z_{ij}=0, (1\leq j\leq \underline{\tau}_i) \wedge (\overline{\tau}_i\leq j\leq T),\forall i\in\mathbb{I}.
\end{align*}
This problem is NP-hard in general~\cite{wolsey1998integer}. In fact, the computational complexity of solving this problem might grow exponentially in the number of the considered trucks and their flexibility $\overline{\tau}_i-\underline{\tau}_i$ for all $i$. 

\begin{remark}[Decomposition of the scheduling problem] For the case where the flexibility in scheduling is limited $\overline{\tau}_i\ll T$ (which is highly likely in practice because long delays might result in undesirable outcomes), the large integer program in~\eqref{eqn:optim_coordination_1} can be decomposed into smaller problems featuring only subsets of vehicles. This is because even if some of the vehicles delay their departure times by the largest possible amount, they cannot meet the vehicles that depart long after them or might arrive at their common links much later than them (due to long travel times across the country). Therefore, with little loss of generality, the scheduling problem in~\eqref{eqn:optim_coordination_1} can be solved for fewer vehicles for which there exists a possibility to cooperate. 
\end{remark}

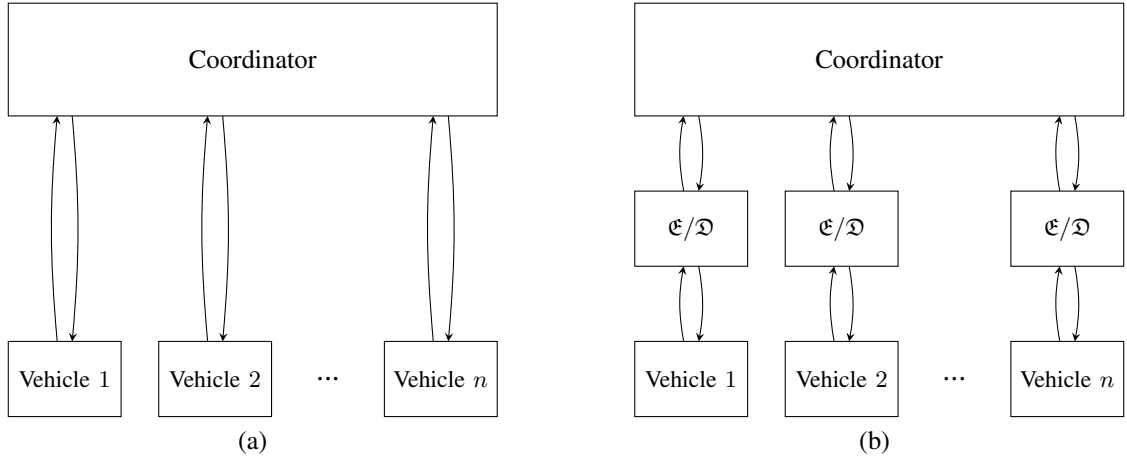
\begin{figure}[t]
\centering
\begin{tabular}{cc}
\begin{tikzpicture}[>=stealth]
\node[rectangle,draw,minimum height=1cm,minimum width=1.5cm] (1) at (-3,-2) {\small Vehicle $1$};
\node[rectangle,draw,minimum height=1cm,minimum width=1.5cm] (2) at (-1,-2) {\small Vehicle $2$};
\node[rectangle,draw,minimum height=1cm,minimum width=1.5cm] (3) at (+2,-2) {\small Vehicle $n$};
\node[] at (0.5,-2) {\large ...};
\node[rectangle,draw,minimum width=6.5cm,minimum height=1.5cm] (0) at (-.5,+2.25) {Coordinator};
\path[->] (-3.1,-1.50) edge[bend left=5] (-3.1,+1.50);
\path[->] (-2.9,+1.50) edge[bend left=5] (-2.9,-1.50);
\path[->] (-1.1,-1.50) edge[bend left=5] (-1.1,+1.50);
\path[->] (-0.9,+1.50) edge[bend left=5] (-0.9,-1.50);
\path[->] (+1.9,-1.50) edge[bend left=5] (+1.9,+1.50);
\path[->] (+2.1,+1.50) edge[bend left=5] (+2.1,-1.50);
\end{tikzpicture}
&
\hspace{.5in}
\begin{tikzpicture}[>=stealth]
\node[rectangle,draw,minimum height=1cm,minimum width=1.5cm] (1) at (-3,-1.25) {\small Vehicle $1$};
\node[rectangle,draw,minimum height=1cm,minimum width=1.5cm] (2) at (-1,-1.25) {\small Vehicle $2$};
\node[rectangle,draw,minimum height=1cm,minimum width=1.5cm] (3) at (+2,-1.25) {\small Vehicle $n$};
\node[rectangle,draw,minimum height=1cm,minimum width=1.5cm] (1) at (-3,+0.75) {\small $\mathfrak{E}/\mathfrak{D}$};
\node[rectangle,draw,minimum height=1cm,minimum width=1.5cm] (2) at (-1,+0.75) {\small $\mathfrak{E}/\mathfrak{D}$};
\node[rectangle,draw,minimum height=1cm,minimum width=1.5cm] (3) at (+2,+0.75) {\small $\mathfrak{E}/\mathfrak{D}$};
\node[] at (0.5,-1.25) {\large ...};
\node[rectangle,draw,minimum width=6.5cm,minimum height=1.5cm] (0) at (-.5,+3) {Coordinator};
\path[->] (-3.1,-0.75) edge[bend left=10] (-3.1,+0.25);
\path[->] (-2.9,+0.25) edge[bend left=10] (-2.9,-0.75);
\path[->] (-1.1,-0.75) edge[bend left=10] (-1.1,+0.25);
\path[->] (-0.9,+0.25) edge[bend left=10] (-0.9,-0.75);
\path[->] (+1.9,-0.75) edge[bend left=10] (+1.9,+0.25);
\path[->] (+2.1,+0.25) edge[bend left=10] (+2.1,-0.75);
\path[->] (-3.1,+1.25) edge[bend left=10] (-3.1,+2.25);
\path[->] (-2.9,+2.25) edge[bend left=10] (-2.9,+1.25);
\path[->] (-1.1,+1.25) edge[bend left=10] (-1.1,+2.25);
\path[->] (-0.9,+2.25) edge[bend left=10] (-0.9,+1.25);
\path[->] (+1.9,+1.25) edge[bend left=10] (+1.9,+2.25);
\path[->] (+2.1,+2.25) edge[bend left=10] (+2.1,+1.25);
\end{tikzpicture}
\\
(a) & \hspace{.5in}(b)
\end{tabular}
\caption{\label{fig:distributed} The communication structure among the coordinator and the vehicles in the distributed setting without (a) and with (b) homomorphic encryption.  }
\end{figure}

The optimization problem in~\eqref{eqn:optim_coordination_1} can be solved in a distributed manner among all the vehicles. Define $g:\mathbb{T}^{|\mathbb{I}|}\rightarrow \mathbb{R}$, with $\mathbb{R}$ denoting the set of real numbers, such that $g(\tau_1,\dots,\tau_{I})=-\gamma \sum_{t\in\mathbb{T}}\sum_{e\in\mathbb{E}} f(|\{i\,|\,e_i(t-\tau_i)=e\}|)$. This way, the cost function of~\eqref{eqn:optim_coordination_1} becomes $\sum_{i\in\mathbb{I}} h_i(\tau_i)+g(\tau_1,\dots,\tau_{I})$. Note that $g(\cdot,\tau_{-i})$, $\forall i$, can be interpreted as a pricing (or signalling) mechanism employed by the coordinator to nudge vehicle $i$ towards departure times with higher chances of vehicle platooning. Now, consider the following algorithm. At iteration $\beta\in\{0,1,\dots\}$, a vehicle from the set $\mathbb{I}$ is selected randomly with uniform distribution. Vehicle $i$ updates its decision, the delay in its transportation assignment, randomly such that the probability of selecting action $\tau'$ is given by
\begin{align} \label{eqn:distributedupdate}
P\{\tau_i[\beta]=\tau'\}=\frac{\exp(-[h_i(\tau')+g(\tau',\tau_{-i}[\beta-1])]/\varrho)}{\sum_{\tau''=\underline{\tau}_i}^{\overline{\tau}_i}\exp(-[h_i(\tau'')+g(\tau'',\tau_{-i}[\beta-1])]/\varrho)},
\end{align}
where $\varrho>0$ is a constant and $\tau_{-i}:=(\tau_j)_{j\in\mathbb{I}\setminus\{i\}}$. The rest of the vehicles follow the update rule that $\tau_j[\beta]=\tau_j[\beta-1]$ for all $j\neq i$. Figure~\ref{fig:distributed}~(a) demonstrates the communication structure required for implementing this algorithm. The  coordinator (i.e., the cloud computing freight transport service) requires to know the path that each vehicle takes and the delay $\tau_j[\beta-1]$, $\forall j$, to calculate function $\tau\mapsto g(\tau,\tau_{-i}[\beta-1])$ and return it to vehicle~$i$ for following update~\eqref{eqn:distributedupdate}. It can be proved that the stationary distribution of the action profile of the vehicles, irrespective of the initialization of the algorithm, follows
\begin{align} \label{eqn:stationarydistribution}
\lim_{\beta\rightarrow \infty} P\{\tau[\beta]=(\tau'_i)_{i=1}^m\}
&=\frac{\exp(-[\sum_{i\in\mathbb{I}} h_i(\tau'_i)+g(\tau'_1,\dots,\tau'_{I})]/\varrho)}{\sum_{(\tau''_i)_{i\in\mathbb{I}}\in\prod_{i\in\mathbb{I}}\{\underline{\tau}_i,\dots,\overline{\tau}_i\}}\exp(-[\sum_{i\in\mathbb{I}} h_i(\tau''_i)+g(\tau''_1,\dots,\tau''_{I})]/\varrho)}.
\end{align}
The proof follows the following line of reasoning. If each vehicle plans to optimize the cost $U_i(\tau_i,\tau_{-i})=h_i(\tau_i)+g(\tau_i,\tau_{-i})$, the conflict of interest results in a strategic game among all the vehicles. It can be shown that $\Phi(\tau_1,\dots,\tau_I)=\sum_{i\in\mathbb{I}} h_i(\tau_i)+g(\tau_1,\dots,\tau_{I})$ is a potential function for the game, i.e., $\Phi(\tau_i,\tau_{-i})-\Phi(\tau'_i,\tau_{-i})=U_i(\tau_i,\tau_{-i})-U_i(\tau'_i,\tau_{-i})$. Therefore, the log-linear learning dynamics in~\eqref{eqn:distributedupdate} can be used to extract an equilibrium of the game corresponding to the global minimizer of the potential function~\cite{blume1993statistical,marden2012revisiting}. The proof is similar to the scheduling problem considered in~\cite{farokhi2016game}. 

The stationary distribution of the action profile of the vehicles in~\eqref{eqn:stationarydistribution} shows that, upon selecting a small enough $\varrho$, the stationary distribution of the scheduling  concentrates on the global solution of~\eqref{eqn:optim_coordination_1}, i.e., they visit the neighbourhood (in terms of the cost) of the global solution of~\eqref{eqn:optim_coordination_1} infinitely often. 
This is interesting as it makes it possible for the vehicles to coordinate their actions using their mobile devices. However, it is also important to note the effect of $\varrho$ on the numerical stability of the algorithm. That is, if $\varrho$ is selected too small, numerical issues might arise as exponential of large numbers might be required to be calculated.

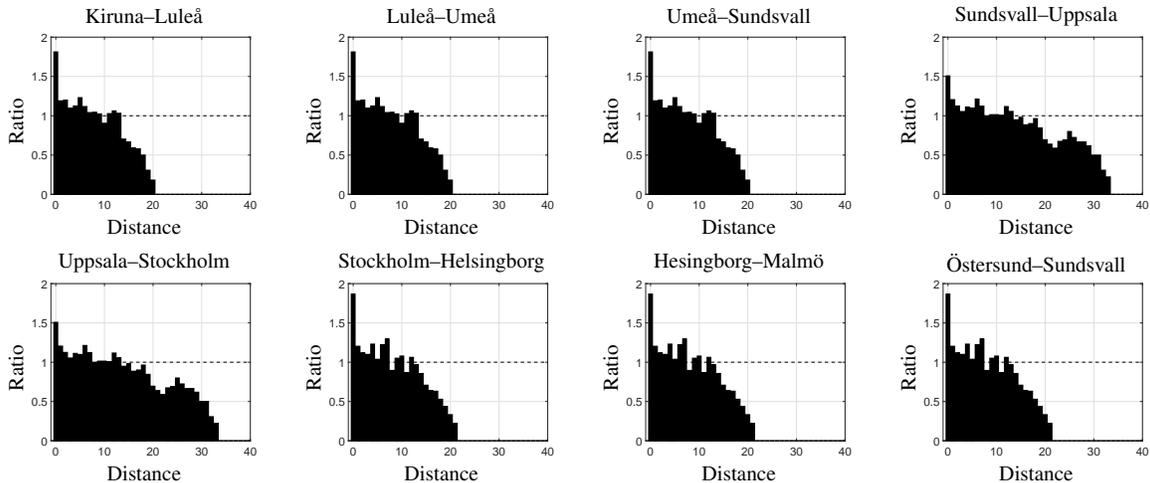
\begin{figure}[t]
\centering
\begin{tabular}{cccc}
\begin{tikzpicture}
\node[] at (0,0) {\includegraphics[width=3cm]{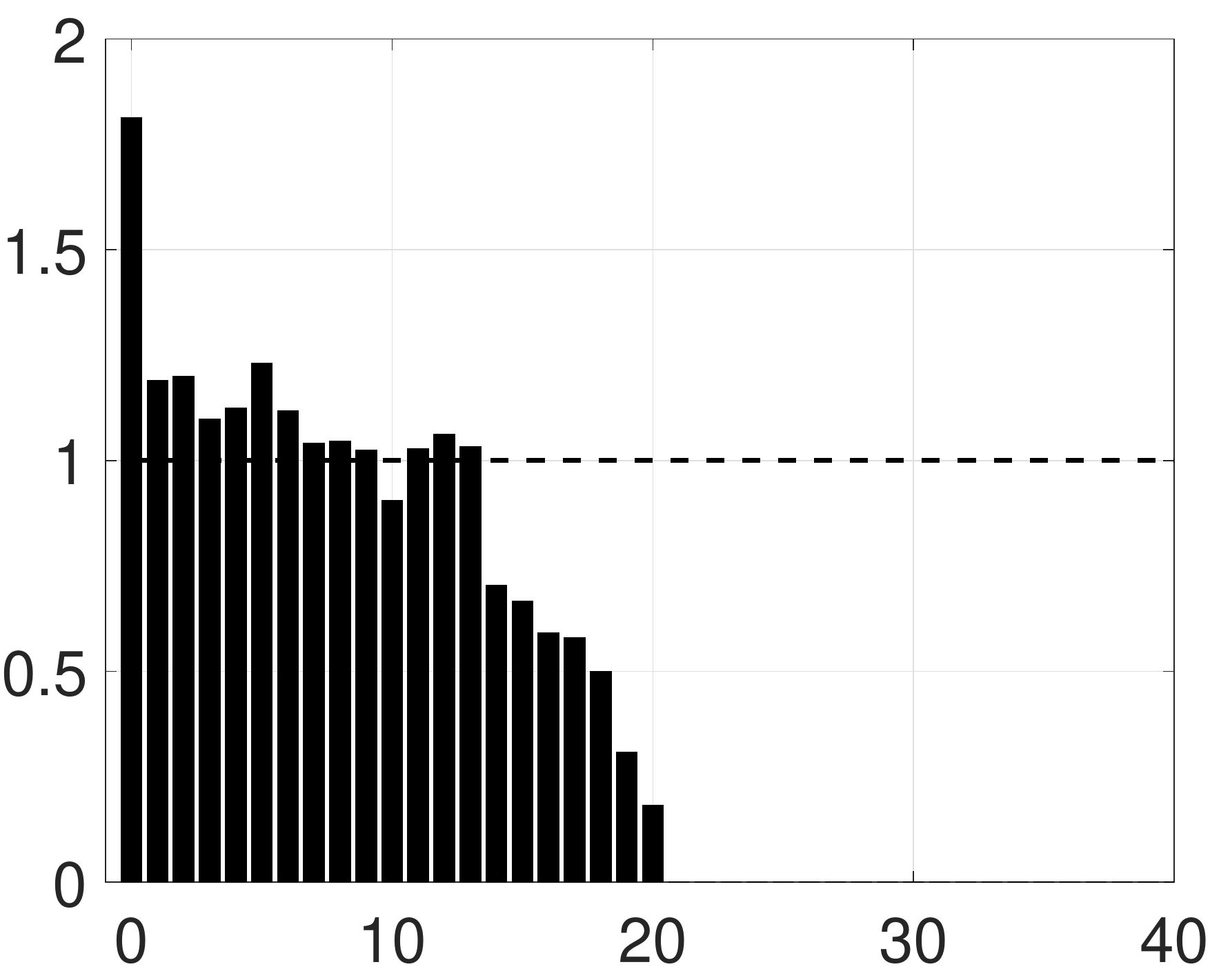}};
\node[rotate=90] at (-1.7,0) {\footnotesize Ratio};
\node[] at (0,-1.4) {\footnotesize Distance};
\node[] at (0,+1.4) {\footnotesize Kiruna--Lule{\aa}};
\end{tikzpicture}
&
\begin{tikzpicture}
\node[] at (0,0) {\includegraphics[width=3cm]{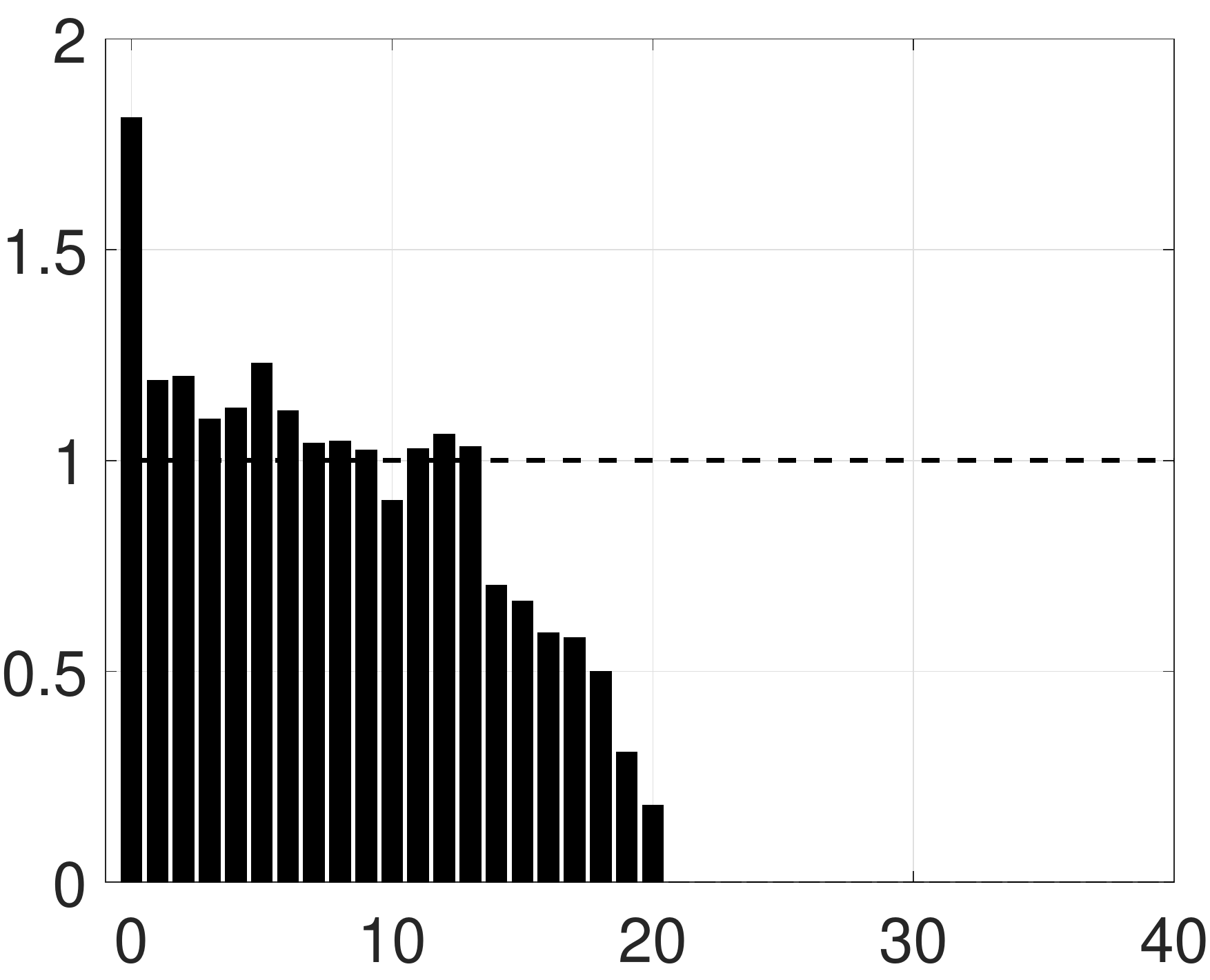}};
\node[rotate=90] at (-1.7,0) {\footnotesize Ratio};
\node[] at (0,-1.4) {\footnotesize Distance};
\node[] at (0,+1.4) {\footnotesize Lule{\aa}--Ume{\aa}};
\end{tikzpicture}
&
\begin{tikzpicture}
\node[] at (0,0) {\includegraphics[width=3cm]{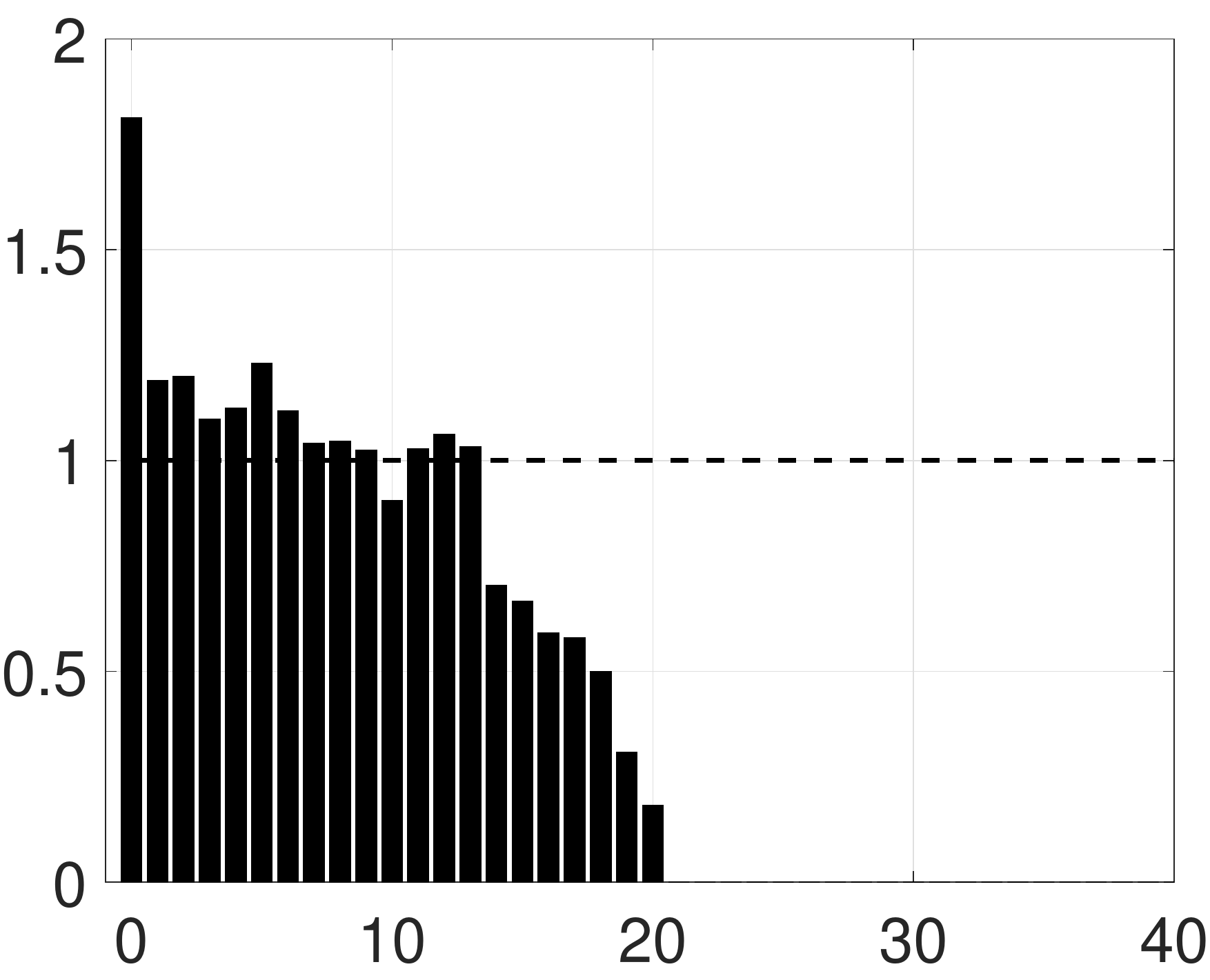}};
\node[rotate=90] at (-1.7,0) {\footnotesize Ratio};
\node[] at (0,-1.4) {\footnotesize Distance};
\node[] at (0,+1.4) {\footnotesize Ume{\aa}--Sundsvall};
\end{tikzpicture}
&
\begin{tikzpicture}
\node[] at (0,0) {\includegraphics[width=3cm]{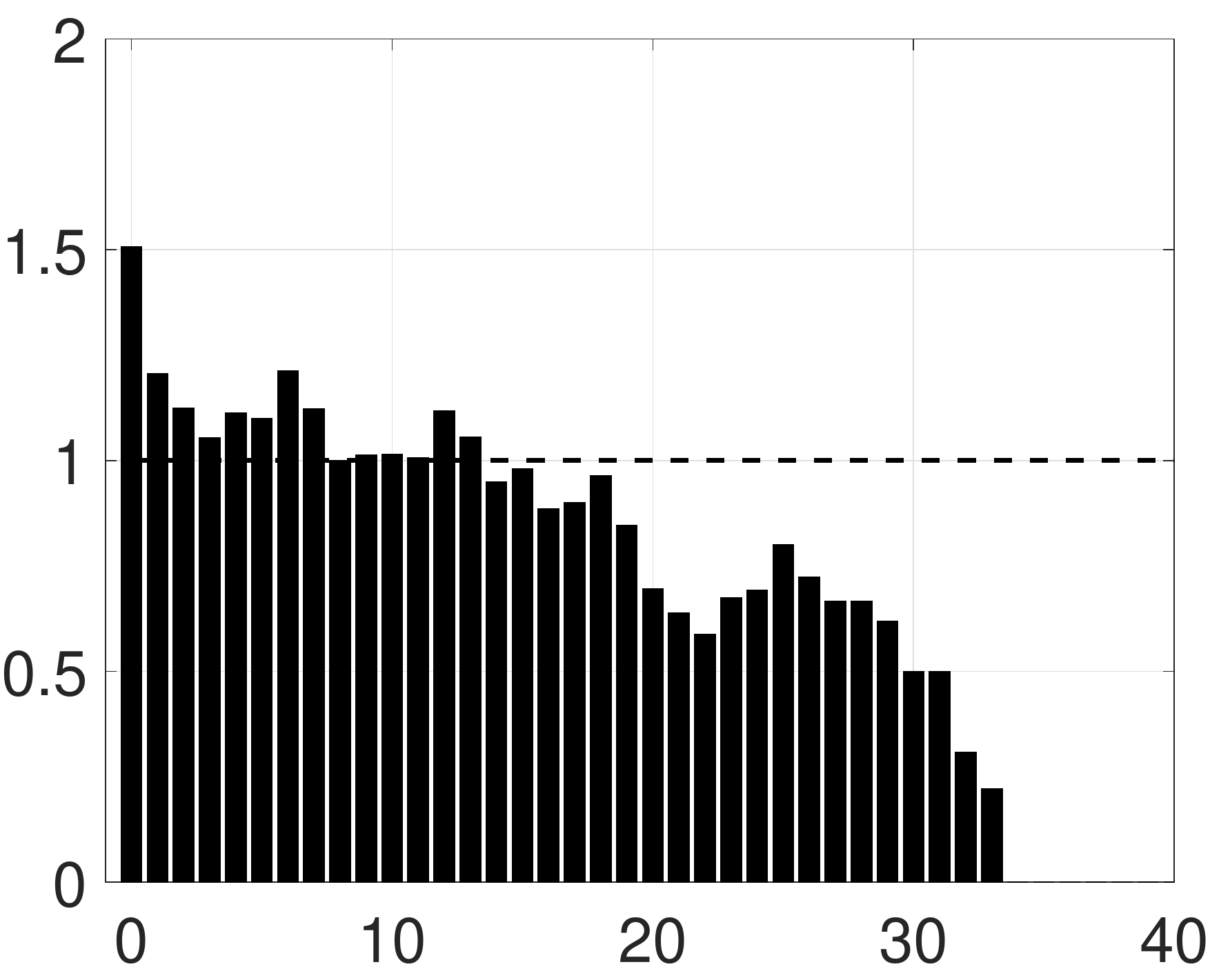}};
\node[rotate=90] at (-1.7,0) {\footnotesize Ratio};
\node[] at (0,-1.4) {\footnotesize Distance};
\node[] at (0,+1.4) {\footnotesize Sundsvall--Uppsala};
\end{tikzpicture}
\\[-.3em]
\begin{tikzpicture}
\node[] at (0,0) {\includegraphics[width=3cm]{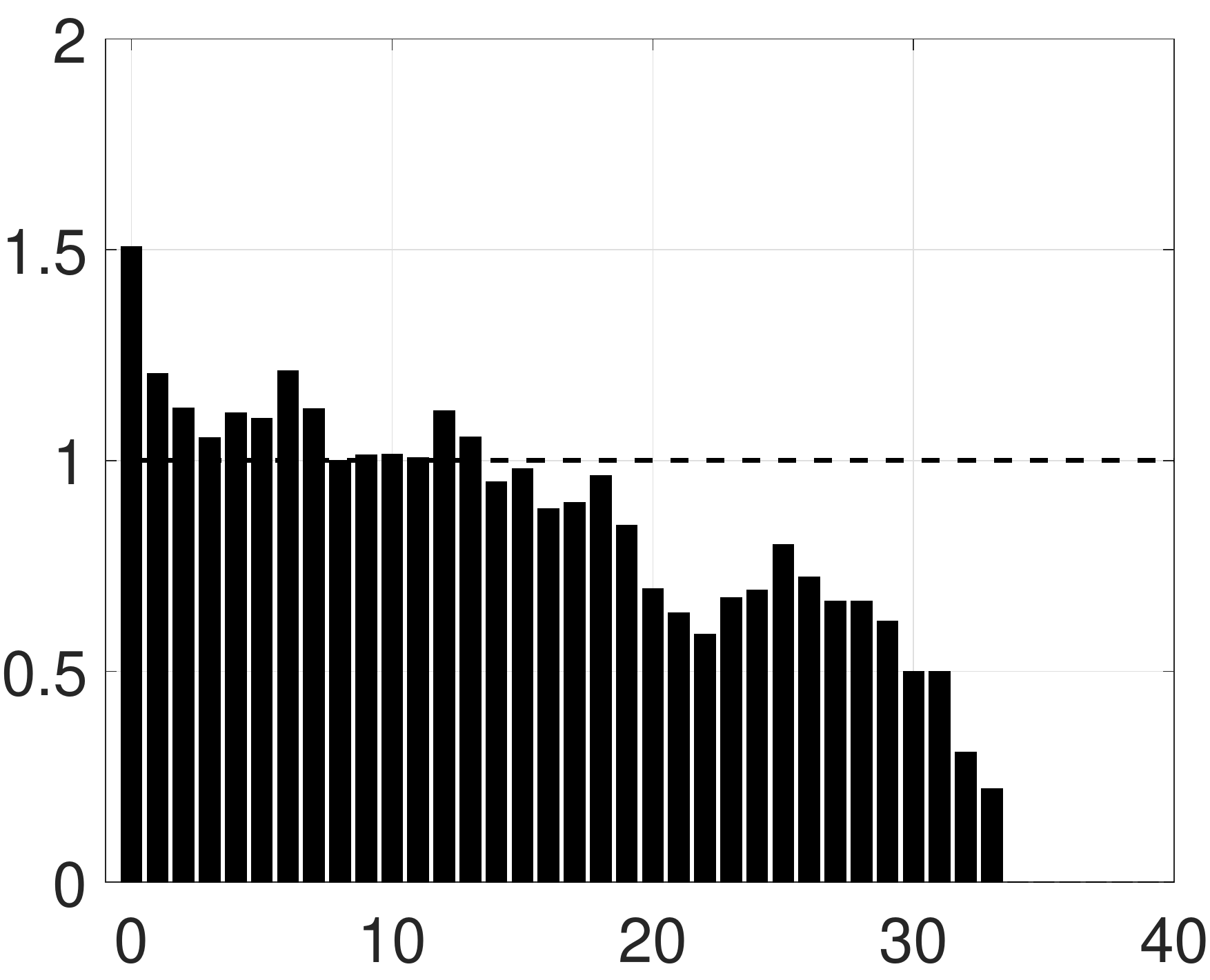}};
\node[rotate=90] at (-1.7,0) {\footnotesize Ratio};
\node[] at (0,-1.4) {\footnotesize Distance};
\node[] at (0,+1.4) {\footnotesize Uppsala--Stockholm};
\end{tikzpicture}
&
\begin{tikzpicture}
\node[] at (0,0) {\includegraphics[width=3cm]{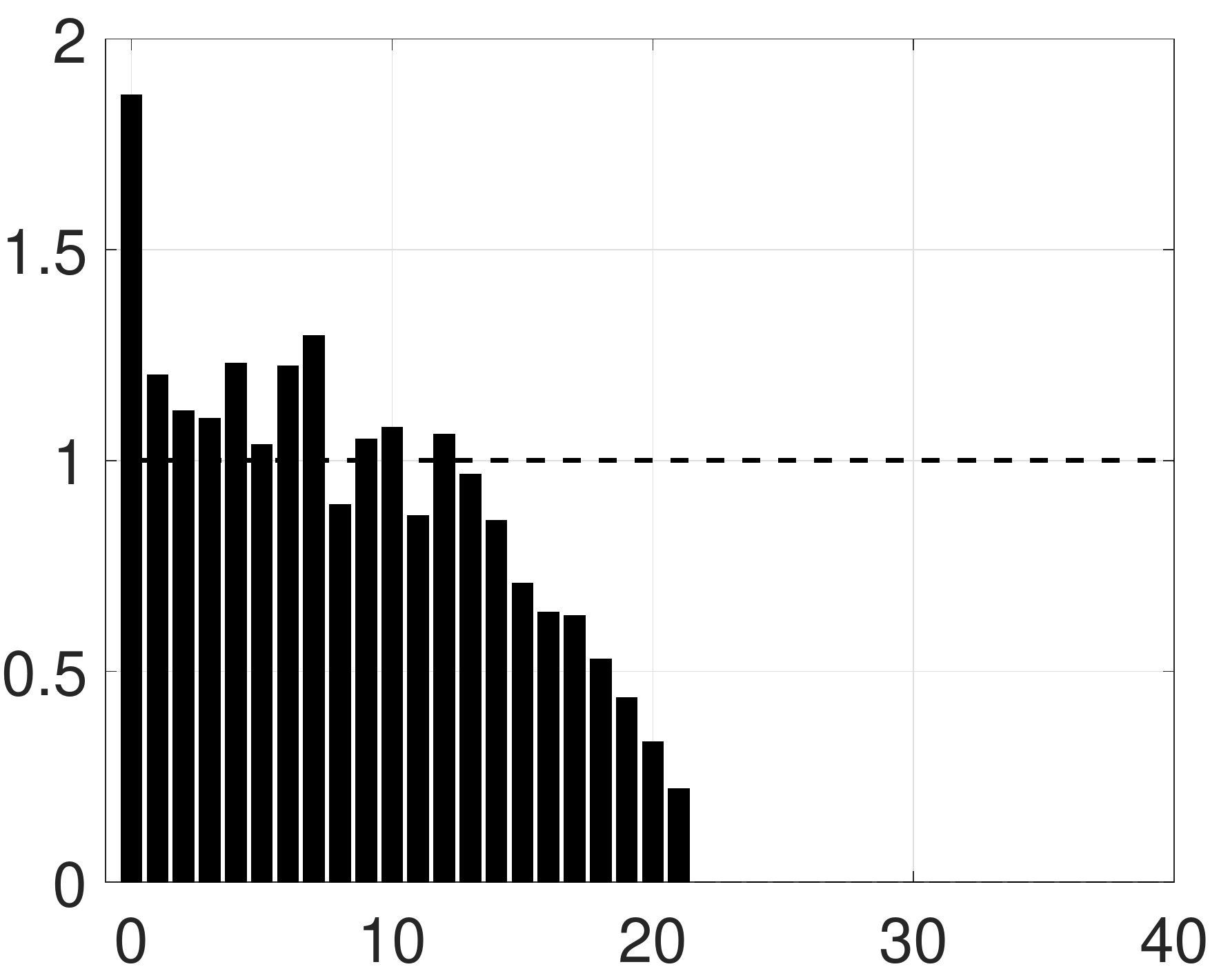}};
\node[rotate=90] at (-1.7,0) {\footnotesize Ratio};
\node[] at (0,-1.4) {\footnotesize Distance};
\node[] at (0,+1.4) {\footnotesize Stockholm--Helsingborg};
\end{tikzpicture}
&
\begin{tikzpicture}
\node[] at (0,0) {\includegraphics[width=3cm]{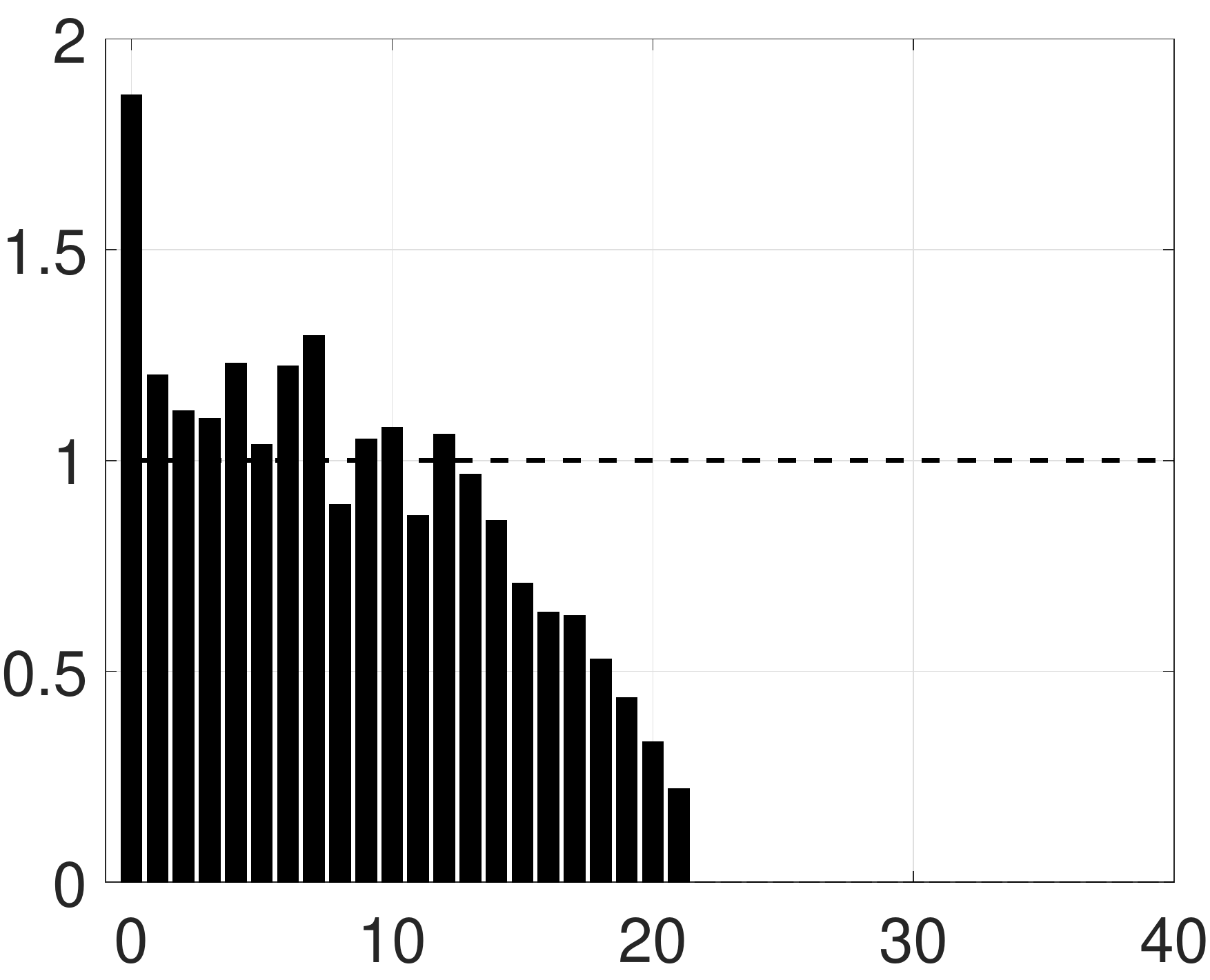}};
\node[rotate=90] at (-1.7,0) {\footnotesize Ratio};
\node[] at (0,-1.4) {\footnotesize Distance};
\node[] at (0,+1.4) {\footnotesize Hesingborg--Malm{\"o}};
\end{tikzpicture}
&
\begin{tikzpicture}
\node[] at (0,0) {\includegraphics[width=3cm]{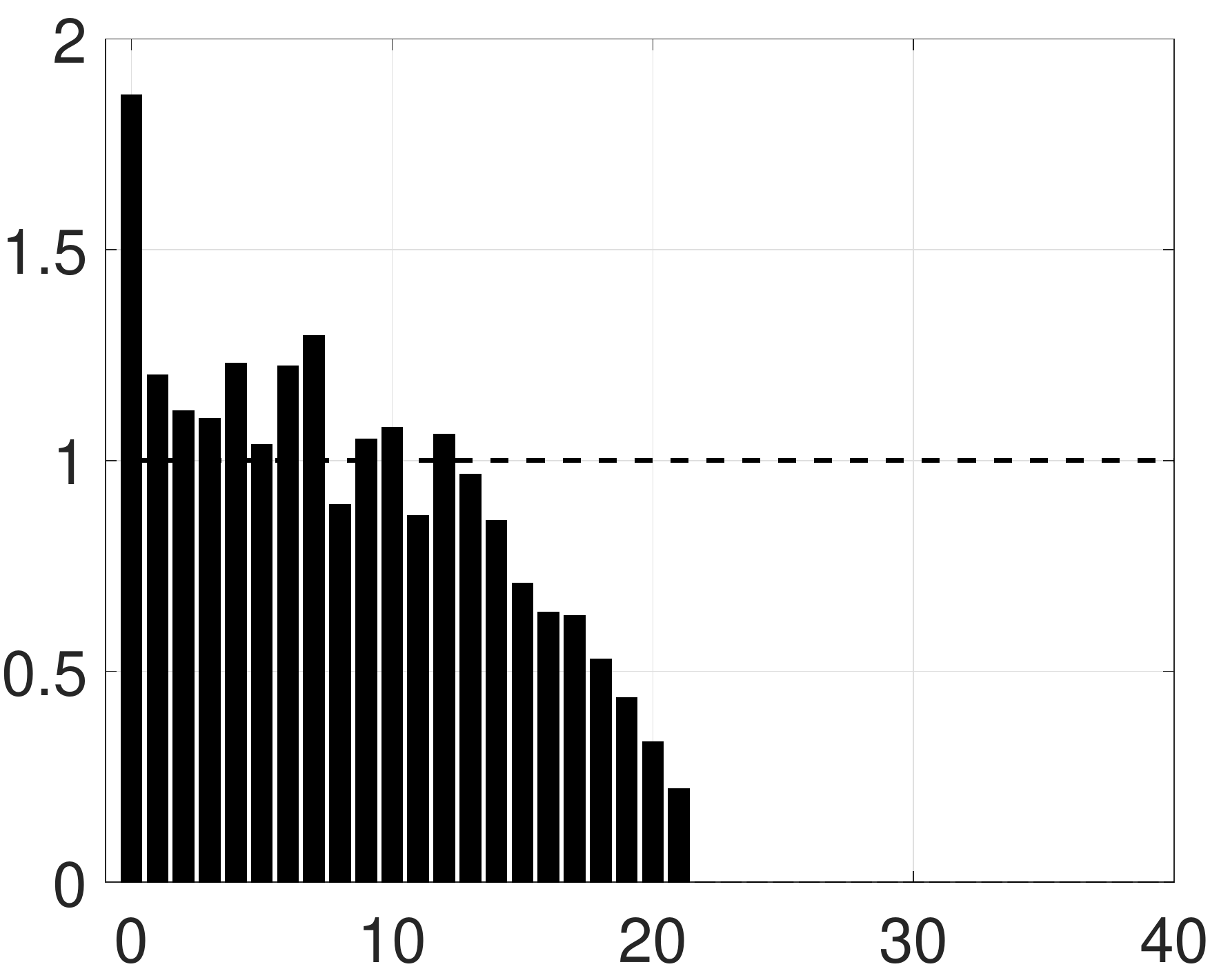}};
\node[rotate=90] at (-1.7,0) {\footnotesize Ratio};
\node[] at (0,-1.4) {\footnotesize Distance};
\node[] at (0,+1.4) {\footnotesize {\"O}stersund--Sundsvall};
\end{tikzpicture}
\end{tabular}
\caption{\label{fig:number_vehicle_road} The ratio of the number of the pairs of vehicles with a given distance for the optimal schedule scaled over the case without any delay (i.e., no scheduling) as a function of the distance. 
}
\end{figure}

\begin{example} We consider $40$ vehicles poised to travel over the  Kiruna--Stockholm path in Sweden; see Figure~\ref{fig:sweden}~(a). The sampling time is selected to be $5\,\mathrm{min}$. The number of time instants that a vehicle stays on an edge of the graph in Figure~\ref{fig:sweden}~(a) is chosen to be proportional to its physical length.
Figure~\ref{fig:sweden}~(b) shows the sequence of the edges used by the vehicles travelling from Kiruna to Stockholm. Their departure time is spread equally between midnight and 2am. In addition, it is assumed that $40$ vehicles are going to travel over {\"O}stersund--Malm{\"o} path (going through Stockholm--Helsingborg connection). 

A similar plan to Figure~\ref{fig:sweden}~(b) can also be devised for these vehicles. The departure time of these vehicles is spread uniformly between 7:00am and 9:00am. Each vehicle can be delayed by at most $15\,\mathrm{min}$. Note that many more vehicles might travel between {\"O}stersund--Malm{\"o} during the day, however, only the ones that depart between 7:00am and 9:00am have a chance to platoon with the vehicles depart from Kiruna between midnight and 2am (due to the short delay constraint and the fact that it takes roughly 10:30 hours to travel over Kiruna--Sundsvall and 2:30 hours to travel over {\"O}stersund--Sundsvall). Let $h_i(\tau_i)=0$ for all $i$ since $15\,\mathrm{min}$ is a relatively short delay (in comparison to the travel time of the paths). Pick $\gamma=1$, $f(z)=z^2$, $\varrho=100$. The weight $w_e$ is also selected to be proportional to the length of the edge $e$ in reality. The distributed algorithm converges relatively fast within roughly 500 iterations (recall that in each iteration the decision of only one of the vehicles can be updated). A pair of vehicles is considered to have distance $d\in\mathbb{N}\cup\{0\}$ from each other on edge $e\in\mathbb{E}$ if they arrive at that edge within $d$ time steps of each other. Evidently, two vehicles that arrive at the same time step have a distance of zero from each other.

Figure~\ref{fig:number_vehicle_road} illustrates the ratio of the number of the pairs of vehicles with a given distance for the optimal schedule (extracted from the convergence point of the suggested algorithm) and the case without any delay (i.e., no scheduling) as a function of the distance. If the ratio is greater than one for a given distance, the number of the pairs of vehicles that have that distance is higher for the optimal schedule. The optimal schedule significantly increases the number of the vehicles with distance zero (noting that the ratio for distance zero is almost two for most of the edges). This greatly improves the chances of platooning as the vehicles with distance zero are only $\pm5\,\mathrm{min}$ away from each other and can easily form platoons. In fact, the \emph{relative} increase in the amount of the average fuel saving facilitated by the increases in the platooning opportunities (assuming a linear increase) under the optimal scheduling in comparison to no scheduling is $68\%$. Note that this improvement is achievable with only $15\,\mathrm{min}$ delays (or correspondingly small change in velocity). By doubling the maximum amount of allowable delay to $30\,\mathrm{min}$, the {relative} increase in amount of the average fuel saving can be pushed to $269\%$, which means that the scheduling service would generate 2--3 times more platooning than the nominal case without the service. 
\end{example}

An important factor, which might keep vehicles from coordinating their platooning, is that fleet owners might be unwilling to share their private business data on vehicle routes and travel times. Such information could reveal their customer base and other competitive information. Therefore, it is of interest to create private and secure match-making services for effective coordination among competing entities to facilitate the adoption of truck platooning as described above. Such a cooperation service under privacy constraints is discussed next. 

\subsection{Cooperation under privacy constraints}
In this section, a private framework for match-making among vehicles traveling across a freight transportation network is developed. First, however, a few definitions and results from cryptography are introduced.

Let the tuple $(\mathbb{P},\mathbb{C},\mathbb{K},\mathfrak{E},\mathfrak{D})$ denote an arbitrary public key encryption technique, where $\mathbb{P}$ is the set of plaintexts, $\mathbb{C}$ is the set of ciphertexts, $\mathbb{K}$ the set of keys, $\mathfrak{E}$ is the encryption algorithm, and $\mathfrak{D}$ is the decryption algorithm. Each $k=(k_p,k_s)\in\mathbb{K}$ is composed of a public key $k_p$ (shared with and used by everyone for encrypting a plaintext) and a private key $k_s$ (maintained only by trusted parties for decryption purposes). Most often, the algorithms $\mathfrak{E}$ and $\mathfrak{D}$ are publicly known while the keys are generated and appropriately used for a specific case. Note the use of the word algorithm, instead of mapping or function, is to emphasize that the possibility of the presence of random or pseudo-random elements in the encryption procedure, which might results in a certain plaintext being mapped to multiple ciphertexts. A necessary requirement of the encryption algorithm is to be invertible, i.e., $\mathfrak{D}(\mathfrak{E}(x,k_p),k_s)=x$ for all allowable plaintexts $x\in\mathbb{P}$ given a key $k=(k_p,k_s)\in\mathbb{K}$. In what follows, it is assumed that there exists operators $\circ$ and $\diamond$ such that $(\mathbb{P},\circ)$ and $(\mathbb{C},\diamond)$ form two groups. 

\begin{Definition}[Homomorphic encryption] A public key encryption $(\mathbb{P},\mathbb{C},\mathbb{K},\mathfrak{E},\mathfrak{D})$ is called homomorphic if for all $x_1,x_2\in\mathbb{P}$ and $k\in\mathbb{K}$, $\mathfrak{E}(x_1,k_p)\diamond\mathfrak{E}(x_2,k_p)=\mathfrak{E}(x_1\circ x_2,k_p)$.
\end{Definition}

In most common encryption algorithms, $\mathbb{P}$ and $\mathbb{C}$ are, respectively, rings of integers $\mathbb{Z}_{n_p}$ and $\mathbb{Z}_{n_c}$, where $\mathbb{Z}_q:=\{0,\dots,q-1\}$ for all positive integers $q\in\mathbb{N}$, $n_p=|\mathbb{P}|$, and $n_c=|\mathbb{C}|$. If $x_1\circ x_2:=(x_1+x_2)\,\modd\,n_p$ and the public key encryption is homomorphic, it is referred to as additively homomorphic. In this case, let $\oplus$ denotes the equivalent operator in the ciphertext domain. On the other hand, if $x_1\circ x_2:=(x_1\times x_2)\,\modd\,n_p$ and the public key encryption is homomorphic, it is referred to as multiplicatively homomorphic. In this case, let $\otimes$ denotes the equivalent operator in the ciphertext domain. If a public key encryption is both additively homomorphic and multiplicatively homomorphic, it is called fully homomorphic. However, if only one of these conditions is satisfied, it is referred to as semi-homomorphic. Homomorphism shows that operations, such as summation and multiplication, can be performed on the plaintexts without the need for decryption of their corresponding ciphertexts. An example of additively homomorphic encryption is the Paillier's encryption method~\cite{Paillier1999}. ElGamal and RSA are examples of multiplicatively homomorphic encryption~\cite{1057074,rivest1978method}. Recently, several fully homomorphic encryption methods have been introduced, e.g.,~\cite{vanDijk2010}. In what follows, we consider public key encryption methods that are at least additively homomorphic; however, extension to multiplicatively homomorphic is straightforward and can be achieved with the same line of reasoning.

We consider the case where vehicle $i$ is selected to update its decision according to~\eqref{eqn:distributedupdate}. To do so, it needs to compute the value of the function $g(\tau_i,\tau_{-i}[\beta])$. Therefore, it needs to inquire how many vehicles are using edges $e$ at time instant $t$. Define $\zeta_{t,e}[\beta]:=|\{j\neq i\,|\,e_j(t-\tau_j[\beta])=e\}|$. Evidently, $|\{j\,|\,e_j(t-\tau_j[\beta])=e\}|=\zeta_{t,e}[\beta]+1$ if  $e_i(t-\tau_i)=e$ and  $|\{j\,|\,e_j(t-\tau_j[\beta])=e\}|=\zeta_{t,e}[\beta]$ otherwise. 
As a result, $g(\tau_i,\tau_{-i}[\beta])=-\gamma \sum_{t\in\mathbb{T}}\sum_{e\in\mathbb{E}} f(\zeta_{t,e}[\beta]+\mathds{1}_{e_i(t-\tau_i)=e})$. Therefore, all that is required for vehicle $i$ is to compute $\zeta_{t,e}[\beta]$ for all $t$ and $e$ in a private and secure manner. How to do that is discussed next. Without loss of generality, it can be assumed that $i=1$ (up to renumbering of the vehicles).

The dilemma is that none of the vehicle wishes to let the other vehicles know the edges over which it is travelling due to the competitive nature of the fleet owners or the privacy constraints imposed by the end user. The privacy is meant to be kept until a match is identified, after which the vehicles are happy to cooperate. 
Consider the case where the vehicle is interested to compute $\zeta_{t,e}[\beta]$ for a given edge $e$ and time instant $t$. Let $\sigma:\mathbb{E}\rightarrow\{1,\dots,|\mathbb{E}|\}$ be a universally known labelling mechanism for the edges (e.g., it could be agreed upon in advance using OpenStreetMaps). The first vehicle selects the key $k=(k_p,k_s)\in\mathbb{P}$ and constructs the encrypted matrix $x^{(1)}\in\mathbb{C}^{|\mathbb{E}|\times |\mathbb{T}|}$ such that elements of $x^{(1)}$ are given by $x^{(1)}_{t,\sigma(e)}=\mathfrak{E}(0,k_p)$. The first vehicle then transmits $x^{(1)}$ to the second vehicle along with the public key $k_p$. The second vehicle computes the matrix $x^{(2)}$, such that $x^{(2)}_{t,\sigma(e)}=x^{(1)}_{t,\sigma(e)}\oplus \mathfrak{E}(\mathds{1}_{e_2(t-\tau_2[\beta])=e},k_p)$, and transmits $x^{(2)}$ and $k_p$ to the third vehicle. For all $1<i<n$, vehicle $i$ computes the matrix $x^{(i)}$, such that $x^{(i)}_{t,\sigma(e)}=x^{(i-1)}_{t,\sigma(e)}\oplus \mathfrak{E}(\mathds{1}_{e_i(t-\tau_i[\beta])=e},k_p)$, and transmits $x^{(i)}$ and $k_p$ to the $(i+1)$-th vehicle. Finally, the $n$-th vehicle computes the matrix $x^{(n)}$, such that $x^{(n)}_{t,\sigma(e)}=x^{(n-1)}_{t,\sigma(e)}\oplus \mathfrak{E}(\mathds{1}_{e_n(t-\tau_n[\beta])=e},k_p)$, and transmits it to the first vehicle. Therefore, it can be deduced that
\begin{align*}
\mathfrak{D}(x^{(n)}_{t,\sigma(e)},k_s)=\mathfrak{D}(x^{(n)}_{t,\sigma(e)}\oplus x^{(n-1)}_{t,\sigma(e)}\oplus \cdots \oplus x^{(1)}_{t,\sigma(e)}) =\mathfrak{D}\bigg(\mathfrak{E}\bigg(\sum_{j=2}^n\mathds{1}_{e=e_j(t-\tau_j[\beta])}\bigg),k_s\bigg)
=\mathfrak{D}(\mathfrak{E}(\zeta_{n,e}[\beta],k_p),k_s)
=\zeta_{n,e}[\beta].
\end{align*}



Semantic security (e.g.,~\cite{Ostrovsky2007}) provides strong security and privacy guarantees for the vehicles using the encrypted platform above for communication. If the public key encryption $(\mathbb{P},\mathbb{C},\mathbb{K},\mathfrak{E},\mathfrak{D})$ is semantically secure, the probability that probabilistic polynomial time-bounded adversary (other than the first vehicle) correctly guesses the transportation assignments of the vehicles based on the communicated messages among the vehicles is inversely proportional to the number of the eligible paths on the freight transportation network plus a negligible function of the key length $|\mathbb{K}|$. This implies that if the key length is selected properly, a computationally constrained adversary cannot do any better than purely guessing which roads are being utilized by the vehicles. It is worth mentioning that, among the introduced homomorphic encryption techniques, the Paillier's encryption and ElGamal are semantically secure~\cite{Paillier1999,1057074} while RSA is not~\cite{boneh1999twenty}. The privacy and security guarantees are however slightly weaker if the adversary is the first vehicle. This is because, by construction, the first vehicle owner can successfully determine if the other vehicles (albeit in a collective manner, not individually) operate on various edges at different time instants. Therefore, some private information is leaked even in the best of situations. In addition, it can be shown that the vehicle can potentially extract more information by not following description of the vector $x^{(1)}$~\cite{farokhi2017private}.  However, in~\cite{farokhi2017private}, it is also proved that the first vehicle, even using sophisticated attacks, can extract a limited amount of information about the collection of the other vehicles. Noting that the vehicles themselves are not identifiable from the procedure, the reverse privacy and security guarantees are also acceptable in this case.


\section{Incentives Created by Cordon Pricing}
\label{sec:case}

One possibility to enforce cooperation according to the framework introduced in Figure~\ref{fig:framework} is to create user incentives through congestion or \emph{cordon pricing.} 
It has been implemented in three major cities: Singapore, London, and Stockholm. It is designed to reduce vehicle presence in a central and typically congested area of a city, by imposing charges on vehicles entering or exiting a designated area (cordon). The first city to introduce cordon pricing was Singapore in 1975 as daily charges for entering the city center; fully automated electronic charging system was completed in 1998, in particular relying on the pioneering ERP system of Singapore. London introduced automated cordon pricing in 2003, whereas such a system was completed and made permanent for Stockholm in 2006. A comprehensive assessment of the ten year impact of cordon pricing on the mobility patterns in Stockholm was recently completed~\cite{Borjesson}. The system is essential for Stockholm to succeed in becoming a completely fossil-free city by 2040~\cite{sthlm2040}. 
\begin{figure}[t]
\centering
\includegraphics[width=0.9\linewidth]{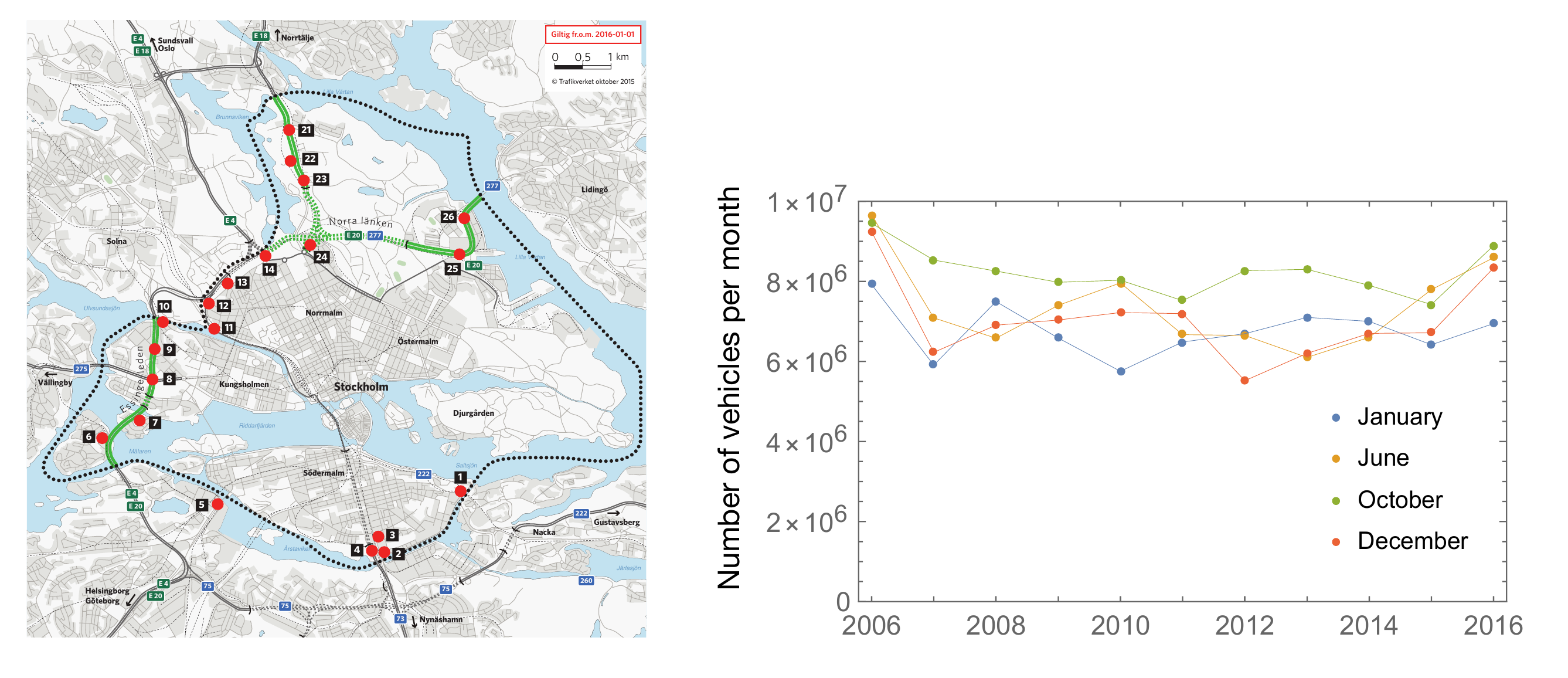}
\caption{\label{fig:FigureX} The cordon boundary for the city of Stockholm (left) with numbered red dots marking the locations at which vehicles are being charged~\cite{TrafikverketKARTA}. The dynamics of vehicle passage for four selected months (right) shown for ten years of cordon pricing operation~\cite{TrafikverketSTATISTICS}. The values for 2006 are averages for respective month before the cordon pricing system was in operation.} 
\end{figure}

The cordon in the city of Stockholm is shown in Figure~\ref{fig:FigureX} where the control points are the numbered red symbols \cite{TrafikverketKARTA}. These consist of gantries on which vehicle identification cameras and other hardware are installed. Most of the vehicles entering or leaving the cordon are charged by a fee that is dependent on the time the day. Charges are implemented on working days only with a maximum charge of about 3.25\euro\ for the morning and afternoon peak hours. The system is fully automated and identifies vehicles by registration plates. 
The time evolution vehicle passages is shown in Figure~\ref{fig:FigureX} for four selected months distributed over a ten year period \cite{TrafikverketSTATISTICS}. The initial value (January 2006) corresponds to the monthly average before cordon pricing was implemented. As can be seen in Figure~\ref{fig:FigureX}, there are systematic differences between for instance January and June, with June being the month with highest number of passages. Although the initial reduction of vehicle passages was substantial (around 20\%), the number of passages has remained relatively steady since, showing a slight increase during 2016 in spite of the fact that the charge was increased from January 2016. Thorough analysis of the ten year data on the effect of cordon pricing vs. costs using a variety of measures, indicates that an expansion of the cordon in Stockholm, increase in the charge, or introduction of the system to other Swedish cities, would not be justified~\cite{Borjesson}. Data show that in spite of a higher charge, passage during peak hours is little affected by cordon pricing, while the low-charge off-peak hours are being more affected. The impact of the cordon pricing could perhaps be enhanced if the charge level was continuously adjusted depending on traffic flow conditions at different parts of the city or cordon, similar to the ``dynamic charging'' in Singapore; no such development of the Stockholm system is currently being planed.   

The experience of Stockholm \cite{Borjesson} as well as of London \cite{LondonStalling} shows the limitation of a non-interactive, gantry-based cordon pricing system as a city-level instrument for achieving better shared use of the road infrastructure. The Singapore  system is currently being transformed into a new gantry-free and highly interactive system~\cite{LTA}. The new system is based on positioning from the Global Navigation Satellite System and is expected to be operational from 2020. An on-board unit with a SIM-card will enable personalised portals for each driver or vehicle. All motorised vehicles will effectively be tracked with full anonymity guaranteed. The system will be capable of monitoring driving speeds, providing the basis for parking charges, distance-based charging, charging based on real-time congestion, dynamic pricing, as well as combining distance and congestion charging. It is also expected to facilitate the development of diverse apps, creating an added-value, rich ecosystem of transportation-related services to both drivers and the transportation planning authority. Intelligent transportation with congestion pricing is likely to evolve from novel combinations of top-down strategies with bottom-up choices by participating users. Physical constraints for traffic as flow on networks where drivers have comparatively few choices due to limited information~\cite{garavello2006traffic,garavello2012cauchy} need to be combined with application-based systems where individual drivers have significantly broader choices based on real-time processing of vast amounts of information on traffic conditions, congestion pricing, minimal  transportation times, etc. To incorporate such choices in a predictive manner, the behavior of drivers as human-social agents who do not necessarily make rational choices, becomes increasingly important~\cite{Leu:2012}. 

\section{Conclusions}
\label{sec:concl}
We have presented a novel framework to model and analyze how mobility management services influence traffic patterns. We reasoned about how specific information structures imposed by these services can have a profound impact on the traffic behavior. Non-local PDE models were introduced to illustrate the generality of the framework, and in particular its ability to integrate models with non trivial mathematical features.  It was shown how the approach can be used to handle time-dependent traffic flows over complex networks and how they allow the integration of available routing information as splits at the junctions. Based on the models, traffic behaviors for routed and non-routed users were discussed in detail and some basic principles for how the users act were discussed. The framework proposed in the article can adjust easily to more complex models, and in particular the problems of networked LWR PDEs will be of interest, as it presents open issues with complexity going beyond this specific article (in particular incorporating weak boundary conditions).
As an illustration for how parts of the introduced framework can already be used, we presented a cooperative freight platooning service, where freight flows follow similar principles as routed users but with the   objective of maximizing fuel-saving platooning, instead of minimizing travel time. As part of this freight service, we showed how the continuous traffic flow models introduced in this article can be used to study the formation of platoons on a single road by controlling the truck flow velocity over that road. Over a large freight transport network, we have shown how the continuous model can be linked to a discrete model to optimize the coordination of trucks and their average velocities or departure times. Finally, we presented the cordon pricing system of Stockholm and related that to the social planning layer of our framework. 

For future work, some main directions can be identified. The framework for the analysis and design of mobility services outlined in this article does not yet contain all details of the underlying mathematical models or synthesis tools for the development of actual services. We believe that our proposal is a first step in a major future research thrust. For example, additional theoretical well-posedness results are needed for the introduced traffic flow models when built on arbitrary networks. Proofs of existence and uniqueness of solutions to the non-local PDEs have to be detailed for such cases, resulting in a better understanding of the properties of the model and how to implement and calibrate it appropriately. The general framework we have presented can also be used with other underlying traffic flow models, such as the Merchant-Nemhauser model based on ordinary differential equations or the link-delay model. It would be interesting to make a comparison study about the strengths and weaknesses of these models and our PDE models when used for the mobility service framework. In addition to simplified illustration of a freight mobility service, it would be interesting to study a case with real data and perhaps even a practical implementation and evaluation. Such experiments will allow us to determine which of the introduced principles for routed and non-routed traffic flows is more prominent and how these routing principles evolve over time. 

\section*{Acknowledgements}
This research was partially supported by the NSF grant 1239166, ``CPS: Frontiers:
Collaborative Research: Foundations of Resilient Cyber-Physical Systems.'' A. Keimer was funded by the program ''FitWeltweit`` of the DAAD. N. Laurent-Brouty was supported by Minist\`ere de l'Environnement, de l'\'Energie et de la Mer, France. F. Farokhi was supported by McKenzie Fellowship from the University of Melbourne. V. Cvetkovic and K. H. Johansson were supported by VINNOVA through Digital Demo Stockholm. K. H. Johansson was also supported by the Knut and Alice Wallenberg Foundation, the Swedish Strategic Research Foundation, and the Swedish Research Council.
\bibliographystyle{IEEEtran}

\end{document}